\documentstyle[epsfig]{elsart}

\begin{document}
\begin{frontmatter}
\title{Predictability: a way to characterize Complexity}
\author{G.Boffetta $^{(a)}$, M. Cencini $^{(b,c)}$, 
M. Falcioni$^{(c)}$,  and A. Vulpiani $^{(c)}$}
\address{$^{(a)}$~Dipartimento di Fisica Generale, Universit\`a di Torino,\\
Via Pietro Giuria 1, I-10125 Torino, Italy \\and 
Istituto Nazionale Fisica della Materia, Unit\`a dell'Universit\`a di Torino}
\address{$^{(b)}$~Max-Planck-Institut f\"ur Physik komplexer Systeme, 
N\"othnitzer Str. 38  , 01187 Dresden, Germany}
\address{$^{(c)}$~Dipartimento di Fisica, Universit\`a di  Roma
"la Sapienza",\\ Piazzale Aldo Moro 5, 00185 Roma, Italy \\and
Istituto Nazionale Fisica della Materia, Unit\`a di Roma 1}

\vskip 1.5cm
\tableofcontents
\begin{keyword}
\end{keyword}
\begin{abstract}
\noindent
Different aspects of the predictability problem in dynamical systems
are reviewed. The deep relation among Lyapunov exponents,
Kolmogorov-Sinai entropy, Shannon entropy and algorithmic complexity
is discussed. In particular, we emphasize how a characterization of
the unpredictability of a system gives a measure of its complexity.
Adopting this point of view, we review some developments in 
the characterization of the predictability of systems showing different 
kind of complexity: from low-dimensional systems to high-dimensional
ones with spatio-temporal chaos and to fully developed turbulence.
A special attention is devoted to finite-time and finite-resolution effects
on predictability, which can be accounted with suitable generalization 
of the standard indicators. The problems involved in systems 
with intrinsic randomness is discussed, with emphasis on the important 
problems of distinguishing chaos from noise and of modeling    
the system. The characterization of irregular behavior in 
systems with discrete phase space is also considered.

\noindent
PACS numbers: 
\end{abstract}

\end{frontmatter}

This review is dedicated to our masters and to our friends; in particular to
Andrei N. Kolmogorov (1903-1987) and Giovanni Paladin (1958-1996)
whose influence on our work runs deeper than we can know. \\

{\it All the simple systems are simple in the same way, each complex
system has its own complexity} \\ (freely inspired by {\it Anna Karenina}
by Lev N. Tolstoy)
\newpage
\section{Introduction}
\label{intro}
\setcounter{equation}{0}
The ability to predict the future state of a system, given the present
one, stands at the foundations of scientific knowledge with relevant
implications from a conceptual and applicative point of view.  The
knowledge of the evolution law of the system may induce one to
conclude that this aim has been attained. This is the classical
deterministic point of view as clearly stated by Laplace \cite{L1814}: once the
evolution laws of the system are known, the state at a certain time
$t_0$ completely determines the subsequent states for every time $t >
t_0$\footnote{In this review we shall always consider the usual setting where
a system is studied by an external observer, so as to avoid the problem of the
self-prediction \cite{Popper}.}. 
However it is well established now that this cannot be
accomplished in practice.

One limitation occurs in systems with many degrees of freedom, namely
the impossibility to manage the huge amount of data required for a
detailed description of a single state of a macroscopic body. This
aspect, which is not discussed in this review, has led to the
development of statistical mechanics.

Another source of difficulty, which arises even in low dimensional
systems, is related to the unavoidable uncertainty in the initial
condition.  As clearly stated by Poincar\'e, this implies that one can
make long-time predictions only if the evolution law does not amplify
the initial uncertainty too rapidly.  This aspect had a relevant role
in the development of the theory of dynamical chaos.

Therefore, from the point of view of predictability, we need to know
how an error in the initial state of the system grows in time.  In
deterministic chaotic systems, i.e., with sensitive dependence on
initial condition, one has an exponential growth of errors and,
consequently, a severe limitation on the ability to predict the future
states.  In addition, since the details of the evolution laws are not
completely known (or, at least, cannot be specified with an arbitrary
accuracy) or some degrees of freedom cannot be resolved, one has
another unavoidable source of unpredictability. This is also true 
for systems with discrete states.

A branch of the theory of dynamical systems has been developed with
the aim of formalizing and quantitatively characterizing the
sensitivity to initial conditions. The Lyapunov exponent and the
Kolmogorov-Sinai entropy are the two indicators for measuring the rate
of error growth and information produced by the dynamical system.  A
complementary approach has been developed in the context of
information theory, data compression and algorithmic complexity
theory.  Nowadays it is rather clear that the latter point of view is
closely related to the dynamical systems one.  If a system is chaotic
then the predictability is limited up to a time which is related to
the first Lyapunov exponent, and the time sequence generated from one of
its chaotic trajectories cannot be compressed by an arbitrary factor,
i.e. is algorithmically complex. On the contrary, a regular
trajectory can be easily compressed (e.g., for a periodic trajectory it is
sufficient to have the sequence for a period) so it is ``simple''.

In this review we will discuss how these points of view are related and
how they complete each other in giving a quantitative understanding
of complexity arising in dynamical systems. In particular, we
shall consider the extension of this approach, nowadays well
established in the context of low dimensional systems and for
asymptotic regimes, to high dimensional systems with attention to
situations far from asymptotic (i.e. finite time and finite
observational resolution).

It is worth remarking that the approach to complexity here
discussed is not able to cover the many aspects of what in recent
years has been indicated under this term \cite{BP97}. Indeed {\it
complexity} has been considered in many different fields of science
and its meaning has become (sometimes) vague.  A challenging problem
in this direction is the definition of indicators which are able to
fulfill the intuitive idea of complexity, namely one looks for
quantities which give a low complexity value both for pure random
sequences and completely regular ones \cite{G86}.  Even if very
interesting this last issue is not addressed here: from the point of
view of predictability both a chaotic system and a purely random
one are highly complex, i.e. unpredictable.

The review is organized as follows.  Sect.~\ref{sec:1} is devoted to
the introduction of the basic concepts and ideas of dynamical systems,
information theory and algorithmic complexity. In particular, we
discuss the relations among Lyapunov exponents, Kolmogorov-Sinai
entropy and algorithmic complexity and their relevance for
predictability. All these quantities are properly defined only in
specific asymptotic limits, that are: very long times and arbitrary
accuracy.  Since in realistic situations one has to deal with finite
accuracy and finite time --- as Keynes said, ``in the long run we
shall all be dead'' --- it is appealing to treat the predictability
problem by taking into account these limitations. This is
the subject of Sect.~\ref{sec:2} where, relaxing the request of
infinite time, we discuss the relevance of the finite time
fluctuations of the ``effective'' Lyapunov exponent. In addition,
relaxing the limit of infinitesimal perturbations, we introduce
suitable tools, such as the Finite Size Lyapunov Exponent (FSLE) and the
$\epsilon$-entropy, for the treatment of non arbitrary high accuracy,
i.e.  non infinitesimal perturbations.

Sects.~\ref{sec:3} and \ref{sec:4} focus on high dimensional dynamical
systems which deserve particular attention. Indeed because of the many
degrees of freedom, and its interest in applications (e.g. in weather
forecasting), it is necessary to consider the detailed behavior of
perturbations and not only the asymptotic features (i.e. long time and
infinitesimal amplitudes). Sect.~\ref{sec:4} is devoted to fully
developed turbulence (here introduced as an important prototype of
high dimensional system) and its finite resolution properties in the
inertial range.

In Sect.~\ref{sec:5} we consider the effects of uncertainties on the
evolution laws and we discuss systems containing some randomness.  In
such a situation there are two ways to approach the predictability: by
considering either two trajectories generated with the same
realization of randomness, or two trajectories evolving with different
realizations.  Both approaches are physically relevant in different
contexts, and the results can be very different in presence of strong
intermittency.

For the sake of completeness in Sect.~\ref{sec:6} we discuss dynamical
systems with discrete states, e.g., Cellular Automata. 

Sect.~\ref{sec:7} is dedicated to a discussion on data analysis. In
particular we discuss the use of $\epsilon$-entropy and FSLE for a
pragmatic classification of signals.

Sect.~\ref{sec:8} reports some concluding remarks.
In the Appendices we discuss some more technical details.
\section{ Two points of view}
\setcounter{equation}{0}
\label{sec:1}
\subsection{ Dynamical systems approach}
\label{sec:1.1}

Two standard -- tightly linked -- indicators are largely used to
quantify the behavior of a dynamical system with respect to the
asymptotic evolution of an infinitesimal uncertainty: the largest
Lyapunov exponent (LE) and the Kolmogorov-Sinai (or metric) entropy 
\cite{ER85}.

\subsubsection{ Characteristic Lyapunov exponents}
\label{sec:1.1.1}
The characteristic Lyapunov exponents are somehow an extension of
the linear stability analysis to the case of aperiodic motions.
Roughly speaking, they measure the typical rate of exponential
divergence of nearby trajectories. In this sense they give information
on the rate of growth of a very small error on the initial state of a
system.

Consider a dynamical system with an evolution law given, in the case
of continuous time, by the differential equation
\begin{equation}
\label{eq:1-1}
{ d {\bf x} \over dt } =  {\bf F} ({\bf x}),
\end{equation}
or, in the case of discrete time, by the map 
\begin{equation}
\label{eq:1-2}
{\bf x} (t+1)  =  {\bf G} ({\bf x}(t))\,.
\end{equation}
In both cases, for simplicity, we suppose that a vector ${\bf x} \in
{\mathrm I\!R}^d$ uniquely specifies one state of the system.  We also
assume that ${\bf F}$ and ${\bf G}$ are differentiable functions, that
the evolution is well-defined for time intervals of arbitrary
extension, and that the motion takes place in a bounded region of the
phase space.  We intend to study the separation between two
trajectories, ${\bf x}(t)$ and ${\bf x}'(t)$, starting from two close
initial conditions, ${\bf x}(0)$ and ${\bf x}'(0)= {\bf x}(0) + \delta
{\bf x}(0)$, respectively.

As long as the difference between the trajectories, $\delta {\bf x}(t)
= {\bf x}'(t) - {\bf x}(t)$, remains small (infinitesimal, strictly
speaking), it can be regarded as a vector, ${\bf z}(t)$, in the
tangent space. The time evolution of ${\bf z}(t)$ is given by the
linearized differential equations:
\begin{equation}
\label{eq:1-3}
   {d z_i(t)\over d t}= \sum_{j=1}^d\,
   \left. {\partial F_i  \over \partial x_j} 
   \right|_{ {\bf x}(t)}\, z_j(t)  
\end{equation}
or, in the case of discrete time maps: 
\begin{equation}
\label{eq:1-4}
   z_i(t+1)= \sum_{j=1}^d\,
  \left. {\partial G_i \over \partial x_j} 
  \right|_{{\bf x}(t)}\, z_j(t) .
\end{equation}
Under rather general hypothesis, Oseledec \cite{O68} proved that for
almost all initial conditions ${\bf x}(0)$, there exists an
orthonormal basis $\lbrace {\bf e}_i \rbrace$ in the tangent space
such that, for large times,
\begin{equation}
\label{eq:1-5}
{\bf z}(t) = \sum _{i=1}^{d} c_i {\bf e}_i e^{\lambda_i \, t} \,, 
\end{equation}
where the coefficients $\{c_i\}$ depends on ${\bf
z}(0)$.  The exponents $\lambda_1 \ge \lambda_2 \ge\cdots\ge
\lambda_d$ are called {\it characteristic Lyapunov exponents}.  If the
dynamical system has an ergodic invariant measure, the spectrum of LEs
$\lbrace \lambda_i \rbrace $ does not depend on the initial condition,
except for a set of measure zero with respect to the natural invariant
measure.

Loosely speaking, (\ref{eq:1-5}) tells us that in the phase space,
where the motion evolves, a $d$-dimensional sphere of small radius
$\epsilon$ centered in ${\bf x}(0)$ is deformed with time into an
ellipsoid of semi-axes $\epsilon _i (t) = \epsilon \exp (\lambda _i
t)$, directed along the $ {\bf e}_i $ vectors.  Furthermore, for a
generic small perturbation $\delta {\bf x}(0)$, the distance between a
trajectory and the perturbed one behaves as
\begin{equation}
\label{eq:1-6}
|\delta {\bf x} (t)|\sim |\delta {\bf x} (0)|\,e^{\lambda_1\, t}\, 
    \left[
     1 + O\left(e^{-(\lambda_1-\lambda_2)t}\right)  \right].
\end{equation}
If $\lambda_1 > 0$ we have a rapid (exponential) amplification of an
error on the initial condition. In such a case, the
system is chaotic and, {\it de facto}, unpredictable on the long
times. Indeed, if we put $\delta_0 = |\delta {\bf x} (0)|$ for the
initial error, and we want to predict the states of the system with a
certain tolerance $\Delta$ (not too large), then the prediction is
possible just up to a {\it predictability time} given by 
\begin{equation}
\label{Tpred}
T_p \sim {1 \over \lambda_1} \ln \left({\Delta \over \delta_0}\right)\, .
\end{equation}
This equation shows that $T_p$ is basically determined by the
largest Lyapunov exponent, since its dependence on $\delta_0$ and
$\Delta$ is very weak. Because of its preeminent role, very often one
simply refers to $\lambda_1 $ as ``the Lyapunov exponent'', and one
indicates it with $\lambda$.

Eq.~(\ref{eq:1-6}) suggests how to numerically compute $ \lambda_1 $. 
We introduce the response, after a time $t$, to a perturbation 
on ${\bf x} (\tau)$, defined as follows:
\begin{equation}
\label{eq:1-7}
   R_{\tau}(t)\equiv { |{\bf z}(\tau+t)| \over |{\bf z}(\tau)| }
   = { |\delta  {\bf x}(\tau+t)| \over |\delta  {\bf x}(\tau)| },
\end{equation}
where, again, $|\delta {\bf x}(\tau)|$ and $|\delta {\bf x}(\tau +
t)|$ are infinitesimal. The LE $\lambda_{1}$ is obtained by 
averaging the logarithm of the response over the initial 
conditions or along the trajectory:
\begin{equation}
\label{eq:1-8}
\lambda_1= \lim_{t\to\infty} {1\over t} \langle \ln R_{\tau}(t) \rangle,
\end{equation}
where $ \langle \cdot \rangle $ denotes the time average
$\lim_{T\to\infty} (1/T)\int_{\tau_0}^{\tau_0+T} (\cdot) d \tau$.
The Oseledec's theorem implies that $(1/t)\ln R_{\tau}(t)$, for large
$t$, is a non-random quantity, i.e. for almost all the initial
conditions its value does not depend on the specific initial
condition. Therefore, for large times, the average in (\ref{eq:1-8}) 
can be neglected.

As the typical growth rate of a generic small segment in phase space is
driven by the largest LE, the sum of the first $n$ ($\leq d$) Lyapunov
exponents controls the variations of small $n$-dimensional volumes in
phase space.  This gives us a way to compute the sub-leading
Lyapunov exponents.  After the selection of $n\leq d$ non parallel
tangent vectors $[{\bf z}^{(1)}(0), \ldots,{\bf z}^{(n)}(0)]$, one
introduces the $n$-order response $R_{\tau}^{(n)}(t)$ \cite{BGGS80}
\begin{equation}
\label{eq:1-9}  
R_{\tau}^{(n)}(t)\equiv {
 |{\bf z}_1(t+\tau)\times {\bf z}_2(t+\tau)\times\cdots\times
 {\bf z}_n(t+\tau)|  \over
       |{\bf z}_1(\tau)\times {\bf z}_2(\tau)\times\cdots
  \times{\bf z}_n(\tau)|  } \,.
\end{equation}
Analogously to the LE, it can be shown that
\begin{equation}
\label{eq:1-10}  
\sum_{i=1}^{n}\lambda_i= \lim_{t\to\infty} {1\over t} 
   \langle \ln R_{\tau}^{(n)}(t) \rangle .
\end{equation}

Let us stress that the Lyapunov exponents give information on the typical
behaviors along a generic trajectory, followed for infinite time
and keeping the perturbation infinitesimally small. In this respect,
they are global quantities characterizing fine-grained properties of a
system.

\subsubsection{ Kolmogorov-Sinai entropy }
\label{sec:1.1.2}

The LE, $\lambda$,  gives
a first quantitative information on how rapidly we loose the ability
of predicting the evolution of a system. A state, initially determined
with an error $\delta {\bf x} (0)$, after a time enough larger than
$1/\lambda$, may be found almost everywhere in the region of
motion. In this respect, the Kolmogorov-Sinai (KS)
entropy, $h_{KS}$, supplies a more refined information. The error
on the initial state is due to the maximal resolution we use for
observing the system.  For simplicity, let us assume the same
resolution $\epsilon$ for each degree of freedom. We build a partition
of the phase space with cells of volume $\epsilon ^d$, so that the
state of the system at $t=t_0$ is found in a region of volume $V_0 =
\epsilon ^d$ around ${\bf x}(t_0)$. Now we consider the trajectories
starting from $V_0$ at $t_0$ and sampled at discrete times $t_j =j\,
\tau$ ($j=1,2,3, \dots , t$); in the case of a map one can put
$\tau=1$. Since we are considering motions that evolve in a bounded
region, all the trajectories visit a finite number of different cells, each one
identified by a symbol. In this way a unique sequence of symbols
$\lbrace s(0), s(1), s(2), \dots \rbrace$ is associated with a given
trajectory. In a chaotic system, although each evolution ${\bf x}(t)$
is univocally determined by ${\bf x}(t_0)$, a  great number
of different symbolic sequences originates by the same initial cell,
because of the divergence of nearby trajectories. The total number of
the admissible symbolic sequences, ${\widetilde N} (\epsilon, t)$,
increases exponentially with a rate given by the topological entropy
\begin{equation}
\label{topentro}  
h_T = \lim _{\epsilon \to 0}\lim _{t \to \infty} 
{1 \over t} \ln {\widetilde N} (\epsilon, t) \,.
\end{equation}
However, if we consider only the number of sequences $N_{eff}
(\epsilon,t)\leq {\widetilde N} (\epsilon,t)$ which appear with very high
probability
in the long time limit -- those that can be numerically or
experimentally detected and that are associated with the natural
measure -- we arrive at a more physical quantity called the
Kolmogorov-Sinai or metric entropy \cite{ER85}:
\begin{equation}
\label{metentro}  
h_{KS} =\lim _{\epsilon \to 0} \lim _{t \to \infty} 
{1 \over t} \ln  N_{eff} (\epsilon, t) \leq h_T .
\end{equation}
$h_{KS}$ quantifies the long time exponential rate of growth of the
number of the effective coarse-grained trajectories of a system. This
suggests a link with information theory where the Shannon entropy
measures the mean asymptotic growth of the number of the typical
sequences -- the ensemble of which has probability almost one --
emitted by a source.
In the following we will discuss in more detail the KS-entropy and its
relation with the information theory. Here we obtain, by means of a
heuristic reasoning, the relation among $h_{KS}$ and Lyapunov
exponents. 

We may wonder what is the number of cells where, at a time
$t>t_0$, the points that evolved from $V_{0}$ can be found, i.e. we
wish to know how big is the coarse-grained volume $V (\epsilon, t)$,
occupied by the states evolved from $V_0$, if the minimum volume we
can observe is $\epsilon^d$.  As stated at the end of the preceding
subsection, we have $V(t) \sim V_0 \exp (t\,\sum_{i=1}^d \lambda_i)$.
However, this is true only in the limit $\epsilon \to 0$.  In this
(unrealistic) limit, $V (t) = V_0$ for a conservative system (where
$\sum_{i=1}^d \lambda_i=0$) and $V (t) < V_0$ for a dissipative system
(where $\sum_{i=1}^d \lambda_i<0$).  As a consequence of limited
resolution power, in the evolution of the volume $V_0 = \epsilon ^d$
the effect of the contracting directions (associated with the negative
Lyapunov exponents) is completely lost. We can experience only the
effect of the expanding directions, associated with the positive
Lyapunov exponents. As a consequence, in the typical case, 
the coarse grained volume behaves as
\begin{equation}
\label{eq:1-11}  
V (\epsilon, t)  \sim V_0 \, 
e ^{ (\sum_{\lambda_i >0} \lambda_i) \, t } ,  
\end{equation}
when $V_0$ is small enough.  Since $N_{eff}(\epsilon,t) \propto
V(\epsilon,t)/V_{0}$, one has
\begin{equation}
h_{KS} = \sum_{\lambda_i >0} \lambda_i \,.
\label{eq:1-12}  
\end{equation}
This argument can be made more rigorous with a proper mathematical 
definition of the metric entropy. In this case one
derives the Pesin relation \cite{P76,ER85}
\begin{equation}
\label{eq:1-13}  
h_{KS} \leq  \sum_{\lambda_i >0} \lambda_i .
\end{equation}
Because of its relation with the Lyapunov exponents -- or by the
definition (\ref{metentro}) -- it is clear that also $h_{KS}$ is a
fine-grained and global characterization of a dynamical system. 

The metric entropy is an invariant characteristic quantity of a
dynamical system \cite{K58,S59}, i.e. given two systems with invariant
measures, their KS-entropies exist and they are equal if the systems
are isomorphic \cite{bill65}. This intrinsic quantity may be properly defined
by means of tools borrowed from the mathematical theory of
communication.

\subsection{ Information theory approach}
\label{sec:1.2}
In experimental investigations of physical processes, we typically
have access to the system only trough a measuring device which
produces a time record of a certain observable, i.e. a sequence of
data. In this regard a system, whether or not chaotic, generates messages and
may be regarded as a source of information.  This observation opens
the possibility to study dynamical systems from a very interesting
point of view.

Information has found a proper characterization in the framework of
the theory of communication to cope with the practical problem of
transmitting a message in the cheapest way without losing
information.  The characterization of the information contained in a
sequence can be approached by two very different points of view. The
first one, that of information theory \cite{S48}, is a statistical
approach, i.e., it does not consider the transmission of a specific
message (sequence) but refers to the statistical properties of all the
messages emitted by the source.  Information theory approach
characterizes the source of information, so that it gives us a
powerful method to characterize chaotic systems.

The second point of view considers the problem of characterizing a
single sequence. This latter has led to the theory of algorithmic
complexity and algorithmic information theory \cite{Ch66,K65,S64}.

\subsubsection{Shannon entropy }
\label{sec:1.2.1}
At the end of forties Shannon \cite{S48} introduced rather powerful
concepts and techniques for a systematic study of sources emitting
sequences of discrete symbols (e.g. binary digit sequences). 
Originally  information theory was introduced in the
practical context of electric communications, nevertheless in a few
years it became an important branch of both pure and applied
probability theory with strong relations with other fields as computer
science, cryptography, biology and physics \cite{zurek}.

For the sake of self-consistency we briefly recall the basic concepts
and ideas about the Shannon entropy. Consider a source that can output
$m$ different symbols; denote with $s(t)$ the symbol emitted by the
source at time $t$ and with $P(C_N)$ the probability that a given word
$C_N=(s(1),s(2),\dots,s(N))$, of length $N$, is emitted:
\begin{equation}
P(C_N)=P(s(1),s(2),\dots,s(N))\,.
\label{eq:pword}
\end{equation}
We assume that the source is stationary, so that for the sequences
$\{s(t)\}$ the time translation invariance holds: 
$P(s(1),\dots,s(N))=P(s(t+1),\dots,s(t+N))$.

Now we introduce the $N$-block entropies
\begin{equation}
H_N=-\sum_{\{C_N\}} P(C_N)\ln P(C_N)\,,
\label{eq:block}
\end{equation} 
and the differential entropies 
\begin{equation}
h_N=H_{N+1}-H_N\,,
\end{equation} 
whose meaning  is the average information supplied by the
$(N+1)$-th symbol, provided the $N$ previous ones are known.  One
can also say that $h_N$ is the average uncertainty about the 
$(N+1)$-th symbol, provided the $N$ previous ones are given. For a 
stationary source the limits in the following equations exist, are 
equal and define the Shannon entropy $h_{Sh}$:
\begin{equation}
h_{Sh}=\lim_{N \to \infty} h_N= \lim_{N \to \infty} {H_N \over N} .
\label{eq:shannon}
\end{equation}
The $h_N$ are non increasing quantities: $h_{N+1} \leq h_{N}$; that is:
the knowledge of a longer past history cannot increase the
uncertainty on the next outcome.  In the case of a $k$-th order Markov
process $h_N=h_{Sh}$ for all $N\geq k$. This is because a $k$-th order Markov
process has the property that the conditional probability to have a
given symbol only depends on the results of the last $k$ times, i.e.
\begin{eqnarray}
P(s(t)|s(t-1),s(t-2),\dots)=P(s(t)|s(t-1),s(t-2),\dots,s(t-k))\nonumber\,.
\\
\end{eqnarray}

The Shannon entropy is a measure of the ``surprise'' the source emitting
the sequences can reserve to us, since it quantifies the richness (or
``complexity'') of the source. This can be precisely expressed by the
first theorem of Shannon-McMillan \cite{K57} that applies to stationary 
ergodic sources: 
\\ If $N$ is large enough, the ensemble of $N$-long subsequences can be 
partitioned in two classes, $\Omega_1(N)$ and $\Omega_0(N)$
such that all the words $C_N \in
\Omega_1(N)$ have the same probability $P(C_N)\sim \exp(-Nh_{Sh})$ and
\begin{equation}
\sum_{C_N \in \Omega_1(N)} P(C_N) \to 1  \qquad {\mbox{ for }} N\to \infty
\end{equation}
while
\begin{equation}
\sum_{C_N \in \Omega_0(N)} P(C_N) \to 0  \qquad {\mbox{ for }} N\to \infty \,.
\end{equation}
The meaning of this theorem is the following.  An $m$-states process
admits in principle $m^N$ possible sequences of length $N$, but
the number of typical sequences $N_{eff}(N)$ (those ones in
$\Omega_1(N)$) effectively observable is
\begin{equation}
N_{eff}(N) \sim \exp(Nh_{Sh})\,.
\label{eq:wordstypical}
\end{equation}
Note that $N_{eff}\ll m^N$ if $h_{Sh}<\ln m$. Moreover the entropy per
symbol, $h_{Sh}$, is a property of the source. Because of the ergodicity 
it can be obtained by analyzing just one single sequence in the ensemble of
the typical ones, and it can also be viewed as a property of each one
of the typical sequences. Therefore, as in the following, one may
speak about the Shannon entropy of a sequence.

The above theorem in the case of processes without 
memory is nothing but the law of large
numbers. Let us observe that (\ref{eq:wordstypical}) is somehow the
equivalent in information theory of the Boltzmann equation in
statistical thermodynamics: $S \propto \ln W$, being $W$ the number of
possible microscopic configurations and $S$ the thermodynamic
entropy. This justifies the name ``entropy'' for $h_{Sh}$. Under
rather natural assumptions it is possible to prove that the Shannon
entropy, apart from a multiplicative factor, is the unique quantity
which characterizes the ``surprise'' \cite{K57}.

Let us now mention another important result about the Shannon entropy.
It is not difficult to recognize that the quantity $h_{Sh}$ sets
 the maximum compression rate of a
sequence $\{s(1),s(2),s(3), \dots\}$. Indeed a theorem of Shannon
states that, if the length $T$ of a sequence is large enough, one
cannot construct another sequence (always using $m$ symbols), from
which it is possible to reconstruct the original one, whose length is
smaller than $(h_{Sh} / \ln m)T$ \cite{S48}.  In other words
$h_{Sh}/\ln m$ is the maximum allowed compression rate. 

The relation between Shannon entropy and the data compression problem
is well highlighted by considering the Shannon-Fano code to map ${\cal
N}$ objects (e.g. the $N$-words $C_N$) into sequences of binary digits
$(0,1)$ \cite{welsh}. The main goal in building a code is to define the
most efficient coding procedure, i.e. the one which generates the
shortest possible (coded) sequence.  The Shannon-Fano code is as
follows. At first one orders the ${\cal N}$ objects according to their
probabilities, in a decreasing way, $p_1,p_2,\dots,p_{{\cal N}}$.
Then the passage from the ${\cal N}$ objects to the symbols $(0,1)$ is
obtained by defining the coding $E(r)$, of binary length $\ell(E(r))$,
of the $r$-th object with the requirement that the expected total
length, $\sum_{r}p_r \ell_r$, be the minimal one. This can be realized
with the following choice
\begin{equation}
-\ln_2 p_r \leq \ell(E(r)) < 1-\ln_2 p_r \,.
\end{equation}
In this way highly probable objects are mapped into short code words
while the low probability ones are mapped to longer code words. 
So that the average code length is bounded by
\begin{equation}
{H_N \over \ln 2} \leq \sum_r p_r \ell(E(r)) \leq {H_N +1 \over \ln 2} \,, 
\end{equation}
and in the limit $N \to \infty$ one has
\begin{equation}
\lim_{N \to \infty} {\langle \ell_N \rangle \over N}=\lim_{N \to \infty} 
{\sum_r p_r \ell(E(r)) \over N}={h_{Sh} \over \ln 2} \,,
\label{eq:fano}
\end{equation}
i.e., in a good coding, the mean length of a $N$-word is equal to $N$
times the Shannon entropy (apart from a multiplicative factor, due to
the fact that in the definition (\ref{eq:shannon}) of $h_{Sh}$ we used
the natural logarithm and here we want to work with a two symbol
code).

An alternative coding method, based on variable length words, is due
to Ziv and Lempel \cite{LZ76}. Remarkably it is very efficient for
data compression and gives the same asymptotic result of the
Shannon-Fano code.

\subsubsection{Again on the Kolmogorov-Sinai entropy}
\label{sec:1.2.2}

After the introduction of the Shannon entropy we can give a more
precise definition of the KS-entropy. Consider a trajectory, ${\bf
x}(t)$, generated by a deterministic system, sampled at the times $t_j
= j \, \tau$, with $j=1, 2,3, \dots$. Perform a finite partition
${\cal A}$ of the phase space. With the finite number of symbols
$\lbrace s \rbrace_{{\cal A}}$ enumerating the cells of the partition,
the time-discretized trajectory ${\bf x}(t_j)$ determines a sequence
$\lbrace s(1), s(2), s(3), \dots \rbrace$, whose meaning is clear: at
the time $t_j$ the trajectory is in the cell labeled by $s(j)$.  To
each subsequence of length $N\cdot\tau$ one can associate a word of
length $N$: $ W^N_j ({\cal A})= \left( s(j), s(j+1) , \dots,
s(j+(N-1)) \right) $. If the system is ergodic, as we suppose, from
the frequencies of the words one obtains the probabilities by which
one calculates the block entropies $H_N ({\cal A})$:
\begin{equation}
H_{N} ({\cal A}) = - \sum _{ \lbrace W^{N}({\cal A})\rbrace } 
P(W^{N}({\cal A}))  \ln P(W^{N}({\cal A})).
\end{equation} 
It is important to note that the probabilities $P(W^{N}({\cal A}))$,
computed by the frequencies of $W^{N}({\cal A})$ along a trajectory,
are essentially dependent on the stationary 
measure selected by the trajectory. This
implies a dependence on this measure of all the quantities defined
below, $h_{KS}$ included. We shall always understand to consider the
natural invariant measure and do not indicate this kind of
dependence. The entropy per unit time of the trajectory with respect
to the partition ${\cal A}$, $h ({\cal A} )$, is defined as follows:
\begin{eqnarray}
h_N({\cal A})&=& {1 \over \tau} \lbrack H_{N+1} ({\cal A})
-H_N ({\cal A}) \rbrack \; ,\\
h({\cal A}) &=& \lim _{N \to \infty} h_N({\cal A}) = 
{1 \over \tau} \lim _{N \to \infty} {1 \over N} H_{N} ({\cal A})\; .
\end{eqnarray} 
Notice that, for the deterministic systems we are considering, the
entropy per unit time does not depend on the sampling time $\tau$
\cite{bill65}. The KS-entropy, by definition, is the supremum of $h
({\cal A})$ over all possible partitions \cite{bill65,ER85}
\begin{equation}
h_{KS} = \sup _{{\cal A}} h ({\cal A}) \, .
\end{equation}
It is not simple at all to determine $h_{KS}$ according to this
definition. A useful tool in this respect would be the
Kolmogorov-Sinai theorem, by means of which one is granted that
$h_{KS} = h({\cal G})$ if ${\cal G}$ is a generating partition. A
partition is said to be generating if every infinite sequence $\lbrace
s(n) \rbrace_{n=1,\dots,\infty} $ individuates a single initial
point. However the difficulty now is that, with the exception of very
simple cases, we do not know how to construct a generating
partition. We only know that, according to the Krieger theorem
\cite{krieger}, there exists a generating partition with $k$ elements
such that $e^{h_{KS}} < k \leq e^{h_{KS}} + 1$. Then, a more tractable
way to define $h_{KS}$ is based upon considering the partition ${\cal
A}_{\epsilon}$ made up by a grid of cubic cells of edge $\epsilon$,
from which one has
\begin{equation}
h_{KS} = \lim_{\epsilon \to 0} 
h ({\cal A}_{\epsilon}) \, .
\end{equation}
We expect that $h ({\cal A}_{\epsilon})$ becomes independent of 
$\epsilon$ when ${\cal A}_{\epsilon}$ is so fine to be ``contained''
in a generating partition.

For discrete time maps what has been exposed above is still valid, with 
$\tau =1$ (however, Krieger's theorem only applies to invertible maps). 

The important point to note is that, for a truly stochastic 
(i.e. non-deterministic) system, with continuous states, 
$h ({\cal A}_{\epsilon})$ is not bounded and $ h_{KS} = \infty$.

\subsubsection{Algorithmic complexity}
\label{sec:1.2.3}

We saw that the Shannon entropy puts a limit on how efficiently
the ensemble of the messages emitted by a source
can be coded. We may wonder about the compressibility 
properties of a single sequence. This problem can be addressed 
by means of the notion of algorithmic complexity, that is concerned
with the difficulty in reproducing a given string of symbols. 

Everybody agrees that the binary digits sequence
\begin{equation}
0111010001011001011010...
\label{eq:seq1}
\end{equation}
is, in some sense, more random than 
\begin{equation}
1010101010101010101010...
\label{eq:seq2}
\end{equation}  
The notion of algorithmic complexity, independently introduced by
Kolmogorov \cite{K65}, Chaitin \cite{Ch66,Ch90} and 
Solomonov \cite{S64}, is a way to formalize the intuitive
idea of randomness of a sequence.

Consider a binary digit sequence (this does not constitute a
limitation) of length $N$, $(i_1,i_2,\dots,i_N)$, generated by a
certain computer code on some machine ${\cal M}$. One defines the
algorithmic complexity, or algorithmic information content, $K_{\cal
M}(N)$ of a $N$-sequence as the bit length of the shortest computer
program able to give the $N$-sequence and to stop afterward. Of
course, such a length depends not only on the sequence but also on the
machine. However, a result of Kolmogorov~\cite{K65} proves the
existence of a universal computer, ${\cal U}$, that is able to perform the
same computation a program $p$ makes on ${\cal M}$ with a modification
of $p$ that depends only on ${\cal M}$.  This implies that for all
finite strings:
\begin{equation}
K_{\cal U}(N) \leq K_{\cal M}(N)
+ C_{\cal M}\, , 
\label{eq:kolmocomplex}
\end{equation}
where $K_{\cal U}(N)$ is the complexity with respect to the universal
computer and $C_{\cal M}$ depends only on the machine ${\cal M}$. At
this point we can consider the algorithmic complexity with respect to
a universal computer -- and we can drop the machine dependence in the
symbol for the algorithmic complexity, $K(N)$. The reason is that we
are interested in the limit of very long sequences, $N \to \infty$,
for which one defines the algorithmic complexity per unit symbol:
\begin{equation}
{\cal C}=\lim_{N \to \infty} {K(N) \over N}\,,
\label{eq:acomplexity}
\end{equation}
that, because of (\ref{eq:kolmocomplex}), is an intrinsic quantity,
i.e. independent of the machine.

Now coming back to the two $N$-sequences (\ref{eq:seq1}) and
(\ref{eq:seq2}), it is obvious that the second one can be obtained
with a small-length minimal program, i.e.
\begin{equation}
\nonumber
{\rm "PRINT }\;10 \;\; {N \over 2}\;\;{\rm TIMES"}\,.
\end{equation}
The bit length of the above program is $O(\ln N)$ and therefore when
taking the limit $N \to \infty$ in (\ref{eq:acomplexity}), one obtains
${\cal C}=0$.  Of course $K(N)$ cannot exceed $N$, since the sequence
can always be obtained with a trivial program (of bit length $N$)
\begin{equation}
\nonumber
{\rm "PRINT }\;i_1,i_2,\dots,i_N {\rm "}\,.
\end{equation}
Therefore, in the case of a very irregular sequence, e.g.,
(\ref{eq:seq1}) one expects $K(N) \propto N$ i.e. ${\cal C}\neq 0$.  In
such a case one calls the sequence complex (i.e. of non zero
algorithmic complexity) or random.

Algorithmic complexity cannot be computed. Since the algorithm which
computes $K(N)$ cannot have less than $K(N)$ binary digits and since
in the case of random sequences $K(N)$ is not bounded in the limit $N
\to \infty$ then it cannot be computed in the most interesting cases.
The un-computability of $K(N)$ may be understood in terms of G\"odel's
incompleteness theorem \cite{Ch74,Ch82,Ch90}. Beyond the problem of whether
or not $K(N)$ is computable in a specific case, the concept of
algorithmic complexity brings an important improvement to clarify the
vague and intuitive notion of randomness.  For a systematic treatment
of algorithmic complexity, information theory and data compression see
\cite{LV97}.

There exists a relation between the Shannon entropy, $h_{Sh}$, and the
algorithmic complexity ${\cal C}$. It is possible to show that
\begin{equation}
\lim_{N \to \infty} {\langle K(N) \rangle \over H_N}={1 \over \ln 2}\,,
\label{eq:38}
\end{equation}
where $\langle K(N) \rangle =\sum_{C_N} P(C_N) K_{C_N}(N)$, being
$K_{C_N}(N)$ the algorithmic complexity of the $N$-words. Therefore the
 expected complexity $\langle K(N) /N\rangle$ is asymptotically equal to the
Shannon entropy (apart the $\ln 2$ factor).

Eq.~(\ref{eq:38}) stems from the results of the Shannon-McMillan
theorem about the two classes of sequences (i.e. $\Omega_1(N)$ and
$\Omega_0(N)$).  Indeed in the limit of very large $N$, the probability
to observe a sequence in $\Omega_1(N)$ goes to $1$, and the entropy of
such a sequence as well as its algorithmic complexity equals the
Shannon entropy. Apart from the numerical coincidence of the values of
${\cal C}$ and $h_{Sh}/\ln 2$ there is a conceptual difference between
the information theory and the algorithmic complexity theory. The
Shannon entropy essentially refers to the information content in a
statistical sense, i.e.  it refers to an ensemble of sequences
generated by a certain source. On the other hand, the algorithmic
complexity defines the information content of an individual
sequence \cite{G86}. Of course, as noted above, the fact that it is possible to
use probabilistic arguments on an individual sequence is a consequence
of the ergodicity of the system, which allows to assume good
statistical properties of arbitrary long $N$-words.

For  a dynamical system one can define the notion
of algorithmic complexity of the trajectory starting from the point
${\bf x}$, ${\cal C} ({\bf x})$. This requires the introduction of
finite open coverings of the phase space, the consideration of
symbolic sequences thus generated, for the given trajectory, sampled
at constant time intervals, and the searching of the supremum of the
algorithmic complexity per symbol at varying the coverings
\cite{AY81}. Then it can be shown -- Brudno's and White's theorems 
\cite{brudno,white} -- 
that for almost all ${\bf x}$
(we always mean with respect to the natural invariant measure)
one has:
\begin{equation}
{\cal C} ({\bf x}) = {h_{KS} \over \ln 2} \, ,
\label{eq:dscomplex}
\end{equation} 
where, as before, the factor $\ln 2$ is a conversion factor 
between natural logarithms and bits.  

This result says that the KS-entropy quantifies not only the richness, 
or surprise, of a dynamical system but also the difficulty of describing
(almost) everyone of its typical sequences. 

\subsection{Algorithmic complexity and Lyapunov Exponent}
\label{sec:1.3}

Summing up, the theorem of Pesin together with those of Brudno and
White show that a chaotic dynamical system
may be seen as a source of messages that cannot be described in a
concise way, i.e. they are complex. We expose here two examples that
may help in understanding the previous conclusion and the relation
between the Lyapunov exponent, the $KS$-entropy and the algorithmic
complexity.

Following Ford \cite{F83,F86}, 
let us consider the shift map 
\begin{equation}
\label{fordmap}
x(t+1) = 2 \, \, x(t) \qquad  {\rm mod} \,  1 ,
\end{equation}
which has $\lambda = \ln 2$.  If one writes an initial condition in
binary representation, i.e., $ x(0) = \sum_{j=1} ^\infty a_j \, 2^{-j}
$, such that $a_j = 0$ or $1$, it is clearly seen that the action of
the map (\ref{fordmap}) on $x(0)$ is just a shift of the binary
coordinates:
\begin{equation}
x(1) = \sum_{j=1} ^\infty a_{j+1} \, 2^{-j} \qquad  \cdots \qquad  
x(t) = \sum_{j=1} ^\infty a_{j+t} \, 2^{-j} .
\end{equation}
With this observation it is possible to verify that $K(N) \simeq N$
for almost all the solutions of (\ref{fordmap}). Let us consider
$x(t)$ with accuracy $2^{-k}$ and $x(0)$ with accuracy $2^{-l}$, of
course $l=t+k$. This means that, in order to obtain the $k$ binary
digits of the output solution of (\ref{fordmap}), we must use a
program of length no less than $l=t+k$. Basically one has to specify
$a_1, a_2, \dots, a_l$. Therefore we are faced with the problem of the
algorithmic complexity of the binary sequence $( a_1, a_2, \dots,
a_{\infty} )$ which determines the initial condition
$x(0)$. Martin-L\"of \cite{M-L66} proved a remarkable theorem stating
that, with respect to the Lebesgue measure, almost all the binary
sequences $( a_1, a_2, \dots, a_{\infty} )$, which represent a real
number in $[0, 1]$, have maximum complexity, i.e. $K(N) \simeq N$. In
practice no human being will ever be able to distinguish the typical
sequence $( a_1, a_2, \dots, a_{\infty} )$ from the output of a fair
coin toss.

Finally let us consider a $1d$ chaotic map 
\begin{equation}
x(t+1)=f(x(t))\,.
\label{eq:mappa}
\end{equation}
If one wants to transmit
to a friend on Mars the sequence $\{ x(t),\;t=1,2,\dots,T\}$ accepting
only errors smaller than a tolerance $\Delta$, one can use the
following strategy \cite{PSV95}:
\begin{enumerate}
\item Transmit the rule 
      (\ref{eq:mappa}): for this task one has to use a number of bits      
      independent of the length of the sequence $T$.
\item Specify the initial condition $x(0)$ with a precision        
      $\delta_0$ using a finite number of bits which is independent    
      of  $T$.
\item Let the system evolve till the first time $\tau_1$ such that
      the distance between two trajectories, that  was initially
      $\delta x(0)=\delta_0$,
      equals $\Delta$ and then specify again the new   
      initial condition $x(\tau_1)$ with precision $\delta_0$.
\item Let the system evolve and repeat the procedure (2-3), i.e.
      each time the error acceptance tolerance is reached specify
      the initial conditions, 
      $x(\tau_1+\tau_2), \,\;x(\tau_1+\tau_2+\tau_3)\,\dots$, with
      precision $\delta_0$. The times $\tau_1,\tau_2,\dots$
      are defined as follows: putting $x^{'}(\tau_1)=x(\tau_1)+  
      \delta_0$, $\tau_2$ is given by the minimum time such that
      $|x^{'}(\tau_1+\tau_2)-x(\tau_1+\tau_2)| \geq \Delta$ and so  
      on.
\end{enumerate}
By following the steps $(1-4)$ the friend on Mars can reconstruct with
a precision $\Delta$ the sequence $\{x(t)\}$ simply iterating on
a computer the system (\ref{eq:mappa}) between $1$ and
$\tau_1-1$, $\tau_1$ and $\tau_1+\tau_2-1$, and so on.

Let us now compute the amount of bits necessary to implement the above
transmission (1-4). For simplicity of notation we introduce the
quantities
\begin{equation}
\gamma_i={1 \over \tau_i} \ln {\Delta \over \delta_0} \,
\label{eq:mars1}
\end{equation}
which can be considered as a sort of {\it effective} Lyapunov exponents (see
Sect.~\ref{sec:2.1}).  The LE $\lambda$ can be written in terms of
$\{\gamma_i\}$ as follows
\begin{equation}
\lambda=\langle \gamma_i \rangle={\sum_i \tau_i \gamma_i \over \sum_i
\tau_i}= {1 \over {\overline {\tau}}} \ln {\Delta \over \delta_0}
\label{eq:liapT}
\end{equation}
where 
$$
{\overline {\tau}} = {1 \over N}\sum \tau_i\,,
$$
is the average time after which we have to transmit the new initial
condition (let us observe that to obtain $\lambda$ from the
$\gamma_i$'s one has to perform the average (\ref{eq:liapT}) because
the transmission time, $\tau_i$, is not constant).  If $T$ is large
enough the number of transmissions, $N$, is $T/{\overline {\tau}}\simeq
\lambda T/ \ln(\Delta/\delta_0)$.  Therefore, noting that in each
transmission for the reduction of the error from $\Delta$ to
$\delta_0$ one needs to use $\ln_2 (\Delta/\delta_0)$ bits, the total
amount of bits used in the whole transmission is 
\begin{equation}
{T \over {\overline{ \tau}}} \ln_2 {\Delta \over \delta_0}= {\lambda
\over \ln 2} T \,.
\label{eq:bits}
\end{equation}
In other words the number of bits for unit time is proportional to $\lambda$.

In  more than one dimension,
we have simply to replace $\lambda$ with $h_{KS}$ in (\ref{eq:bits}).
To intuitively understand this point one has to repeat the above 
transmission procedure in each of the expanding directions.

In this section, 
we briefly discussed how the limit of predictability, the fact that a
sequence cannot be arbitrarily compressed, and the information
contained in a signal are deeply related. In the following we will
mainly discuss the dynamical system point of view,
i.e.,  in terms of Lyapunov exponents,  Kolmogorov Sinai entropy 
and their generalizations for less ideal cases.

\section{Limits of the Lyapunov exponent for predictability}
\label{sec:2}
\setcounter{equation}{0}

We saw how a first obvious way for quantifying the predictability of a
physical system is in terms of the {\it predictability time}
$T_p$, i.e. the time interval on which one can typically forecast the
system.  A simple argument previously suggested
\begin{equation}
T_{p} \sim {1 \over \lambda} \ln \left({\Delta \over \delta_0}\right) \,.
\label{eq:2.1-1}
\end{equation}
However, in any realistic system, relation (\ref{eq:2.1-1}) is too
naive to be of actual relevance.  Indeed, it does not take into
account some basic features of dynamical systems:
\begin{description}
\item{-} The Lyapunov exponent (\ref{eq:1-8}) is a global quantity:
it measures the average rate of divergence of nearby trajectories.
In general there exist finite-time fluctuations and the 
probability distribution functions (pdf) of these fluctuations is important
 for the characterization of
predictability. The {\it generalized Lyapunov exponents} have
been introduced with the purpose to take into account these
fluctuations \cite{BPPV85,Fu83}.

\item{-} The Lyapunov exponent is defined for the linearized dynamics,
i.e., by computing the rate of separation of two infinitesimally close
trajectories. On the other hand,  for measuring the predictability time
(\ref{eq:2.1-1}) one is interested in a finite tolerance $\Delta$,
because the initial error $\delta_0$ is typically finite. A recent
generalization of the Lyapunov exponent to {\it finite size} errors
extends the study of the perturbation growth to the nonlinear regime,
i.e. both $\delta_0$ and $\Delta$ are not infinitesimal
\cite{ABCPV96,ABCPV97}.
\end{description}

\subsection{Characterization of finite-time fluctuations}
\label{sec:2.1}

Let us consider the linear response, at a delay $t$, 
to an infinitesimal perturbation $\delta {\bf x}(0)$:
\begin{equation}
R(t) = {|\delta {\bf x}(t)| \over |\delta {\bf x}(0)|} \, ,
\label{eq:2.1-2}
\end{equation}
from which the LE is computed according to
(\ref{eq:1-8}).  In order to take into account the finite-time
fluctuations, we can compute the different moments $\langle R(t)^q
\rangle$ and introduce the so-called generalized Lyapunov exponents (of
order $q$) \cite{BPPV85,Fu83}:
\begin{equation}
L(q) = \lim_{t \to \infty} {1 \over t} \ln \langle R(t)^q \rangle \, 
\label{eq:2.1-3}
\end{equation}
where $\langle ... \rangle$ indicates the time average along the
trajectory (see Sect.~\ref{sec:1}).  It is easy to show that
\begin{equation}
\lambda_{1} = \left. {d L(q) \over d q} \right|_{q=0} \, .
\label{eq:2.1-4}
\end{equation}

In the absence of fluctuations, $\lambda_{1}$ completely characterizes the
error growth and we have $L(q)=\lambda_{1} q$, while in the general
case $L(q)$ is concave in $q$ \cite{PV87}.  Before discussing the
properties of the generalized Lyapunov exponents, let us
consider a simple example with a non trivial $L(q)$. The model is the
one-dimensional map
\begin{equation}
x(t+1) = \left\{ \begin{array}{ll}
        {x(t)\over a} & \mbox{for $0 \le x \le a$} \\
        {1 - x(t)\over 1-a} & \mbox{for $a < x \le 1$}
        \end{array} \right. \,,
\label{eq:2.1-5}
\end{equation}
which for $a=1/2$ reduces to the tent map. For $a \ne 1/2$ the system
is characterized by two different growth rates. The presence of
different growth rates makes $L(q)$ non-linear in $q$.  Since the map
(\ref{eq:2.1-5}) is piecewise linear and with a uniform invariant
density, the explicit computation of $L(q)$ is very easy. The moments
of the response after a time $t$ are simply given by
\begin{equation}
\langle R(t)^q \rangle = \left[a \left({1 \over a}\right)^q + 
(1-a) \left({1 \over 1-a}\right)^q \right]^t \, .
\label{eq:2.1-6}
\end{equation} 
From (\ref{eq:2.1-3}) and (\ref{eq:2.1-6}) we have:
\begin{equation}
L(q) = \ln \left[ a^{1-q} + (1-a)^{1-q} \right] \,,
\label{eq:2.1-7}
\end{equation}
which recovers the non intermittent limit $L(q)=q \ln 2$ in the
symmetric case $a=1/2$. In the general case, assuming
$0 \le a < 1/2$, we have that for $q \to +\infty$, 
$L(q)$ is dominated by the less probable, most unstable
contributions and $L(q)/q \simeq -\ln(a)$. 
In the opposite limit, $q \to -\infty$,
we obtain $L(q)/q \simeq -\ln(1-a)$.

We now show how $L(q)$ is related to the fluctuations
of $R(t)$ at finite time $t$. Define an ``effective'' Lyapunov exponent
$\gamma(t)$ at time $t$ by
\begin{equation}
R(t) \sim e^{\gamma(t) t} \, .
\label{eq:2.1-8}
\end{equation}
In the limit $t \to \infty$, the Oseledec theorem \cite{O68} assures that, for
typical trajectories,  $\gamma(t) = \lambda_{1}=-a \ln a -(1-a)
\ln (1-a)$.  Therefore, for large $t$, the probability density of
$\gamma(t)$ is peaked at the most probable value $\lambda_1$.  Let us
introduce the probability density $P_{t}(\gamma)$ of observing a given
$\gamma$ on a trajectory of length $t$. Large deviation theory
suggests
\begin{equation}
P_{t}(\gamma) \sim e^{-S(\gamma) t}\, ,
\end{equation}
where $S(\gamma)$ is the Cramer function \cite{V84}.  The
Oseledec theorem implies that $\lim_{t \to
\infty} P_{t}(\gamma)=\delta(\gamma-\lambda_1)$, this gives a
constraint on the Cramer function, i.e.  $S(\gamma=\lambda_1)=0$ and
$S(\gamma)>0$ for $\gamma\neq\lambda_1$.

The Cramer function $S(\gamma)$ is related to the generalized Lyapunov
exponent $L(q)$ trough a Legendre transform. Indeed, at large $t$, one
has
\begin{equation}
\langle R(t)^q \rangle = \int d \gamma P_{t}(\gamma) e^{q \gamma t}
\sim e^{L(q) t}\,,
\label{eq:2.1-9}
\end{equation}
by a steepest descent estimation one obtains
\begin{equation}
L(q) = \max_{\gamma} \left[ q \gamma - S(\gamma) \right]\,.
\label{eq:2.1-10}
\end{equation}
In other words  each value of $q$ selects a particular $\gamma^{*}(q)$
given by 
\begin{equation}
\left. {d S(\gamma) \over d \gamma}\right|_{\gamma^{*}}=q\, .
\label{eq:2.1-11}
\end{equation}

We have already discussed that, for negligible fluctuations of the
``effective'' Lyapunov exponents, the LE completely characterizes the 
error growth and $L(q)=\lambda_{1} q$. 
In presence of fluctuations, the probability distribution for $R(t)$
can be approximated by a log-normal distribution. This can be
understood assuming weak correlations in the response function so that
(\ref{eq:2.1-2}) factorizes in several independent contributions
and the central limit theorem applies.  We
can thus write
\begin{equation}
P_{t}(R) = {1 \over R \sqrt{2 \pi \mu t}} 
\exp\left({-{(\ln R -\lambda_1 t)^2 \over 2 \mu t}}\right)\,,
\label{eq:2.1-12}
\end{equation}
where $\lambda_1$ and $\mu$ are given by
\begin{eqnarray}
\lambda_{1}&=&\lim_{t\to \infty}{1 \over t} \langle \ln R(t) \rangle 
\nonumber\\
\mu&=&\lim_{t\to \infty} {1 \over t}\left(\langle \ln R(t)^2\rangle 
- \langle \ln R(t) \rangle^2 \right)\,.
\end{eqnarray}

\begin{figure}[hbt]
\centerline{\epsfig{figure=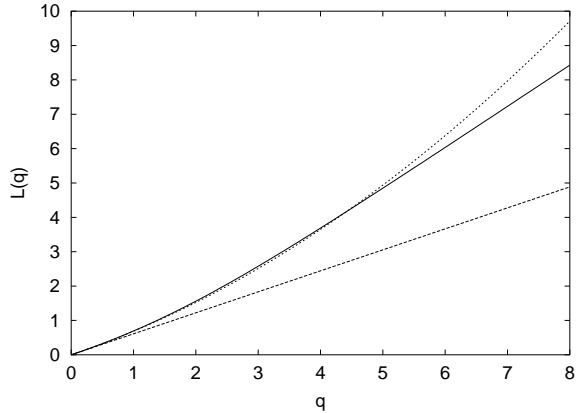,width=8cm, angle=0}}
\caption{\label{fig:2.1-1}Generalized Lyapunov exponent, $L(q)$ 
for the map (\ref{eq:2.1-5}) with $a=0.3$ (solid line) 
compared with  the linear non intermittent approximation, 
$\lambda_1 q$ (dashed line), and with the log-normal one 
Eq.~(\ref{eq:2.1-14}) (dotted line).}
\end{figure}

The log-normal distribution for $R$ corresponds to 
a Gaussian distribution for $\gamma$ 
\begin{equation}
S(\gamma)={\left(\gamma -\lambda_1\right)^2 \over 2 \mu}\, ,
\label{eq:2.1-13}
\end{equation}
and to a quadratic in $q$ generalized Lyapunov exponent: 
\begin{equation}
L(q)=\lambda_{1} q + {1 \over 2} \mu q^2 \, .
\label{eq:2.1-14}
\end{equation}
Let us remark that, in general, the log-normal distribution
(\ref{eq:2.1-12}) is a good approximation for non extreme events,
i.e. small fluctuation of $\gamma$ around $\lambda_1$, so that the
expression (\ref{eq:2.1-14}) is correct only for small $q$ (see
Fig.~\ref{fig:2.1-1}). This is because the moments of the log-normal
distribution grow too fast with $q$ \cite{O70}. Indeed from
(\ref{eq:2.1-11}) we have that the selected $\gamma^{*} (q)$ is given
by $\gamma^{*}(q)=\lambda_1+\mu q$ and thus $\gamma^{*}(q)$ is not
finite for $q \to \infty$.  This is unphysical because $\gamma^{*}(\infty)$
is the fastest error growth rate in the system and, we may reasonably suppose 
that it is finite.

Let us consider again the map (\ref{eq:2.1-5}). In this case we have
$\lambda_1=L'(0)=-a \ln(a)-(1-a)\ln(1-a)$ and
$\mu=L''(0)=a(1-a)\left(\ln(a)-\ln(1-a)\right)^2$, which are nothing
but the coefficients of the Taylor expansion of (\ref{eq:2.1-7})
around $q=0$. For large $q$ the log-normal approximation gives $L(q)/q
\simeq q \mu/2$ while the correct limit is the constant $L(q)/q \simeq
-\ln(a)$.

Nevertheless, $\lambda_{1}$ and $\mu$ are the two basic parameters
for characterizing the intermittency of a system. To be more specific,
let us remark that $P_{t}(R)$ (\ref{eq:2.1-12}) reaches its maximum
for
\begin{equation}
R_{max}(t)=e^{(\lambda_{1}-\mu)t} \, ,
\label{eq:2.1-15}
\end{equation}
so that for $t \to \infty$:
\begin{equation}
\begin{array}{ll}
R_{max} \to \infty & \mbox{if $\mu/\lambda_1<1$} \nonumber \\
R_{max} \to 0 & \mbox{if $\mu/\lambda_1>1$} \nonumber \,.
\end{array}
\label{eq:2.1-16}
\end{equation}

Thus in the weak intermittency limit, $\mu/\lambda_1<1$,
the most probable response function $R_{max}(t)$ follows the 
correct behavior (with the corrected exponent $\lambda_1-\mu$).
In the strong intermittent limit, $\mu/\lambda_1>1$,
the most probable estimation breaks down because it
predicts an asymptotic stable phase $R_{max}(t) \to 0$
instead of the chaotic exponential growth.

As in the case of the first LE, it is possible to introduce higher
order generalized Lyapunov exponents.  Starting from the $n$-order
response function $R^{(n)}(t)$ (\ref{eq:1-9}), we define
\begin{equation}
L^{(n)}(q) = \lim_{t \to \infty} {1 \over t} \ln \langle
R^{(n)}(t)^q \rangle\,,
\label{eq:2.1-17}
\end{equation}
where $L^{(1)}(q)=L(q)$.
From (\ref{eq:1-10}) we have
\begin{equation}
\sum_{i=1}^{n} \lambda_{i} = \left. {d L^{(n)}(q) \over d q} \right|_{q=0}\,.
\label{eq:2.1-18}
\end{equation}

The generalized $L^{(n)}(q)$ represents the fluctuations of the
exponential divergence of a $n$-dimensional volume in phase space
\cite{PV87}.  The properties of $L^{(n)}(q)$ are analogous to the
properties of $L(q)$, i.e. $L^{(n)}(q)$ is a concave function of $q$
for any $n$ and for a non-intermittent behavior they are linear in
$q$.

\subsection{Renyi entropies}
\label{sec:2.1.1}

In Section~\ref{sec:1.1.2} we defined the Kolmogorov-Sinai entropy
(\ref{metentro}) and discussed its relation with the Lyapunov
exponents by means of the Pesin relation (\ref{eq:1-13}). Analogously
to the generalized LE, it is possible to introduce a generalization of
the Kolmogorov-Sinai entropy in order to take into account the
intermittency.

Let us recall the definition of Kolmogorov-Sinai entropy
\begin{equation}
h_{KS} = - \lim_{\tau \to 0} \lim_{\epsilon \to 0} \lim_{N \to \infty}
{1 \over N \tau} \sum_{\lbrace W^{N}({\cal A}_{\epsilon})\rbrace} 
P(W^{N}({\cal A}_{\epsilon}))  \ln P(W^{N}({\cal A}_{\epsilon}))
\label{eq:2.1-20}
\end{equation}
where ${\cal A}_{\epsilon}$ is a partition of the phase space in cells
of size $\epsilon$ and $W^{N}({\cal A}_{\epsilon})$ indicates a
sequence of length $N$ in this partition.  The generalized Renyi
entropies \cite{PV86,PV87}, $K_q$, can be introduced by observing that
(\ref{eq:2.1-20}) is nothing but the average of $-\ln P(W^{N})$ with
the probability $P(W^{N})$:
\begin{equation}
K_{q} = - \lim_{\tau \to 0} \lim_{\epsilon \to 0} \lim_{N \to \infty}
{1 \over N \tau (q-1)} \ln 
\left( \sum_{\lbrace W^{N}({\cal A}_{\epsilon})\rbrace} 
P(W^{N}({\cal A}_{\epsilon}))^q \right) \, .
\label{eq:2.1-21}
\end{equation}
As in (\ref{eq:2.1-4}) one has $h_{KS} = \lim_{q \to 1} K_{q} =
K_{1}$; in addition from general results of probability theory, one
can show that $K_{q}$ is monotonically decreasing with $q$.

It will not be surprising that the generalized {\it Renyi entropies}
are related to the generalized Lyapunov exponents $L(q)$.  Introducing
the number of non-negative Lyapunov exponents $n^{*}$
(i.e. $\lambda_{n^*}\geq 0$, $\lambda_{n*+1}<0$), the Pesin relation
(\ref{eq:1-13}) can be written as
\begin{equation}
h_{KS} = \sum_{i=1}^{n^*} \lambda_{i} = 
\left. {d L^{(n^{*})}(q) \over d q} \right|_{q=0} \,.
\label{eq:2.1-19}
\end{equation}
Moreover, one has \cite{PV86}:
\begin{equation}
K_{q+1} = {L^{(n^*)}(-q) \over - q} \,.
\label{eq:2.1-22}
\end{equation}

\subsection{The effects of intermittency on predictability}
\label{sec:2.2}

We have seen that intermittency can be described, at least at a
qualitative level, in terms of $\lambda_1$ and $\mu$, which are the
two parameters characterizing the log-normal approximation.  We
discuss now the relevance of the log-normal approximation for the
predictability time $T_{p}$.

The predictability time $T_{p}$ is defined as the time it takes 
for the error of initial size $\delta_{0}$ to grow up to the tolerance 
$\Delta$ 
\begin{equation}
T_{p} = \min \left[t \,:\,\, R(t) \ge {\Delta \over \delta_{0}} \right] \, .
\label{eq:2.2-1}
\end{equation}

In the framework of the log-normal approximation, we can write
\begin{equation}
\ln R(t) = \lambda_1 t + \sqrt{\mu} w(t)\,,
\label{eq:2.2-2}
\end{equation}
where $w(t)$ is a Wiener process with $w(0)=0$, $\langle
w(t)\rangle=0$ and $\langle w(t) w(t') \rangle = \min(t,t')$. In this
case the computation of $T_p$ is reduced to a first exit problem,
which is well known in stochastic process theory \cite{B74,F70}. The
solution gives the pdf of $T_p$ \cite{CJPV93}:
\begin{equation}
p(T_p) = {\ln(\Delta/\delta_0) \over \sqrt{2 \pi \mu T_p^3}}
\exp \left[- {(\lambda_1 T_p - \ln(\Delta/\delta_0))^2
\over 2 \mu T_p} \right] \, .
\label{eq:2.2-3}
\end{equation}
Notice that the (\ref{eq:2.2-3}) is not normalized, since there is 
a finite probability of ``no exit''. Of course, this is an artifact
of the approximation in terms of the stochastic process (\ref{eq:2.2-2}).
In non pathological chaotic systems one expects that $p(T_p)$ is normalized.

\begin{figure}[htb]
\centerline{\epsfig{figure=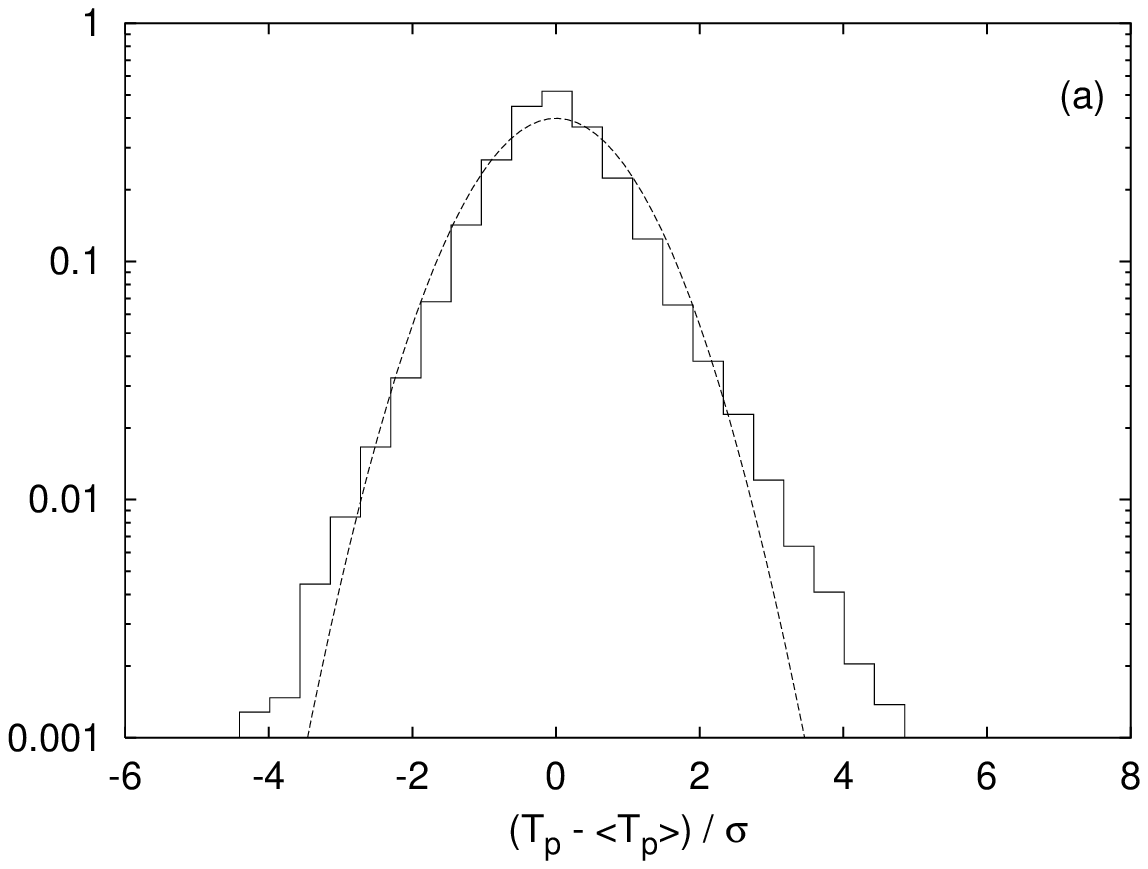,width=9cm, angle=0}}
\centerline{\epsfig{figure=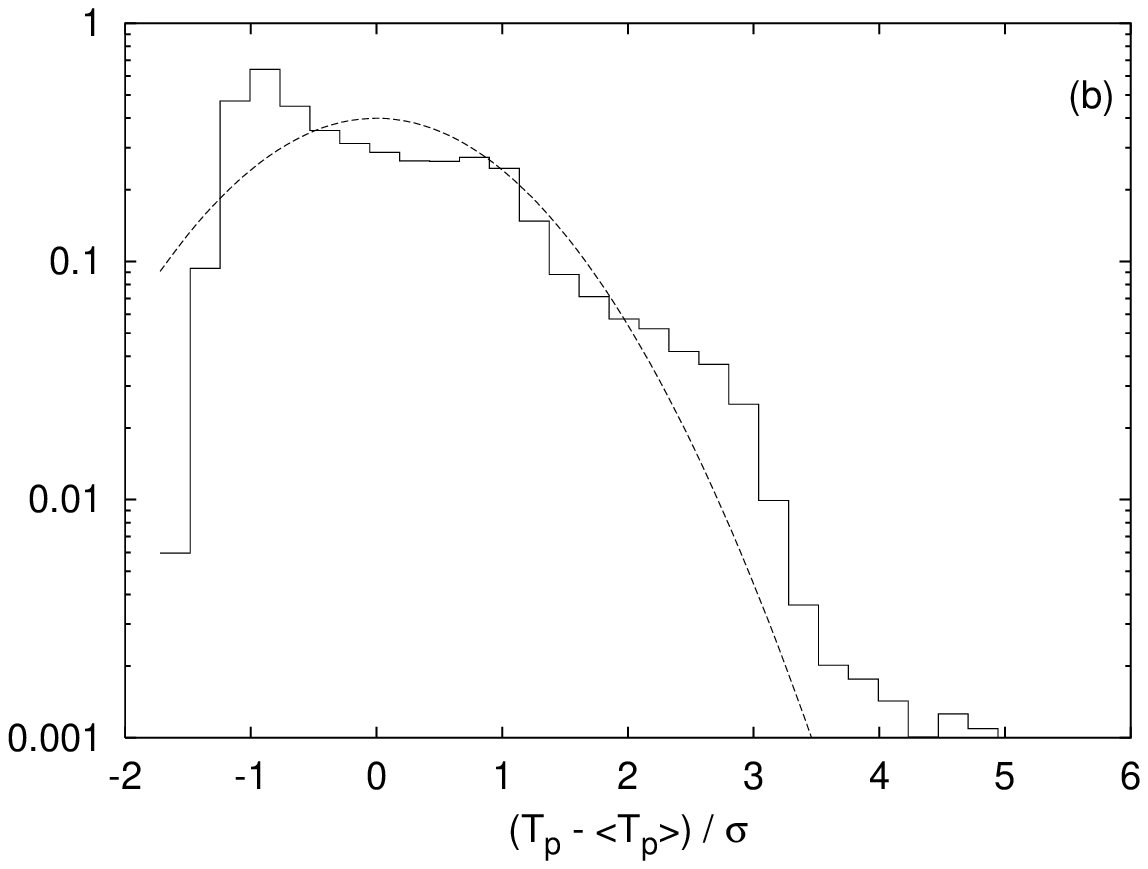,width=9cm, angle=0}}
\caption{Rescaled pdf, $p(T_p)\sigma$, of the
predictability time $T_p$ for the Lorenz model (\ref{eq:2.2-5}):  
(a) with $r=28$  (weak intermittency) the average predictability time is 
    $\langle T_p\rangle=10.84$ and its variance 
    is $\sigma^2= 3.12$  while $\lambda=0.90$, $\mu= 0.06 \pm 0.01$;
(b) with $r=166.3$ (strong intermittency) and 
     $\langle T_p\rangle=8.2385$ and $\sigma^2= 19.75$, while 
     $\lambda=1.18$ and $\mu=3.9\pm 1$. 
    The dashed line is the Gaussian distribution.}
\label{fig:2.2-1}
\end{figure}

In the limit of weak intermittency $\mu/\lambda_1 \ll 1$, $p(T_p)$ is
almost Gaussian and the mean value $\langle T_p \rangle$ is close to
the most probable value of (\ref{eq:2.2-3}) corresponding to the naive
estimation (\ref{eq:2.1-1}).  On the contrary, in the strong
intermittent limit, $\mu/\lambda_1 \gg 1$, the pdf of $T_p$ shows an
asymmetric ``triangular shape'' and the most probable value is
\begin{equation}
T_p = {1 \over 3 \mu} \ln(\Delta/\delta_0)^2 \, .
\label{eq:2.2-4}
\end{equation}

In order to see the effects of intermittency on the predictability time,
let us consider as an example the Lorenz system \cite{L63}:
\begin{equation}
\left\{
\begin{array}{lll}
\displaystyle{d x \over d t} & = & \sigma(y-x) \\

\displaystyle{d y \over  d t} & = & x (r - z) - y \\

\displaystyle{d z \over  d t} & = & x y -b z
\end{array}
\right.
\label{eq:2.2-5}
\end{equation}
with the standard values $\sigma=10$ and $b=8/3$.  For $r=28$, the
Lorenz model is very weakly intermittent, $\mu /\lambda \simeq 7
\times 10^{-2}$, and the pdf of the predictability time is very close
to a Gaussian (Fig.~\ref{fig:2.2-1}). On the contrary, for $r=166.3$
the Lorenz model becomes strongly intermittent \cite{PM80},
$\mu/\lambda \simeq 3.3$ and the pdf of the predictability time
displays a long exponential tail responsible for the deviation from
(\ref{eq:2.1-1}).

This qualitative behavior is typical of intermittent systems.  In
Section~\ref{sec:4.3} we will see a more complex example in the
context of turbulence.

\subsection{Growth of non infinitesimal perturbations}
\label{sec:2.3}

In realistic situations, the initial condition of a system is known
with a limited accuracy.  In this case the Lyapunov exponent is of
little relevance for the characterization of predictability and new
indicators are needed.  To clarify the problem, let us consider the
following coupled map model:
\begin{equation}
\left\{\begin{array}{ll} {\bf x}(t+1) &= {\bf R} \, {\bf x}(t) +
\varepsilon {\bf h}(y(t)) \nonumber \\ y(t+1) &= G(y(t)) \nonumber \,,
\end{array}
\right.
\label{eq:2.3-1}
\end{equation}
where ${\bf x} \in {\mathrm I\!R}^{2}$, $y \in {\mathrm I\!R}^{1}$,
${\bf R}$ is a rotation matrix of arbitrary angle $\theta$, ${\bf h}$
is a vector function and $G$ is a chaotic map.  For simplicity we
consider a linear coupling ${\bf h}(y)=(y,y)$ and the logistic map
$G(y)=4 y (1-y)$.

For $\varepsilon=0$ we have two independent systems:  a regular
and a chaotic one. Thus the Lyapunov exponent of the ${\bf x}$
subsystem is $\lambda_{x}(\varepsilon=0)=0$, i.e., it is
completely predictable. On the contrary, the $y$ subsystem is
chaotic with $\lambda_{y}=\lambda_1=\ln 2$.

If we now switch on the (small) coupling ($\varepsilon>0$) we are
confronted with a single three-dimensional chaotic system with a
positive global Lyapunov exponent
\begin{equation}
\lambda = \lambda_{y} + O(\varepsilon) \, .
\label{eq:2.3-2}
\end{equation}
A direct application of (\ref{eq:2.1-1}) would give
\begin{equation}
T_{p}^{(x)} \sim T_{p} \sim {1 \over \lambda_{y}} \, ,
\label{eq:2.3-3}
\end{equation}
but this result is clearly unacceptable: the
predictability time for ${\bf x}$ seems to be independent of the value
of the coupling $\varepsilon$. Let us underline that this is not due
to an artifact of the chosen example. Indeed, one can use the same
argument in many physical situations \cite{BPV96}. A well known
example is the gravitational three body problem with one body
(asteroid) much smaller than the other two (planets).  If one neglects
the gravitational feedback of the asteroid on the two planets
(restricted problem) one has a chaotic asteroid in the regular field
of the planets. As soon as the feedback is taken into account
(i.e. $\varepsilon>0$ in the example) one has a non-separable
three body system with a positive LE. Of course, intuition correctly
suggests that it should be possible to forecast the motion of the
planets for very long times if the asteroid has a very small mass
($\varepsilon \to 0$).
\begin{figure}[htb]
\centerline{\epsfig{figure=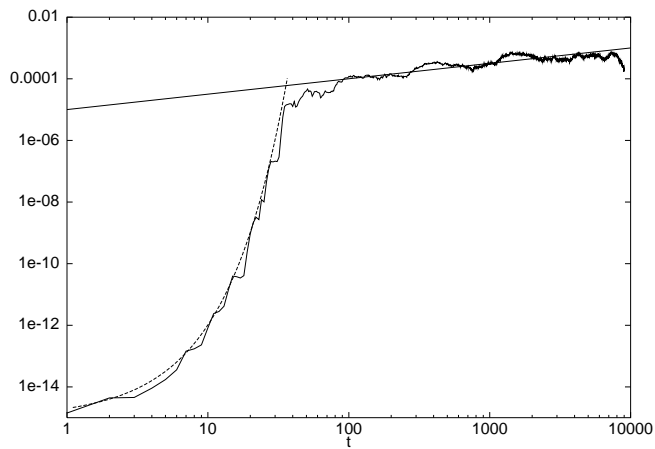,width=6cm, angle=270}}
\vspace{.4cm}
\caption{Growth of error $|\delta {\bf x}(t)|$ for the coupled map
(\ref{eq:2.3-1}).  The rotation angle is $\theta=0.82099$, the
coupling strength $\varepsilon=10^{-5}$ and the initial error only on
the $y$ variable is $\delta y=\delta_0=10^{-10}$. Dashed line $|\delta
{\bf x}(t)|\sim e^{\lambda_1 t}$ where $\lambda_1=\ln 2$, solid line
$|\delta {\bf x}(t)| \sim t^{1/2}$.  }
\label{fig:2.3-1}
\end{figure}

The apparent paradox arises from the use of (\ref{eq:2.1-1}), which is
valid only for the tangent vectors, also in the non infinitesimal
regime. As soon as the errors become large one has to take into
account the full nonlinear evolution.  The effect is shown for the
model (\ref{eq:2.3-1}) in Figure~\ref{fig:2.3-1}.  The evolution of
$\delta{\bf x}$ is given by
\begin{equation}
\delta {\bf x}(t+1)={\bf R} \delta {\bf x}(t)+
\varepsilon \delta {\bf h}(y) \, ,
\label{eq:2.3-4}
\end{equation}
where, with our choice, $\delta {\bf h}= (\delta y,\delta y)$.  At the
beginning, both $|\delta {\bf x}|$ and $\delta y$ grow
exponentially. However, the available phase space for  $y$
is finite and the uncertainty reaches the saturation value $\delta y
\sim O(1)$ in a time $t \sim 1/\lambda_1$.  At larger times the two
realizations of the $y$ variable are completely uncorrelated and their
difference $\delta y$ in (\ref{eq:2.3-4}) acts as a noisy term. As a
consequence, the growth of the uncertainty on ${\bf x}$ becomes
diffusive with a diffusion coefficient proportional to $\varepsilon^2$
\cite{BPV96}
\begin{equation}
|\delta {\bf x}(t)| \sim \varepsilon t^{1/2} 
\label{eq:2.3-5}
\end{equation}
so that:
\begin{equation}
T_{p}^{(x)} \sim \varepsilon^{-2} \, .
\label{eq:2.3-6}
\end{equation}

This example shows that, even in simple systems, the Lyapunov exponent
can be of little relevance for the characterization of the
predictability.

In more complex systems, in which different scales are present, one is
typically interested in forecasting the large scale motion, while the
LE is related to the small scale dynamics. A familiar example is
weather forecast: the LE of the atmosphere is indeed rather large due
to the small scale convective motion, but (large scale) weather
prediction is possible for about $10$ days \cite{L69,M73}.  It is thus
natural to seek for a generalization of the LE to finite perturbations
from which one can obtain a more realistic estimation for the
predictability time.  It is worth underlining the important fact that
finite errors are not confined in the tangent space but are governed
by the complete nonlinear dynamics. In this sense the extension of the
LE to finite errors will give more information on the system.

Aiming to generalize the LE to non infinitesimal
perturbations let us now define the Finite Size Lyapunov Exponent
(FSLE) \cite{ABCPV96,ABCPV97} (see Appendix~\ref{app:fsle} for the
computational details).  Consider a reference trajectory, ${\bf x}(t)$,
and a perturbed one, ${\bf x}^{'}(t)$, such that $|{\bf x}^{'}(0)- {\bf
x}(0)|=\delta$ ($|\dots|$ is the Euclidean norm but one can also
consider other norms). One integrates the two trajectories and
computes the time $\tau_1(\delta,r)$ necessary for the separation
$|{\bf x}^{'}(t)- {\bf x}(t)|$ to grow from $\delta$ to $r \delta$.
At time $t\!=\!\tau_1(\delta,r)$ the distance between the trajectories is
rescaled to $\delta$ and the procedure is repeated in order to compute
$\tau_2(\delta,r), \tau_3(\delta,r) \dots$.

The threshold ratio $r$ must be $r>1$, but not too large in order to
avoid contributions from different scales in $\tau(\delta,r)$. A
typical choice is $r=2$ (for which $\tau(\delta,r)$ is properly a
``doubling'' time) or $r=\sqrt{2}$. In the same spirit of the discussion
leading to Eq.s (\ref{eq:mars1}) and (\ref{eq:liapT}), we may introduce
an effective finite size growth rate:
\begin{equation}
\gamma_i(\delta,r)={1 \over \tau_i(\delta,r)}\ln r\,.
\end{equation}

After having performed $\cal{N}$ error-doubling experiments, we can define
the FSLE as 
\begin{equation}
\lambda(\delta)=\langle\gamma(\delta,r)\rangle_t=
 \left\langle{1 \over
\tau(\delta,r)}\right\rangle_{t} \ln r = {1 \over \langle
\tau(\delta,r) \rangle_{e}} \ln r \, ,
\label{eq:2.3-10}
\end{equation}
where $\langle \tau(\delta,r)\rangle_e$ is
\begin{equation}
\langle \tau(\delta,r)\rangle_{e} = {1 \over \cal{N}} \sum_{n=1}^{{\cal N}}
\tau_n(\delta,r) 
\, .
\label{eq:2.3-8}
\end{equation}
(see Appendix~\ref{app:fsle} and \cite{ABCPV97} for
details).  In the infinitesimal limit, the FSLE reduces to the Lyapunov
exponent
\begin{equation}
\lim_{\delta \to 0} \lambda(\delta) = \lambda_1 \, .
\label{eq:2.3-11}
\end{equation}
In practice this limit means that $\lambda(\delta)$ displays a
constant plateau at $\lambda_1$ for sufficiently small $\delta$
(Fig.~\ref{fig:2.3-1}). For finite value of $\delta$ the behavior of
$\lambda(\delta)$ depends on the details of the non linear dynamics.
For example, in the model (\ref{eq:2.3-1}) the diffusive behavior
(\ref{eq:2.3-5}), by simple dimensional arguments, corresponds to
$\lambda(\delta)\sim \delta^{-2}$.
\begin{figure}[htb]
\centerline{\epsfig{figure=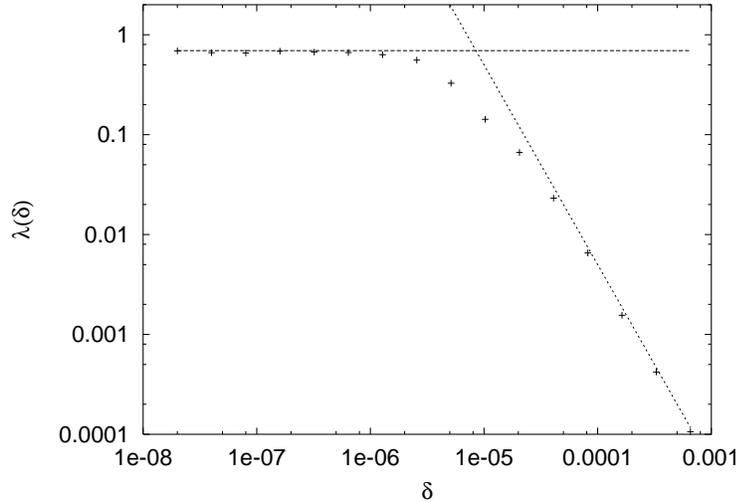,width=10cm, angle=0}}
\vspace{.4cm}
\caption{$\lambda(\delta)$ as a function of $\delta$ for the coupled
map (\ref{eq:2.3-1}) with $\varepsilon=10^{-5}$. The perturbation has
been initialized as in Fig.~\ref{fig:2.3-1}. For $\delta \to 0$,
$\lambda(\delta) \simeq \lambda_1$ (solid line). The dashed line shows
the behavior $\lambda(\delta)\sim \delta^{-2}$.}
\label{fig:2.3-2}
\end{figure}
Since the FSLE measures the rate of divergence of trajectories at
finite errors, one might wonder whether it is just another way to look
at the average response $\langle \ln(R(t)) \rangle$ (\ref{eq:2.1-2})
as a function of time. A moment of reflection shows that this is not
the case. Indeed taking the average at fixed time is not the same as
computing the average doubling time at {\it fixed scale}, as in
(\ref{eq:2.3-10}).  This is particularly clear in the case of strongly
intermittent system, in which $R(t)$ can be very different in each
realization.  In presence of intermittency averaging over different
realizations at fixed times can produce a 
spurious regime due to the superposition of exponential and
diffusive contributions by different samples at the same time
\cite{ABCCV97}. 

The FSLE method can be easily applied for data analysis \cite{BCPPV98}.

For other approaches to address the problem of non-infinitesimal
perturbations see \cite{DF92,TGP95,KL99}.

\subsection{The $\epsilon$-entropy}
\label{sec:2.4}

The Kolmogorov-Sinai entropy, $h_{KS}$ (\ref{metentro}), of a system
measures the amount of information per unit time necessary to record
without ambiguity a generic trajectory of a system. Since the
computation of $h_{KS}$ involves the limit of arbitrary fine
resolution and infinite times (see Sect.~\ref{sec:1.1.2}), it turns
out that, practically, for most systems it is not possible to compute
$h_{KS}$.  Nevertheless, in the same philosophy of the FSLE, by
relaxing the strict requirement of reproducing a trajectory with
arbitrary accuracy, one can introduce the $\epsilon$-entropy which
measures the amount of information for reproducing a trajectory with
accuracy $\epsilon$ in phase-space.  Roughly speaking the 
$\epsilon$-entropy can be
considered the counterpart, in information theory, of the FSLE (as the
$KS$-entropy is for the Lyapunov exponent).
 Such a quantity was originally
introduced by Shannon \cite{S48}, and by Kolmogorov \cite{K56}.
Recently Gaspard and Wang \cite{GW93} made use of this concept to
characterize a large variety of processes.

We start with a continuous (in time) variable ${\bf x}(t) \in {\mathrm
I\!R}^d$, which represents the state of a $d$-dimensional system, we
discretized the time by introducing an interval $\tau$ and we
consider the new variable
\begin{equation}
\label{eq:2-1}
{\bf X}^{(m)}(t)= \left( {\bf x}(t), {\bf x}(t+\tau), \dots, 
{\bf x}(t+(m-1)\tau) \right). 
\end{equation}
Of course ${\bf X}^{(m)}(t) \in {\mathrm I\!R}^{md}$ and it corresponds to 
the trajectory which lasts for a time $T=m \tau$.

In data analysis, the space where the state vectors of the system live
is not known. Moreover, usually only a scalar variable $u(t)$ can be
measured. In such a case, one considers vectors 
$\left( u(t), u(t+\tau), \dots, u(t+m\tau-\tau) \right)$, that live in
${\mathrm I\!R}^m$ and allow a reconstruction of the original phase
space, known as delay embedding in the literature \cite{T80,SYC91}
(see also \cite{A93,A96,KS97,OSY94}), and it is a special case of
(\ref{eq:2-1}).

Introduce now a partition of the phase space ${\mathrm I\!R}^d$, using
cells of edge $\epsilon$ in each of the $d$ directions.  Since the
region where a bounded motion evolves contains a finite number of
cells, each ${\bf X}^{(m)}(t)$ can be coded into a word of length $m$,
out of a finite alphabet:
\begin{equation}
\label{eq:2-2}
{\bf X}^{(m)}(t) \longrightarrow W^{m}(\epsilon, t) =
\left( i(\epsilon, t), i(\epsilon, t+\tau), \dots, 
i(\epsilon, t+m \tau -\tau) \right), 
\end{equation}
where $i(\epsilon, t+j \tau)$ labels the cell in ${\mathrm I\!R}^d$
containing ${\bf x}(t+j \tau)$. From the time evolution of ${\bf
X}^{(m)}(t)$ one obtains, under the hypothesis of ergodicity, the
probabilities $P(W^{m}(\epsilon))$ of the admissible words $\lbrace
W^{m}(\epsilon) \rbrace$. We can now introduce the $(\epsilon ,
\tau)$-entropy per unit time, $h(\epsilon , \tau)$ \cite{S48}:
\begin{eqnarray}
\label{eq:2-3a}
h_m(\epsilon , \tau)&=& {1 \over \tau} \lbrack H_{m+1} (\epsilon,\tau)
-H_m (\epsilon,\tau) \rbrack \\
\label{eq:2-3b}
h(\epsilon , \tau) &=& \lim _{m \to \infty} h_m(\epsilon , \tau) = {1
\over \tau} \lim _{m \to \infty} {1 \over m} H_{m} (\epsilon,\tau) ,
\end{eqnarray} 
where $H_m$ is the block entropy of block length $m$:
\begin{equation}
\label{eq:2-4}
H_{m} (\epsilon,\tau) = - \sum _{ \lbrace W^{m}(\epsilon) \rbrace }
P(W^{m}(\epsilon)) \ln P(W^{m}(\epsilon)).
\end{equation} 
For the sake of simplicity, we ignored the dependence on details of
the partition.  To make $h(\epsilon, \tau)$ partition-independent one
has to consider a generic partition of the phase space $\{{\cal A}\}$
and to evaluate the Shannon entropy on this partition: $h_{Sh}({\cal A},
\tau)$. The $\varepsilon$-entropy is thus defined as the infimum over
all partitions for which the diameter of each cell is less than
$\varepsilon$ \cite{GW93}:
\begin{equation}
h(\varepsilon, \tau)= \inf_{{\cal A}:{\rm  diam}({\cal A}) \leq
\varepsilon} h_{Sh}({\cal A}, \tau)\,.
\label{def:eps}
\end{equation}
Note that the time dependence in (\ref{def:eps}) is trivial for
deterministic systems, and that in the limit $\epsilon \to 0$ one
recovers the Kolmogorov-Sinai entropy
$$
h_{KS} = \lim_{\epsilon \to 0} h(\epsilon, \tau). 
$$

The above entropies $H_m(\epsilon)$ have been  introduced by using a
partition and the usual Shannon entropy; however it is possible to
arrive at the same notion, starting from other entropy-like quantities,
that are numerically more convenient.  For example Cohen and
Procaccia \cite{CP85} proposed to estimate $H_m(\epsilon)$ as follows.
Given a signal composed of $N$ successive records and the embedding
variable ${\bf X}^{(m)}$, let us introduce the quantities:
\begin{equation} 
n_j^{(m)}={1 \over N-m} \sum_{i \neq
j} \Theta(\epsilon-|{\bf X}^{(m)}(i \tau)-{\bf X}^{(m)}(j \tau)|)\, ,
\end{equation}
then the block entropy $H_m(\epsilon)$ is given by 
\begin{equation}
H_m^{(1)}(\epsilon)=-\frac{1}{ (N-m+1)} \sum_j \ln n_j^{(m)}(\epsilon)\,.
\label{H-CP}
\end{equation}
In practice $n_j^{(m)}(\epsilon)$ is an approximation of
$P(W^m(\epsilon))$. From a numerical point of view,
correlation entropies \cite{GP83b,T98} are sometimes more convenient, so
that one studies
\begin{equation}
H_m^{(2)}(\epsilon)=- \ln \left({1 \over N-m+1} \sum_j
n_j^{(m)}(\epsilon)\right) \leq H_m^{(1)}(\epsilon)\,. 
\end{equation} 
This corresponds to  approximate the Shannon by the 
Renyi entropy of order $q=2$ \cite{KS97}.

The $(\epsilon, \tau)$-entropy $h(\epsilon,\tau)$ is well defined also for
stochastic processes. Actually the dependence of $h(\epsilon, \tau)$ on
$\epsilon$ can give some insight into the underlying stochastic process
\cite{GW93}, for instance, in the case of a stationary Gaussian
process with spectrum $S(\omega)\propto \omega^{-2}$ one has
\cite{K56}:
\begin{equation}
\label{eq:2-6}
\lim_{\tau\to 0} h(\epsilon,\tau) \sim {1 \over \epsilon
^2} \,.
\label{eq:kolmo56}
\end{equation}
However, we have to stress that the behavior predicted by
Eq.~(\ref{eq:kolmo56}) may be difficult to be experimentally observed
mainly due to problems related to the choice of $\tau$
\cite{CFKOV00,ABCFVV00a} (see also Appendix~\ref{app:eps}).

\section{Predictability in extended systems}
\label{sec:3}
\setcounter{equation}{0}

Here we consider {\it extended} dynamical systems, whose degrees of
freedom depend on space and time, and which can display
 unpredictable behaviors both in the time
and space evolution, i.e. {\it spatio-temporal chaos}. The inadequacy
of the Lyapunov exponents in characterizing predictability becomes now
well evident.

Following Hohenberg and Shraiman \cite{HS88} (see also \cite{CH93}) we
 give a more precise meaning to the terms {\it spatio-temporal}
chaos and {\it extended} systems.  For a generic system of size $L$,
we can define three characteristic lengths: $\ell_D$, $\ell_E$, $\xi$
respectively associated to the scales of {\it
dissipation} (i.e. the scale at which dissipation becomes effective,
smaller scales can be considered as inactive), {\it excitation}
(i.e. the scale at which energy is injected in the system) and {\it
correlation} (that we assume can be suitably defined). Now one has two
limiting situations.

When all the characteristic lengths are of the same order ($\ell_D,
\ell_E,\xi \sim O(L)$) distant regions of the system are strongly
correlated. Because of the coherence, the spatial nature is not very
important and one speaks of {\it temporal} chaos, i.e. the system is
basically low dimensional.

When $L \gg \xi \gg \ell_D$ distant parts of the system are weakly
correlated so that the number of (active) degrees of freedom and
consequently the number of positive Lyapunov exponents, the
Kolmogorov-Sinai entropy and the attractor dimension, $D_F$, are
extensive quantities, i.e. they increase with the system size, $L$.
Here, spatial aspects are crucial and one speaks of {\it
spatio-temporal chaos}, e.g., Rayleigh-B\'ernad convection for large
aspect ratio \cite{M90}.

The above picture is just an approximative scenario (see \cite{HS88}
for further details) but sufficiently broad to include systems ranging
from fluid dynamics to biological and chemical reactions
\cite{CH93,M90}. In spite of the fact that turbulent flows fit in this
broad definition we shall discuss the predictability problem in
turbulence in the next Section.

For detailed discussions on different physical and chemical systems
which can be included in the above definition see \cite{CH93,BJPV98}.
Here we discuss the predictability problem in a restricted class of
models, which are relatively simple from a computational and
theoretical point of view but, nevertheless, possess the essential
phenomenology of spatio-temporal chaos.

\subsection{Simplified models for extended systems and the 
thermodynamic limit}
\label{sec:3.2}

A great simplification in the study of extended
systems, usually described by partial differential equations (PDE),
can be achieved by considering discrete time and space models, and
introducing the {\it Coupled Map Lattices} (CML) \cite{K84}, i.e.
maps defined on a discrete lattice.  A typical $1$-dimensional CML
(the extension to $d$-dimensions is straightforward) can be written in
the following way:
\begin{equation}
{\bf x}_i(t+1)=(1-\varepsilon_0){\bf f}_a[{\bf x}_i(t)]+{1 \over 2}
\sum_{j=-L/2}^{L/2}  \varepsilon_j \left( 
{\bf f}_a[{\bf x}_{i+j}(t)]+{\bf f}_a[{\bf x}_{i-j}(t)]\right)\,,
\label{eq:3.2}
\end{equation}
with $\varepsilon_0=\sum_{j=1}^{L/2} \varepsilon_j$.  $L$ is the
lattice size, $i=-L/2,\dots,L/2$, ${\bf x}\in {\mathrm I\!R}^n$ is
the state variable which depends on the site and time, and ${\bf f}_a
\in {\mathrm I\!R}^n \to {\mathrm I\!R}^n$ is a non linear map, which
drives the local dynamics and depends on a control parameter $a$.
Usually, periodic boundary conditions ${\bf x}_{i+L}={\bf x}_i$ are
assumed and, for scalar variables ($n=1$), one studies coupled
logistic maps, $f_a(x)=ax(1-x)$ or tent maps, $f_a(x)=a
|1/2-|x-1/2||$.

The parameters $\{\varepsilon_i\}$ rule the strength and the form of
the coupling and they are chosen according to the physical system
under investigation.
For example, with $\varepsilon_j=0$ for $j\geq 2$, i.e. nearest
neighbor coupling, one can mimic PDE's describing reaction diffusion
processes (indeed formally the equation assumes the structure of a
discrete Laplacian). However, it could be misleading to consider CML's
simply as discrete approximation of PDE's.  Indeed, since the local
map ${\bf f}_a$ is usually chaotic, chaos in CML, differently from
PDE, is the result of many interacting chaotic
sub-systems. Hence, the correspondence between the instability
mechanisms in the two type of models is not straightforward
\cite{CH93}.

Other kinds of coupling can be considered to mimic different physical
situations, e.g., asymmetric CML (see Sect.~\ref{sec:3.7}) for
studying convective instabilities \cite{J89,BR91,VFV97}, or mean field
(globally coupled maps) version of (\ref{eq:3.2})
(Sect.~\ref{sec:3.5}) for studying neural network or population
dynamics \cite{K93}.  Further generalizations are quoted in
Ref.~\cite{KK93}.

Lyapunov exponents, attractor dimensions and entropies
can be defined (and, at least the Lyapunov exponents, numerically
computed) also for extended systems.  In particular, for $L<\infty$
the CMLs have finite dimensional phase space and the above quantities
are well defined.  In PDE's some difficulties can rise due to the
problem of the non equivalence of the norms \cite{KF76}: Lyapunov
exponents and consequently the characterization of the predictability
may depend on the chosen norm. We shall see in Sect.~\ref{sec:3.10}
that this is not just a subtle mathematical problem.

In order to build a statistical description of spatio-temporal chaos,
as Ruelle pointed out \cite{R82}, one has to require the existence of
a good thermodynamic limit for the Lyapunov spectrum
$\{\lambda_i(L)\}_{i=1,L}$. This means the existence of the limit
\begin{equation}
\lim_{L \to \infty} \lambda_i(L)=\Lambda(x)\,,
\label{eq:3.6}
\end{equation}
where $x=i/L$ is a continuous index in $[0,1]$, and $\Lambda(x)$ is a
non increasing function.  The function $\Lambda(x)$ can be viewed as a
{\it density} of Lyapunov exponents.  If such limit does not exist, the
possibility to build a statistical description of spatio-temporal
chaos would be hopeless, i.e., the phenomenology of these systems
would depend on $L$.

Once the existence of a Lyapunov density is proved, one can generalize
some results of low dimensional systems \cite{G89,BS93}, namely the
Kaplan-Yorke conjecture \cite{KY79} and the Pesin relation
(\ref{eq:1-13}).  For instance, one can generalize (\ref{eq:1-13}) to
\begin{equation}
{\cal H}_{KS}=\lim_{L\to \infty} {h_{KS} \over L}=\int_{0}^{1} dx 
\Lambda(x)\theta(\Lambda(x))
\label{eq:3.7}
\end{equation}
being $\theta(x)$ the step function. In the same way one can suppose
the existence of a {\it dimension density} ${\cal D}_F$, that is to
say a density of active degrees of freedom, i.e.  ${\cal D}_F=\lim_{L
\to \infty} {D_F / L}$ which by the Kaplan-Yorke \cite{KY79}
conjecture is given by \cite{G89}:
\begin{equation}
\int_{0}^{{\cal D}_F} dx \Lambda(x)=0\,.
\label{eq:3.8}
\end{equation}

The existence of a good thermodynamic limit is supported by numerical
simulations \cite{K86,LPR86} and some exact results \cite{S96}.
Recently Eckmann and Collet \cite{EC99} have proved the existence of a
density of degrees of freedom in the complex Ginzburg-Landau equation.
See also Refs. \cite{G89,BS93} and references therein for a discussion
on such a problem.

\subsection{Overview on the predictability problem in extended systems}
\label{sec:3.4}

In low dimensional systems, no matter how the initial disturbance is
chosen, after a -- usually short -- relaxation time, $T_R$, the
eigendirection with the largest growth rate dominates for almost all
the initial conditions (this, e.g., helps in the numerical estimates
of the Lyapunov exponents \cite{BGGS80}). On the contrary, in
high-dimensional systems this may not be true
\cite{GSO87,P93,PV94,PP98}.
Indeed, in systems with many degrees of freedom there is room for
several choices of the initial perturbation according to the specific
problem under investigation (e.g., localized on certain degrees of
freedom or homogeneous in all the degrees of freedom), and it is not
obvious that for all of them the time $T_R$ needed to align along the
maximally expanding direction is the same.

In general the situation can be very complicated.  For instance, it is
known that, also considering initially homogeneous disturbances, the
Lyapunov vectors can localize (permanently or slowly wandering) on
certain degrees of freedom \cite{K86,FBV91,PP98}.  Of course, this
will severely affect the prediction of the future evolution of the
system. Indeed, regions of large predictability time could coexist
with regions of relatively short predictability time. In
Ref.~\cite{K86,KK93,PP98} one finds an abundance of examples displaying
this phenomenology.  A detailed analysis of this problem is far from
the aims of this review; we just mention that the behavior of the
Lyapunov vectors can range from a strongly localized regime (the
origin of which can be understood by the analogy with  Anderson localization
of the wave function in disordered potential \cite{GP91}) to localized
wandering structures. In particular, in the latter case there is 
strong numerical evidence \cite{PK94c,PP98} that for a large class of
($1$-dimensional) systems the dynamics of the Lyapunov vectors
(actually the logarithm of them) falls into the universality class of
the $1$-dimensional KPZ equation \cite{KPZ86}.

In these situations the main contribution to the predictability time
comes from the time needed for the perturbation to propagate through
the system or to align along the maximally expanding direction, which
can be of the order of the system size \cite{PV94,P93,PP98}.  As a
consequence the predictability time can be much longer than the rough
estimation $T_p\sim 1/\lambda$.

Moreover,  the LE can also be unsatisfactory if one
is interested in perturbations with particular space-time
shapes. Indeed, these features have lead to the introduction of a
number of new indicators; for instance, the {\it temporal} (or {\it
specific}) Lyapunov exponents \cite{PT92,LPT96}, the {\it spatial}
Lyapunov exponents \cite{GP91,LPT96} (which characterize respectively
perturbations exponentially shaped in time and space) or the {\it
comoving} Lyapunov exponents \cite{DK87} for the characterization of
the spatio-temporal evolution of localized perturbation \cite{K86} and
of the convective instabilities \cite{AGRS86,BR91}.

Convectively unstable systems are rather interesting because, even if
the LE (computed in the stationary frame) is negative, some features
of the motion can be highly unpredictable (see Sect.~\ref{sec:3.8}).
It is also worth mentioning the existence of systems with
exponentially long (in the system size) transients during which the
dynamics is essentially unpredictable despite the fact that the LE is
negative \cite{CK88}.  This phenomenon, known under the name of {\it
stable chaos} \cite{PLOK93}, will be briefly discussed in
Sect.~\ref{sec:6.3}.

In high-dimensional systems one is also interested in predicting the
behavior of some average observable to achieve a macroscopic
description of the system.  The coarse grained (hydrodynamic like)
dynamics may be non trivial, and the largest LE, which is related to
the fine grained dynamics, is not relevant for characterizing the
predictability at a coarse grained level (see Sect.~\ref{sec:3.9}).
           
We conclude this brief overview by mentioning another interesting feature:
in spatially distributed systems coherent structures may appear. They
move maintaining for rather long times their shape.  In different
contexts one can be interested in predicting the evolution of such
structures; e.g., cyclonic/anti-cyclonic structures in the atmosphere.
A reasonable question could be the prediction of the center and
orientation of such structures: limiting to these aspects one can
indeed hope to have a rather long predictability time compared to the
rough estimate $T_p\sim O(1/\lambda)$. However, since usually such
phenomena arise in fields, whose evolution is ruled by PDE,
 the non equivalence of the norms makes a
general approach to the problem unfeasible. Therefore, one has to
resort to {\it ad hoc} treatments, based on physical intuition to
identify the most suitable norm to be used for the particular needs
(see Sect.~\ref{sec:3.10}).

\subsection{Butterfly effect in coupled map lattices}
\label{sec:3.5}

In spatially extended systems it is important to understand the way 
an uncertainty initially localized in some region will spread.
Here we study in particular the time needed for a perturbation,
initially seeded in the central site of a lattice of coupled maps, to
reach a preassigned value at the border of the lattice \cite{PV94}
(see also \cite{K86,TGP95,TL97} and Sect.~\ref{sec:3.8}).  In other
terms we wonder about the ``butterfly effect'' starting from the center
of the lattice and arriving up to the boundary.

We shall discuss the properties of such time by varying the
coupling range from local to non local in the
$1$-dimensional CML (\ref{eq:3.2}) with periodic boundary conditions.
We consider two cases: local coupling, i.e. $ \varepsilon_j=0$ if $j
\geq 2$, and non-local coupling, e.g.
\begin{equation}
\varepsilon_1= C_1 \qquad {\rm and} \qquad 
\varepsilon_j = {C_2 \over j^{\alpha} } \qquad {\rm for} \ j \ge 2
\label{eq:3.10}
\end{equation}
where $\alpha$ measures the strength of non-locality. The initial
perturbation is on the central site, i.e.
\begin{equation}
|\delta x_{i}(0)|= \delta_0 \delta_{i,0}\,.
\label{eq:3.11}
\end{equation}
We look at the predictability time $T_p$ needed for the perturbation
to reach a certain threshold $\delta_{max}$ on the boundary of the
lattice, i.e. the maximum time, $t$, such that $|\delta x_{L/2}(t)|\le
\delta_{max}$.

\begin{figure}[htb]
\centerline{\epsfig{figure=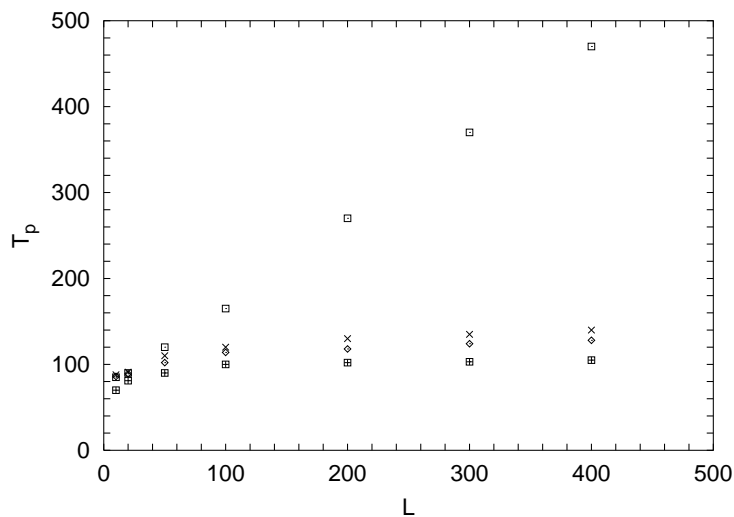,width=10cm, angle=0}}
\caption{Average predictability time $<T_p>$ versus $L$ for a CML of
logistic maps $f_a(x)=a \, x \, (1-x)$ with $a=4$ : local coupling
$\varepsilon_0=0.3$ (squares); non-local coupling (\ref{eq:3.10}) with
$C_1=0.3$, $C_2=0.01$ $\alpha=2$ (crosses) or $\alpha=3$ (diamonds);
mean field coupling $\varepsilon_i=C_2/L$ with $C_2=0.3$ (crosses
squares). The initial perturbation is applied at the center of the
lattice (site $i=0$) and has an amplitude $10^{-14}$; the maximum
admitted error is $\delta_{max}=0.1$.}
\label{fig:3.5-1}
\end{figure}

For nearest neighbor coupling, one has obviously that $\delta
x_{L/2}(t)=0$ for $t < L/2$.  Indeed, by a numerical integration of
(\ref{eq:3.2}) for the short range coupling one observes that $\delta
x_{L/2}(t)=0$ for times $t < t^*\propto \, L$; while for $t>t^*$ the
perturbation, due to the (local) chaotic dynamics, grows as $\delta
x_{L/2}(t) \sim \delta_0 \, \exp[\lambda(t-t^*)]$.  Thus for local
interactions, the predictability is mainly determined by the waiting
time $t^*$, necessary to have $|\delta x_{L/2}| > \delta_0$, which is
roughly proportional to the system size $L$. This is confirmed by
Fig.~\ref{fig:3.5-1}, where it is shown that the average
predictability time $<T_p>$ as a function of $L$ goes as
\begin{equation}
<T_p> = t_1 \, + \, G \, L\,,
\label{eq:3.12}
\end{equation}
where the time $t_1 \sim \lambda^{-1}$ is due to the exponential error
growth after the waiting time and can be neglected in large enough
lattices.  This agrees with the existence of a finite speed for the
perturbation spreading \cite{TGP95}; indeed $G$ is related to the propagation
velocity (see Sect.~\ref{sec:3.8}).

The scenario changes for non-local interactions (\ref{eq:3.10}). Now,
due to the long range coupling, the perturbation (\ref{eq:3.11}) may
propagate without any delay.  The numerical results show that even for
weak non-locality (e.g. $C_2 \, << \, C_1 $ and rather large
$\alpha$-values), the waiting time $t^*$ does not increase (or
increases very slowly) with $L$, so that
\begin{equation}
<T_p> \sim t_1 \sim \lambda^{-1}\,.
\label{eq:3.13}
\end{equation}
As shown in Fig.~\ref{fig:3.5-1}, weakly non-local couplings, and mean
field interactions ($\varepsilon_j=C_2/N$) have the same qualitative
behavior. Very accurate numerical computations have confirmed that
the dependence on $L$ is indeed very weak (only logarithmic), i.e.
$<T_p> \sim t_1 + \alpha \ln L/\lambda_1$ \cite{TL97}.

This example demonstrates that the predictability time is given by two
contributions: the waiting time $t^*$ and the characteristic time $t_1
\sim \lambda^{-1}$ associated with chaos. For non-local interactions,
the waiting time practically does not depend on the system size $L$,
while for local interactions it is proportional to $L$.  Let us
underline that in these results the non-linear terms in the evolution
of $\delta {\bf x}(t)$ are rather important.  One numerically observes
that the waiting time $t^*$ is not just the relaxation time $T_R$ of
$\delta {\bf x}$ on the tangent eigenvector.  Actually, $T_R$ is much
larger than $t^*$.

\subsection{Comoving and Specific Lyapunov Exponents}
\label{sec:3.6}

A general feature of systems evolving in space and time is that a
generic perturbation not only grows in time but also propagates in
space.  Aiming at a quantitative description of such phenomena
Deissler and Kaneko \cite{DK87} introduced a generalization of the LE
to a non stationary frame of reference: the {\it comoving} Lyapunov
exponent.  For the sake of simplicity, we consider again the case of a
1-dimensional CML.

Let us consider an infinitesimally small perturbation initially
different from zero only in one site of the lattice
(\ref{eq:3.11}). By looking at the perturbation evolution along the
line defined by $j(t)=0+[vt]$ (where $[\cdots]$ denotes the integer
part), one expects:
\begin{equation}
|\delta x_j(t)| \approx |\delta x_0(0)| e^{\displaystyle{\lambda(v) t}} \,,
\label{eq:3.14}
\end{equation}
where $\lambda(v)$ is the largest comoving Lyapunov exponent, i.e.
\begin{equation}
\lambda(v)=\lim_{t \to \infty} \lim_{L \to \infty} \lim_{|\delta x_0(0)|\to 0}
{1 \over t} \ln\left({|\delta x_{[vt]}(t)| \over |\delta x_0(0)|}\right)\,.
\label{eq:3.15}
\end{equation}
In Eq.~(\ref{eq:3.15}) the order of the limits is important to avoid
boundary effects. For $v=0$  one recovers the
usual LE. Moreover, one has that $\lambda(v)=\lambda(-v)$ (and the
maximum value is obtained at $v=0$ \cite{PT92}) when a privileged direction
does not exist, otherwise $\lambda(v)$ can be asymmetric and
the maximum can be attained at value $v\neq 0$ (see
Sect.~\ref{sec:3.8}). By writing the response function (\ref{eq:1-7})
in the moving frame one can also introduce the generalized comoving
Lyapunov exponents $L_q(v)$ for studying finite time effects
\cite{FVV99}.

There are other two indicators related to the comoving LE: the local
Lyapunov exponent \cite{P93} and the specific (or temporal) Lyapunov
exponents. Here we only briefly discuss the latter which is indeed
nothing but the Legendre transform of $\lambda(v)$.

The specific Lyapunov exponent, $\Lambda(\mu)$, has been introduced by
Politi and Torcini \cite{PT92} to characterize the growth of
exponentially shaped perturbations, i.e.
\begin{equation}
\delta x_i(t)=\Phi_i(t) e^{-\mu |i|} \qquad i=-{L\over 2}, \dots, {L\over 2} \,,
\label{eq:3.18}
\end{equation}
where $\Phi_i(t)$ gives the fluctuation with respect to the pure
exponential shape. One can see that
$\Lambda(\mu)$ is connected through a Legendre transform to
$\lambda(v)$ \cite{PT92,TGP95}. Indeed, Eq.~(\ref{eq:3.14}) defines a
local exponential profile with $\mu=d\lambda(v)/dv$, which means that
in term of the specific Lyapunov exponents one expects the
perturbation to grow according to
\begin{equation}
\delta x_{i}(t) \sim \exp\{\Lambda(\mu)t -\mu i\}\,, \qquad i=[vt]\,.
\label{eq:3.19}
\end{equation}
Note that for $\mu=0$,
$\Lambda(\mu)$ reduces to the standard LE.
Therefore, the comoving Lyapunov exponent is given by
\begin{equation}
\lambda(v)=\Lambda(\mu)-\mu {d \Lambda(\mu) \over d\mu}\,.
\label{eq:3.20}
\end{equation}
The last equation defines a Legendre transform from $(\lambda(v),v)$
to $(\Lambda(\mu),\mu)$ \cite{PT92}.  By inverting the transformation
(\ref{eq:3.20}) one obtains $v=-d\Lambda(\mu) /d\mu$.

  Working in tangent space by using standard
algorithms \cite{BGGS80}, one computes the specific Lyapunov spectrum
$\Lambda_i(\mu)$ with $i=1,\dots,L$ for each $\mu$. In the limit $L\to
\infty$ a density of such exponents exists and an interpretation of it
is discussed in \cite{LPT96,LPT97}.

\subsection{Convective chaos and spatial complexity}
\label{sec:3.7}

So far we have considered CML's with symmetric spatial coupling,
however there are many physical systems in which a privileged
direction exists, e.g., boundary layers, thermal convection and wind
induced water waves. The term usually employed for denoting such a
class of systems is flow systems. See \cite{AGRS86,AGR88,BR91,J89} for
a discussion of flow systems in different physical contexts.
  
In recent years it has received much attention a simplified model for
flow systems which is able to capture the basic phenomenology
\cite{KK93,RP96,WK94}.  A minimal model is a chain of maps with
unidirectional coupling \cite{AGR88,P89,KK93,VFV97,FVV99}:
\begin{equation}
x_n(t+1)=(1-c)f_a(x_n(t))+ c f_a(x_{n-1}(t)),
\label{eq:3.28}
\end{equation}
where $t$ and $n$ $(=1,\dots,L)$ are 
the discrete time and space respectively; the
map $f_a(x)$ is usually chosen to be the logistic map. One can consider
different boundary conditions, $x_0(t)$. For instance,
$x_0(t)=x^{*}$ with $x^*$ being an unstable fixed point of the map $f_a(x)$,
or more generic time dependent boundary conditions where $x_0(t)$ is
equal to a known function of time $y(t)$, which can be periodic,
quasi-periodic or chaotic.  Here, following Pikovsky
\cite{P89,P91,P92}, we consider a quasi-periodic boundary condition $x_0
(t)=0.5+ 0.4\sin (\omega t)$, with $\omega =\pi(\sqrt{5}-1)$. However,
the results we are going to discuss do not depend too much on the
details of the boundary conditions, i. e. on using $x_0 (t)$
quasi-periodic or chaotic.

A central concept in the study of flow systems is the one of {\it
convective} instability, i.e.  when a perturbation grows exponentially
along the flow but vanishes locally.
\begin{figure}[htb]
\centerline{\epsfig{figure=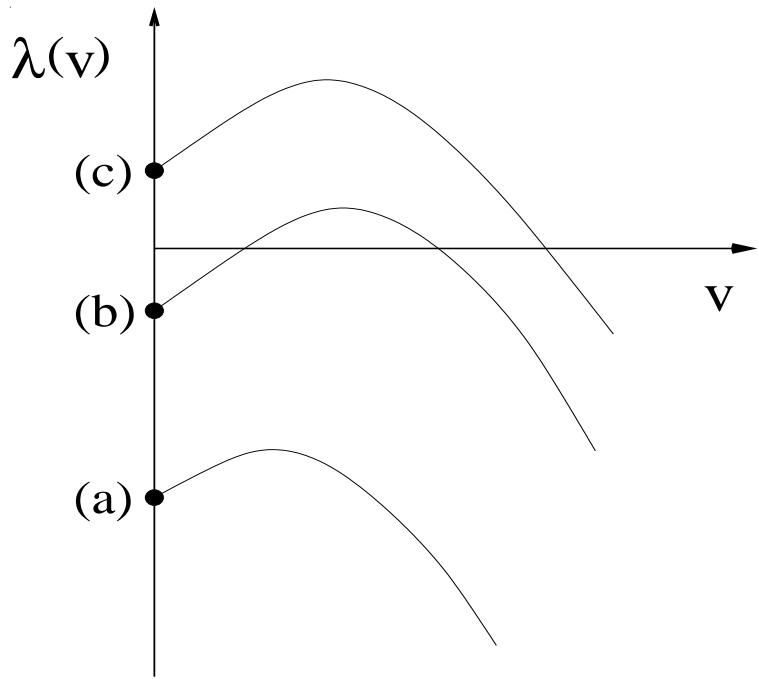,width=10cm, height=9cm, angle=0}}
\caption{Sketch of the behavior of  $\lambda(v)$ for 
(a) an absolutely and convectively stable flow, 
(b) absolutely stable but convectively unstable flow, and 
(c) absolutely unstable flow.}
\label{fig:3.7-0}
\end{figure}

We may give a description of the phenomenology of flow systems in
terms of the largest LE and of the comoving LE. The absolute stability
is identified by the condition $\lambda(v) < 0$ for all $v\geq 0$; the
convective instability corresponds to $\lambda_1=\lambda(v=0) < 0$ and
$\lambda(v)>0$ for some velocities $v>0$ and finally standard chaos
(absolute instability) is present when $\lambda_1=\lambda(v=0) > 0$.
See Fig.~\ref{fig:3.7-0} for a sketch of the possible behaviors.

\begin{figure}[htb]
\centerline{\epsfig{figure=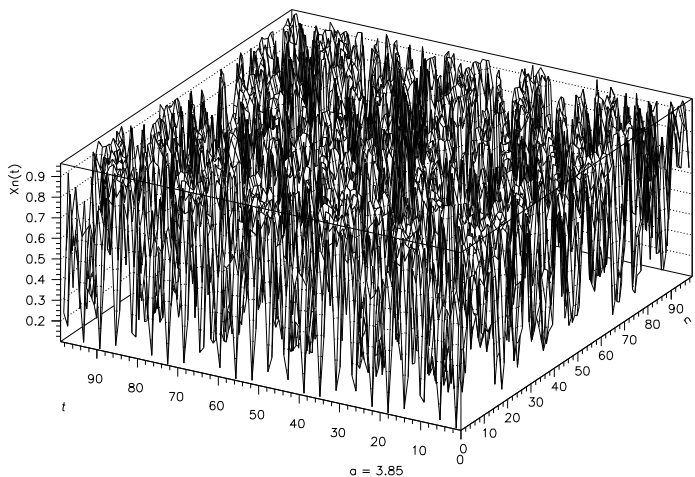,width=10cm, angle=0}}
\caption{Evolution of a state of the system (\ref{eq:3.28}) where
   $f_a(x)$ is the logistic maps, the boundary condition is quasi-periodic,
   $a=3.85$ and $c=0.7$: in this case $\lambda_1 < 0$ but the system is 
   convectively unstable.}
  \label{fig:instconv}
\end{figure}
The convective instability is conceptually very interesting, because
even if the largest LE is negative the behavior of the system may be
very hard to predict, as Figure~\ref{fig:instconv} suggests.

For this kind of spatial ``complexity'' there is not a simple and
systematic characterization.  A first explanation for these features
may be found in the sensitivity of convective unstable systems on
small perturbations at the beginning of the chain (always present in
physical system), which grow exponentially while they are transmitted
along the flow.  This simple intuition can be made more quantitative
defining an indicator which measures the degree of sensitivity on the
boundary conditions\cite{VFV97,FVV99}. We wonder how an uncertainty
$|\delta x_0 (t)|= \delta_0$ in the knowledge of the boundary
conditions will affect the system. We consider only the case of
infinitesimal perturbations, i.e. $\delta x_n$ evolves according to
the tangent space dynamics, and, for the moment we do not consider
 intermittency (i.e. time fluctuations of
the comoving Lyapunov exponents).

The uncertainty $\delta x_n (t)$, on the determination of the variable
at time $t$ and site $n$, is given by the superposition of the evolved
$\delta x_0 (t-\tau)$ with $\tau =n/v$:
\begin{equation}
\label{eq:3.30}
\delta x_n(t) \sim \int \delta x_0 (t-\tau) e ^{\lambda(v)\tau} dv
     = \delta_0 \int e ^{ [\lambda(v)/v] n} dv. 
\end{equation}
Since we are interested in the asymptotic spatial behavior,
i.e. large $n$, we can write:
\begin{equation}
\label{eq:3.31}
\delta x_n(t)\sim \delta_0 e ^{\Gamma n},
\end{equation}
The quantity $\Gamma$ can be considered as a sort of
spatial-complexity-index, an operative definition of which is the
following:
\begin{equation}
\label{eq:3.33}
\Gamma = \lim _{n \to \infty} \frac {1}{n}\Biggl
\langle \ln \frac {|\delta x_n|} {\delta_0} \Biggr \rangle , 
\end{equation}
where the brackets mean a time average.

In the particular case of a non intermittent system, a simple 
saddle-point estimate of Eq.~(\ref{eq:3.30}) gives
\begin{equation}
\label{eq:3.32}
\Gamma=\max _v  \left[{\lambda(v) \over v}\right]. 
\end{equation}
Equation (\ref{eq:3.32}) is a link between the comoving and the
``spatial'' Lyapunov exponent $\Gamma$, i.e. a relation between the
convective instability of a system and its sensitivity to the boundary
conditions.

Eq.~(\ref{eq:3.32}) holds exactly only in absence of intermittency; in
general the relation is more complicated. One can introduce the
effective comoving Lyapunov exponent, $\tilde \gamma _t (v)$, that
gives the exponential changing rate of a perturbation, in the frame of
reference moving with velocity $v$, on a finite time interval $t$.
According to general arguments (see Sect.~\ref{sec:2.1} and \cite{PV87})
one has $\langle \tilde \gamma _t (v) \rangle = \lambda(v)$.  Then,
instead of (\ref{eq:3.30}) one obtains
\begin{equation}
\label{eq:3.34}
\delta x_n(t) \sim \delta_0 
     \int e ^{[\tilde \gamma _t (v)/v] n} dv,  
\end{equation}
and therefore:
\begin{equation}
\label{eq:3.35}
\Gamma = \lim _{n \to \infty} \frac {1}{n}\Biggl
\langle \ln \frac {|\delta x_n|} {\delta_0} \Biggr \rangle =
\lim _{n \to \infty} \frac {1}{n} 
\ln \frac {|\delta x_n^{typical}|} {\delta_0}=\Biggl
\langle \max _v \left[{\tilde \gamma _t (v) \over v} \right]\Biggr \rangle .
\end{equation}
Therefore, because of the fluctuations, it is not possible to write
$\Gamma$ in terms of $\lambda (v)$, although one can obtain a
lower bound~\cite{VFV97}:
\begin{equation}
\label{eq:3.36}
\Gamma \geq \max _v \left[
\frac {\langle \tilde \gamma _t (v) \rangle} {v}\right]=
\max _v \left[\frac {\lambda (v)} {v} \right] \equiv \Gamma^{\ast}\,.
\end{equation}

\begin{figure}[h]
\centerline{\epsfig{figure=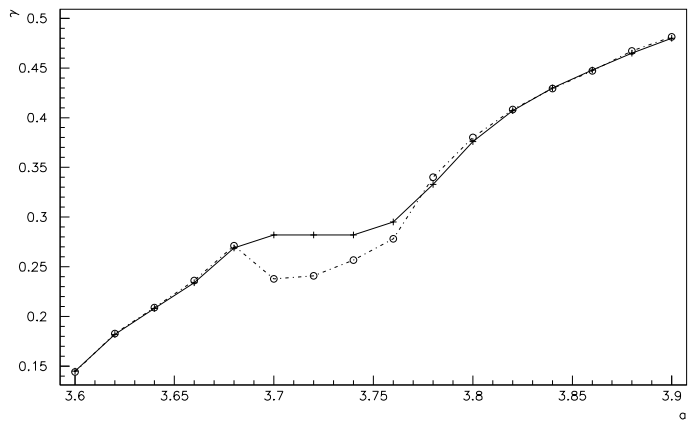,width=10cm, angle=0}}
\caption{$\Gamma$ ($+$) and $\Gamma ^{\ast}$ ($\bigcirc$) vs. $a$ at a
fixed value of $c$ ($c=0.7$), for the system (\ref{eq:3.28}) of
logistic maps with quasi-periodic boundary conditions (the system is
convectively unstable for all the considered values of the
parameters).  }
\label{fig:1old}
\end{figure}

In Fig.~\ref{fig:1old} we show $\Gamma$ and $\Gamma^{\ast}$ vs. $a$
for a fixed value of $c$.  There is a large range of values of the
parameter $a$ for which $\Gamma$ is rather far from $\Gamma^{\ast}$.
This difference is only due to intermittency, as investigations of the
map $f_a(x)=ax$ mod $1$ or the computation of the generalized spatial
Lyapunov exponents $L_s(q)$~\cite{VFV97} confirm.

Concluding, we underline that the spatial complexity displayed by
these systems indicates that the unpredictability of a system 
cannot be completely
reduced to the existence of at least one positive LE.

\subsection{Space-time evolution of localized perturbations}
\label{sec:3.8}

\begin{figure}[htb]
\centerline{\epsfig{figure=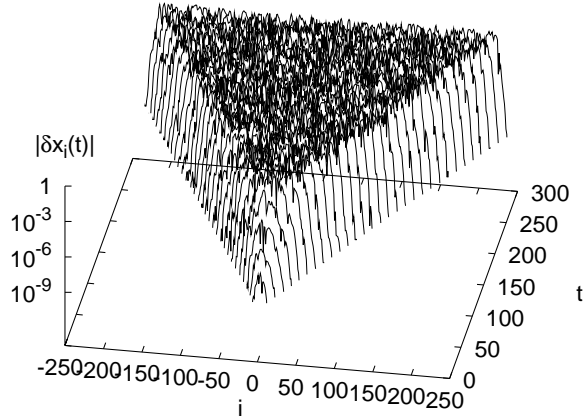,width=10cm,angle=0}(a)}
\centerline{\epsfig{figure=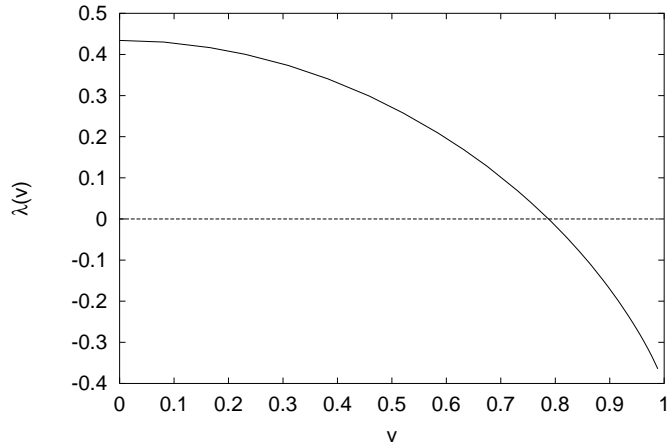,width=9cm,angle=0}\hspace{1cm}(b)}
\protect\caption{(a) Space-time evolution of $|\delta x_i(t)|$ for an
initially localized perturbation (\ref{eq:3.11}) with
$\delta_0=10^{-8}$.  We used a CML of tent maps, $f_a(x) = a(1/2 -
|x-1/2|)$, with $a=2$, $\varepsilon=2/3$ and $L=1001$. (b)
$\lambda(v)$ for $v>0$ for the CML of Fig.~(a). The straight line
indicates the zero and the intersection between the curve $\lambda(v)$
and $0$ indicates the perturbation velocity $V_F \approx 0.78$.}
\label{fig:3dplotcml}
\end{figure}

In Fig.~\ref{fig:3dplotcml} we show the evolution of a perturbation
$|\delta x_i(t)|$ initialized as (\ref{eq:3.11}) as a function of
space and time ($i,t$), for a 1-dimensional lattice of locally coupled
tent maps.  The perturbation grows in time and propagates linearly in
space creating a sort of predictability ``horizon'': this defines a
propagation velocity $V_F$ \cite{K86,PT92,TGP95}.

The velocity $V_F$ is defined in terms of the left and right edges of
the disturbance i.e. the first left (right) site for which at time $t$
the perturbation reaches a preassigned arbitrary threshold. 
Numerically it has been found that $V_F$ is independent
either of the amplitude of the initial perturbation, $\delta_0$,
and of the threshold value, so that it is a well defined
quantity \cite{K86}.

It is easy to realize that $V_F$ is nothing but the velocity of a moving 
frame of reference in which the perturbation is seen neither
to grow nor to decrease (i.e. the frame comoving with the edges of the
perturbation).  Therefore, $V_F$ is given by \cite{K86}
\begin{equation}
\lambda(V_F)=0\,.
\label{eq:3.22}
\end{equation}

The interesting point in Eq.~(\ref{eq:3.22}) is that it gives not only
a definite prescription to derive the propagation velocity but also a
link between the velocity and the stability properties of the system.
From this point of view it is instructive to look at the propagation
velocity in another way \cite{TGP95}.

The perturbation at different times resembles a propagating front,
similar to those encountered in reaction-diffusion processes. But
while in the latter context the front usually separates two stable phases or
a stable from an unstable phase, here one phase is
unstable and the other chaotic \cite{TGP95}.  
Made this analogy one can ask if it
is possible to obtain the propagation velocity as  in
reaction-diffusion phenomena, where we know that the dynamics
spontaneously selects the minimum allowed velocity \cite{KPP37}.

Torcini et al. \cite{TGP95} have shown that this is indeed the case.
They studied the evolution of a perturbation with an exponential
profile (\ref{eq:3.18}) which, according to the definition of the
specific Lyapunov exponent, evolves as in Eq.(\ref{eq:3.19}), i.e.
$\delta x_i(t) \sim \exp \lbrace \Lambda(\mu)t-\mu i\rbrace$. 
This last relation tells us that the velocity of
a front with shape given by $\mu$ is $V(\mu)=\Lambda(\mu) / \mu$.
According to the analogy with reaction-diffusion systems, one expects
that a generic localized perturbation develops an exponential decaying
shape (at the edges) with a definite exponent $\mu_0$ (selected by the
dynamics) \cite{TGP95}.  This means that the propagation velocity
$V_F$ is determined by the relation $V_F=V(\mu_0)$.  Now the problem
is to compute $\mu_0$.

From Eq.~(\ref{eq:3.20}), which relates $\Lambda(\mu)$ and $\lambda(v)$ 
through a Legendre transformation, one obtains
\begin{equation}
{d V \over d \mu} = {1 \over \mu} \left( {d \lambda \over d \mu} -
   {\lambda \over \mu} \right) = -{\lambda(v) \over \mu^2} \,.    
\label{eq:3.25}
\end{equation}
Moreover, since $\lambda(V_F)=0$ (\ref{eq:3.22}) 
one has that $dV/d\mu=0$ at  $\mu_0$ such that
$V(\mu_0)=V_F$, i.e. $\mu_0$ selects the minimal velocity.
Indeed $\Lambda(\mu)$ is convex 
(being a Legendre transform), so that the
minimum is unique and
\begin{equation}
V_F= {\Lambda(\mu_0) \over \mu_0} = \left.
{d \Lambda (\mu)  \over d \mu} \right|_{\mu=\mu_0}\;.
\label{eq:3.26}
\end{equation}
Thus for an infinitesimal perturbation, the selected velocity is the
lowest possible one \cite{TGP95}.

Summarizing, for short range coupling the speed of propagation is
finite and fully determines the spatio-temporal evolution of the
perturbation.  The situation becomes different for long-range coupling
as (\ref{eq:3.10}).  In this case the velocity of propagation is
unbounded \cite{PV94}.  For the sake of completeness, we mention that
the long-range coupling case has been also investigated in terms of a
specific-like Lyapunov exponent which characterizes power law shaped
perturbations \cite{TL97}. The result of this analysis shows that the
perturbation propagates exponentially fast with a rate given by the
ratio of the largest LE and the power of the coupling.

We conclude this Section by mentioning that there are cases in which
the analysis in terms of $\lambda(v)$ or, equivalently, $\Lambda(\mu)$
fails to give the measured propagation velocity.  Indeed, it has
been found that $V_F$ can be larger than the velocity for which
$\lambda(v)=0$.  A finite propagation velocity has been measured
even in systems with $\lambda<0$ (the so-called stable chaos
phenomenon, see Section~\ref{sec:6.3}) for which the above analysis
predicts $V_F=0$ \cite{PT94}.

\begin{figure}[hbt]
\centerline{\epsfig{figure=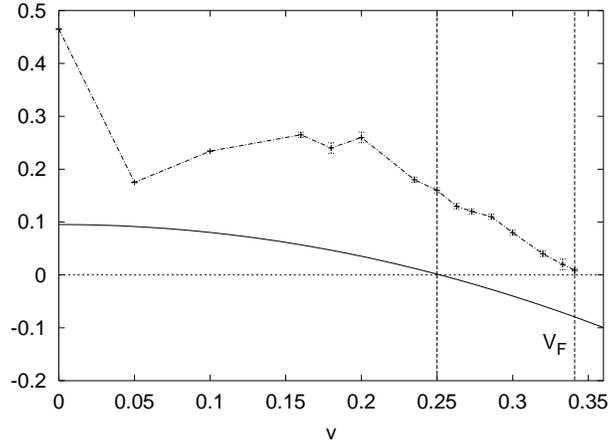,width=9cm,angle=0}}
\caption{$\max_{\delta}[\lambda(\delta,v)]$ (dashed line with points)
versus $v$ for the shift map $f(x)=r x$ mod $1$ with $r=1.1$ and
$\varepsilon_0=1/3$, compared with $\lambda(v)$ (continuous line). The
two vertical lines indicates the velocity obtained by (\ref{eq:3.22})
which is about $0.250$ and the directly measured one $V_F \approx
0.342$. Note that $\max_{\delta}[\lambda(\delta,v)]$ approaches zero
exactly at $V_F$.}
\label{fig:vlvnl}
\end{figure}

This failure is related to the presence of strong non
linearities. Recently, it has been proposed to generalize
(\ref{eq:3.22}) to the non linear regime of the perturbation growth by
the definition of the Finite Size Comoving Lyapunov Exponent \cite{CT99},
$\lambda(\delta,v)$. It measures the  divergence rate of
perturbations of size $\delta$ (not necessarily infinitesimal) in a
moving frame.  The algorithm is a generalization of the FSLE
(Sect.~\ref{sec:2.3}), where now one follows an initially localized
perturbation along the line $[vt]$. In Fig.~\ref{fig:vlvnl} we compare
$\lambda(v)$ with $\lambda(\delta,v)$ for a CML of shift maps, i.e.
$f(x)= r x \;\;{\mbox {mod}}\,1$. The latter goes to zero exactly at
the directly measured propagation velocity $V_F$. Similar results hold
for other maps \cite{CT99}.  These results suggest that a
generalization of Eq.(\ref{eq:3.22}), which is able to take into
account also possible non linear effects, is:
$$
\max_{\delta}\left[ \lambda(\delta,V_F)\right]=0.
$$
The numerical evidences also suggest that the condition which should
be accomplished in order to have deviation from the linear prediction
given by (\ref{eq:3.22}) and (\ref{eq:3.26}) is that
$\lambda(\delta,v=0)> \lambda(0,0)=\lambda$, confirming a conjecture
done in \cite{TGP95}. However, even if interesting such a behavior
seems to be rather non generic.

\subsection{Macroscopic chaos in Globally Coupled Maps}
\label{sec:3.9}

Recently the emergence of non trivial collective behaviors
in high-dimensional dynamical systems has gathered much
attention\cite{CM92,K89,K90,PK94}. A
limit case of macroscopic coherence is the global synchronization of
all the parts of the system.  Beyond synchronization there exist other
 interesting phenomena, among which we just mention: 
clustering\cite{K89,KK91,CFV96} and
collective motion in globally coupled
maps (GCM)\cite{K95,SK97,PK94}. 
The latter behavior, in the case that we call 
{\it macroscopic chaos} \cite{CFVV99,SK98} (see below), 
is the subject of this Section.

Let us consider a globally coupled map (GCM) defined as follows
\begin{equation}
\label{eq:3.38}
x_n(t+1)=(1-\varepsilon)f_a(x_n(t))+ 
  \frac {\varepsilon} {N} \sum_{i=1}^{N}  f_a(x_{i}(t)),
\end{equation}
where $N$ is the total number of elements.

The evolution of a macroscopic variable, e.g., the center of mass
\begin{equation}
\label{eq:3.39}
m(t)= \frac {1}{N} \sum_{i=1}^{N}  x_{i}(t) ,
\end{equation}
upon varying $\varepsilon$ and $a$ in Eq.~(\ref{eq:3.38}), displays
different behaviors \cite{CFVV99}:
\begin{description}
\item{(a)} {\it Standard Chaos}: $m(t)$ obeys a Gaussian statistics
with a standard deviation $\sigma _{N}=\sqrt{\langle m(t)^2\rangle -
\langle m(t) \rangle ^2} \sim N^{-1/2}$;
\item{(b)} {\it Macroscopic Periodicity}: $m(t)$ 
is a superposition of a periodic function and small fluctuations 
$O(N^{-1/2})$; 
\item{(c)} {\it Macroscopic Chaos}: $m(t)$ displays an irregular 
motion as it can be seen by looking at the plot of $m(t)$ vs. $m(t-1)$ that 
appears as a structured function (with thickness $\sim N^{-1/2}$), and 
suggests a chaotic motion for $m(t)$. 
\end{description}

Phenomena (a) and (b) also appear in CML with local coupling in
high  enough dimensional lattices \cite{CM92}, for the interesting case (c), as
far as we know, there is not a direct evidence in finite dimensional
CMLs.

In the case of {\it macroscopic chaos} one can expect that the center of
mass evolves with typical times longer than the characteristic time of
the full dynamics (i.e.  the microscopic dynamics); the order of
magnitude of the latter time may be estimated as
$1/\lambda_1$. Indeed, conceptually, macroscopic chaos for GCM can be
thought of as the analogous of the hydro-dynamical chaos for molecular
motion, where, in spite of a huge microscopic Lyapunov exponent
($\lambda_1 \sim 1/\tau_c\sim 10^{11} s^{-1}$, $\tau_c$ being the
collision time), one can have rather different behaviors at a
hydro-dynamical (coarse grained) level, i.e.: regular motion
($\lambda_{hydro}\leq 0$) or chaotic motion ($0<\lambda_{hydro}\ll
\lambda_1$). In principle, if one knows the hydrodynamic equations,
it is possible to characterize the macroscopic behavior by means of
standard dynamical system techniques. However, in generic CML there
are no general systematic methods to build up the macroscopic
equations, apart from particular cases \cite{K95,PK94}. Therefore, here
we discuss the macroscopic behavior of the system relying upon the
full microscopic level of description.

The microscopic Lyapunov exponent cannot give a characterization of
the macroscopic motion. To this purpose, recently
 different approaches have been proposed based on the evaluation of the
self-consistent 
Perron-Frobenius (PF) operator \cite{K95,PC92,PK94} and on the FSLE
\cite{CFVV99,SK98}. Despite the conceptual interest of the former (in
some sense the self-consistent PF-operator plays a role similar to that of the
Boltzmann equation for gases \cite{CFVV99}), here we shall only
discuss the latter which seems to us more appropriate to address the
predictability problem.

We recall that for chaotic systems, in the limit of infinitesimal
perturbations $\delta \to 0$, one has $\lambda(\delta) \to
\lambda_{1}$, i.e.  $\lambda(\delta)$ displays a plateau at the value
$\lambda_{1}$ for sufficiently small $\delta$. However, for non
infinitesimal $\delta$, one can expect that the $\delta$-dependence of
$\lambda(\delta)$ may give information on the characteristic
time-scales governing the system, and, hence, it could be able to
characterize the macroscopic motion.  In particular, at large scales,
i.e.  $\delta \gg 1/\sqrt{N}$, one expects the (fast) 
microscopic components to saturate
and $\lambda(\delta) \approx \lambda_M$, where $\lambda_M$ can be
fairly called the ``macroscopic'' Lyapunov exponent.

The FSLE has been determined by looking at the evolution of
$|\delta m(t)|$, which has been initialized at the value $\delta
m(t)=\delta_{min}$ by shifting all the elements of the unperturbed
system by the quantity $\delta_{min}$ (i.e. $x^{'}_{i}(0)
=x_{i}(0)+\delta_{min}$), for each realization.  The computation has
been performed by choosing  the tent map as local map, but similar results
can be obtained for other  maps \cite{SK98,CFVV99}.

\begin{figure}[hbt]
\centerline{\epsfig{figure=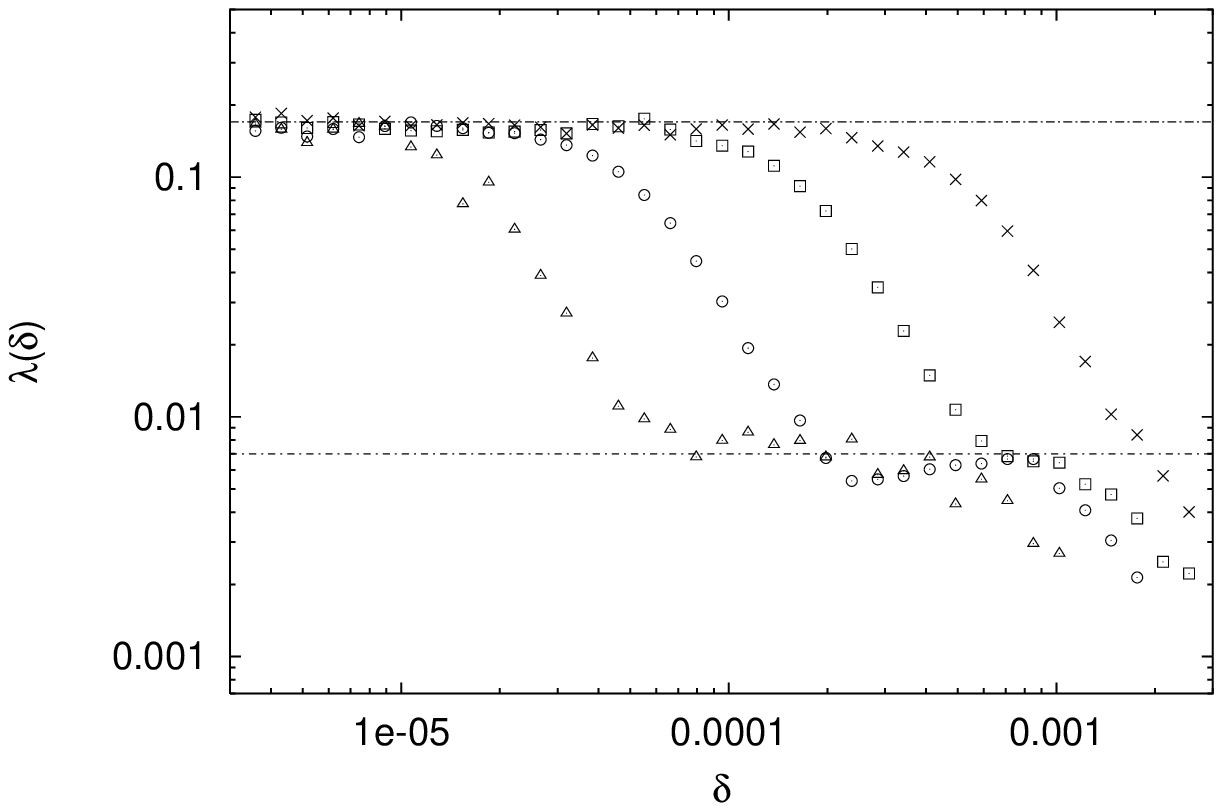,width=9cm,angle=0}(a)}
\centerline{\epsfig{figure=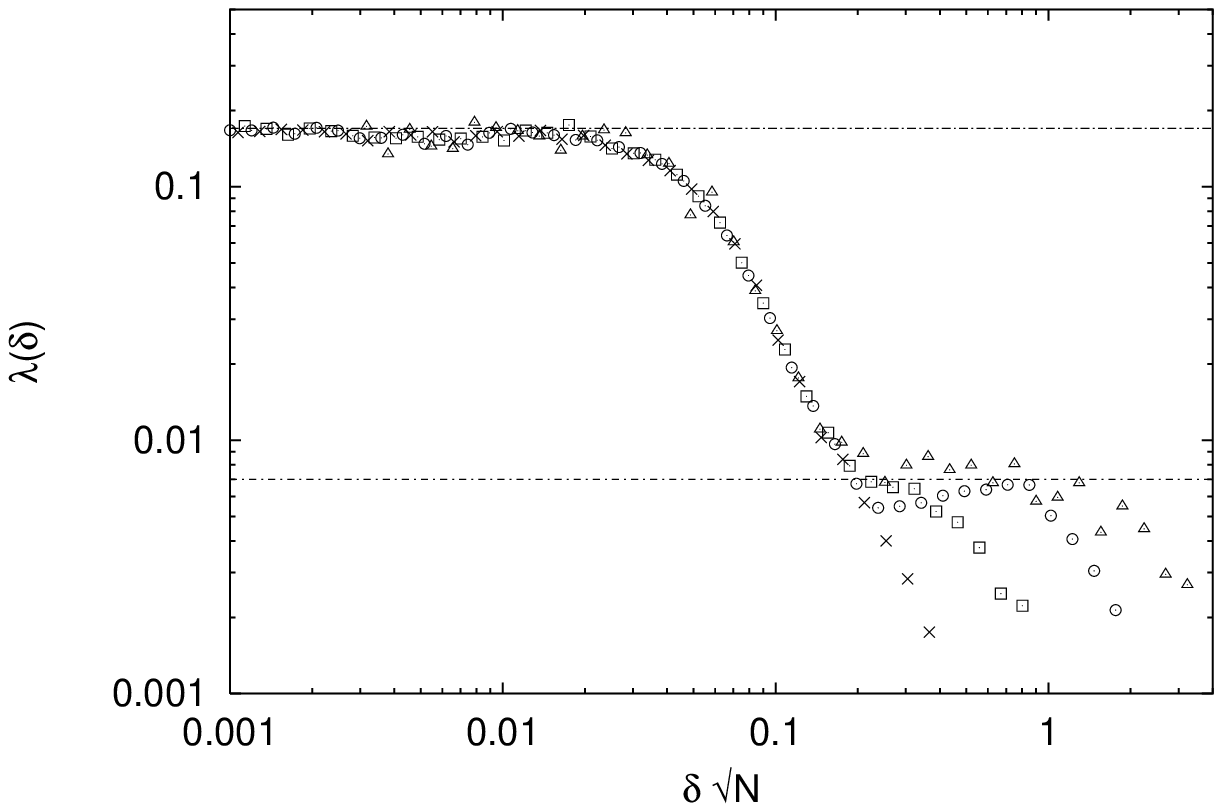,width=9cm,angle=0}(b)}
\caption{$\lambda(\delta)$ versus $\delta$ for the system
(\ref{eq:3.38}) with $a=1.7$, $\varepsilon=0.3$ for $N=10^4$
($\times$), $N=10^5$ ($\Box$), $N=10^6$ ($\odot$) and $N=10^{7}$
($\triangle$) .  The first plateaus corresponds to the microscopic
Lyapunov exponent $\lambda_{micro} \approx 0.17$ and the second one to
the macroscopic Lyapunov exponent $\lambda_{macro}\approx 0.007$.  The
average is over $2\cdot 10^3$ realizations for
$N=10^{4},\,10^5,\,10^6$ and $250$ realizations for $N=10^7$. 
 (b) The same as (a) rescaling the
$\delta-$axis with $\sqrt{N}$.}
\label{fig:macro}
\end{figure}

Figure~\ref{fig:macro}a shows $\lambda(\delta)$ versus $\delta$ in the
case of macroscopic chaos. One has two plateaus: at small values of
$\delta$ ($\delta \leq \delta_{1}$), as expected from general
considerations, $\lambda(\delta)=\lambda_{1}$; for $\delta \geq
\delta_{2}$ one has another plateau, the ``macroscopic'' Lyapunov
exponent, $\lambda(\delta)=\lambda_{M}$.  Moreover, $\delta_1$ and
$\delta_2$ decrease at increasing $N$: indeed, by looking at
Fig.~\ref{fig:macro}b one can see that $\delta_{1},\;\delta_{2} \sim
1/\sqrt{N}$. It is important to observe that the macroscopic plateau,
being almost non-existent for $N=10^4$, becomes more and more resolved
and extended on large values of $\delta \sqrt{N}$ at increasing $N$ up
to $N=10^7$. Therefore we can argue that the macroscopic motion is
well defined in the limit $N \rightarrow \infty$ and one can
conjecture that in this limit the microscopic signature in the
evolution of $\delta m(t)$ completely disappears in favor of the
macroscopic behavior. In the case of standard chaos ($\lambda_{M}<0$)
one has only the microscopic plateau and then a fast decreasing of
$\lambda(\delta)$\cite{CFVV99}.

\subsection{Predictability in presence of coherent structures}
\label{sec:3.10}

Here we discuss some problems which arise in characterizing the
predictability in continuous systems, described by PDE.  In this case
the norms are not equivalent \cite{KF76} and the computation of
the LE can give different results.  Rather than discussing the problem
in general terms, we consider here two-dimensional turbulence as a
specific example.  The choice of this example is due to several
reasons.  First of all, two-dimensional turbulence is a 
continuous system of relevance in atmospheric physics, and it 
 has been extensively investigated in
the last years \cite{SY93,T94,PT97,BCV00}.  The statistical theory for
two-dimensional turbulence has been developed by Kraichnan and
Batchelor \cite{KM80} on a similar basis to the Kolmogorov theory for
three-dimensional turbulence. The main formal difference 
 is the existence of a second
inviscid invariant, the enstrophy (average square vorticity). As a
consequence, in the limit of high Reynolds numbers, the energy cannot
be dissipated by viscosity and one expects a direct cascade of
enstrophy. With an input source at intermediate scales, the energy
injected into the system is transferred to large scales by an {\em
inverse} cascade. A large numbers of numerical simulations
\cite{SY93,BCV00} and experiments \cite{PT97} have demonstrated the
universality of the inverse cascade with spectral index very close to
the predicted Kolmogorov exponents.

The situation is much less clear for what concerns the direct cascade.
The predicted spectral slope (Kraichnan--Batchelor spectrum) is seldom
observed and even universality with respect to the forcing or to the
form of dissipation is questioned \cite{McW84}.  The freely decaying
evolution is characterized by the emergence of coherent structures
\cite{McW84} which eventually dominate the dynamics. Coherent
structures are weakly dissipative, rather regular regions of fluids in
the turbulent background flow whose interactions can be
approximately described by a conservative dynamics \cite{LSB88}. The
spontaneous emergence of coherent structures makes two-dimensional
turbulence a prototype model for geophysical flows \cite{L90} and,
most important for our purpose, gives a natural example for
illustrating the effects of choosing different error norms.

The equation for describing two-dimensional turbulence is the
Navier--Stokes equation written for the scalar vorticity $\omega=\nabla \times
{\bf v}$
as \cite{BLSB81,McW84}
\begin{equation}
{\partial \omega \over \partial t} +
{\partial (\psi,\omega) \over \partial (x,y)} =
(-1)^{p+1} \nu_{p} \triangle^{p} \omega
\label{eq:3.4-1}
\end{equation}
where $\psi$ is the stream function such that ${\bf v}=(\partial_y \psi,
-\partial_x \psi)$ and $\triangle \psi = -
\omega$.  As customary in direct numerical simulations, the
dissipation is modified by employing high order viscosity $p>1$ in
order to achieve larger Reynolds numbers. The numerical results
discussed below are obtained by integrating (\ref{eq:3.4-1}) by means
of a standard pseudo--spectral code on a periodic computational domain
with resolution $N \times N$.

The classical theory of predictability in turbulence \cite{L71,LK72}
studies the evolution of a difference (or error) field, defined as
\begin{equation}
\delta \omega({\bf x},t) = {1 \over \sqrt{2}} \left(\omega'({\bf x},t) -
\omega({\bf x},t)\right)
\label{eq:3.4-2}
\end{equation}
where $\omega$ and $\omega'$ are solutions of (\ref{eq:3.4-1}) started
from slightly different initial conditions.  The ``error'' is
computed from $\delta \omega$ and measured in terms of 
a given norm which is the
subject of our discussion. 
Indeed the method used for defining the distance between the reference and
perturbed field is a delicate point for continuous systems such as
Navier--Stokes equations. 
 Classical norms are based on the invariants
of (\ref{eq:3.4-1}) in the inviscid limit $\nu_p=0$, i.e. enstrophy and energy norms \cite{LK72,KYO90}
\begin{eqnarray}
Z_{\delta}(t) &=&  {1 \over 2} \int d^{2}x\, |\delta \omega({\bf x},t)|^2
               = \int_{0}^{\infty} dk\, Z_{\delta}(k,t)
\label{eq:3.4-3}\\
E_{\delta}(t) &=& \int_{0}^{\infty} dk\, k^{-2}\, Z_{\delta}(k,t).
               =  \int_{0}^{\infty} dk\, E_{\delta}(k,t)
\label{eq:3.4-4}
\end{eqnarray}
where we have also introduced the enstrophy ($Z_{\delta}$) 
and energy  ($E_{\delta}$) error spectra.
It is also natural to introduce the relative errors, defined as
\begin{equation}
r(t) = {E_{\delta}(t) \over E(t)} , \qquad
z(t) = {Z_{\delta}(t) \over Z(t)}
\label{eq:3.4-5}
\end{equation}
where $E(t)=1/2\int v^2(x)dx$ and $Z=1/2\int \omega^2(x)dx$, 
and the relative error spectrum
\begin{equation}
r(k,t) = {E_{\delta}(k,t) \over E(k,t)}
= {Z_{\delta}(k,t) \over Z(k,t)}\,,
\label{eq:3.4-6}
\end{equation}
This issue was already addressed in
\cite{KYO90} where the infinitesimal (linear) error growth was
computed using several ``Eulerian norms'' as (\ref{eq:3.4-3})-(\ref{eq:3.4-4}).

We will consider an initial error given by complete uncertainty at
small scales, i.e. $r(k,0) = 0$ for $k < k_0$ and $r(k,0)=1$ for
$k>k_0$. This assumption is physically justified by the finite
resolution of any measurement device and/or the numerical simulation
scheme.  For an infinitesimal perturbation, the error is expected to
grow exponentially with the largest LE $\lambda_1$.  Because we are
dealing with a dissipative system which ultimately collapses on the
trivial fixed point $\omega=0$, $\lambda_1$ is formally negative.
However, this is only a formal problem. Indeed in high Reynolds number
turbulence the dissipation time scale is much longer than the dynamical
time and we can make use of the effective LE $\gamma(t)$
(\ref{eq:2.1-8}). For $t$ much smaller than the dissipation time, we
can consider, from any point of view, $\gamma(t)$ as the Lyapunov
exponent of the decaying turbulence.

\begin{figure}[hbt]
\centerline{\epsfig{figure=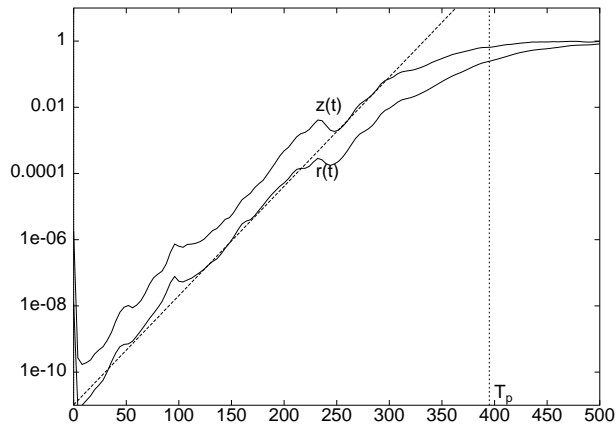,width=8cm,angle=0}}
\vspace{0.4truecm}

\caption{Relative energy ($r$) and enstrophy ($z$) error growth for a
$512^2$ simulation. $T_p$ indicate the predictability time defined as
$r(T_p)=1/4$. The dashed line represents the exponential regime
$r(t)\sim \exp(0.08 t)$. }
\label{fig:3.10-1}
\end{figure}

The exponential growth regime ends at times much smaller than the
dissipative time, as soon as the separation of the two fields cannot
be any more considered infinitesimal and the finite error regime sets
in.  The predictability time is defined by means of the accepted
tolerance $\Delta$ or, which is equivalent, by a threshold for the
relative errors (\ref{eq:3.4-5}). We will follow the classical
prescription for the predictability time $r(T_p)=1/4$ \cite{LK72}. In
Figure~\ref{fig:3.10-1} we plot relative errors (\ref{eq:3.4-5}) as
functions of time for a $512^2$ resolution simulation
\cite{BCCV97}. For small times ($t<250$) we can see an exponential
growth for both $r(t)$ and $z(t)$ with effective LE $\gamma \simeq
0.08$.  At larger times the error curves bend and a predictability time
estimation with energy norm gives $T_p \simeq 395$.  From 
Figure~\ref{fig:3.10-1} we learn at least two lessons.  First (but not
surprisingly) about half of the predictability time is governed by
non-exponential error growth behavior. This is another demonstration
of the little relevance of LE for characterizing predictability in
realistic complex systems.  The second observation is that the
different norms $r(t)$ and $z(t)$ give qualitatively similar
results. Because the error is initially confined to small
scales $k>k_0$, the vorticity-based norm is always larger than the 
energy-based norm, but the predictability time is essentially independent 
of the norm used.  It is not difficult to understand that any Eulerian
norm would give comparable result. Because the error propagates from
small to large scales, a norm which emphasizes small scale features (as
the enstrophy norm) saturates earlier than a large scale based norm
(energy, in our example), but the results remain essentially the same.

\begin{figure}[htb]
\centerline{\epsfig{figure=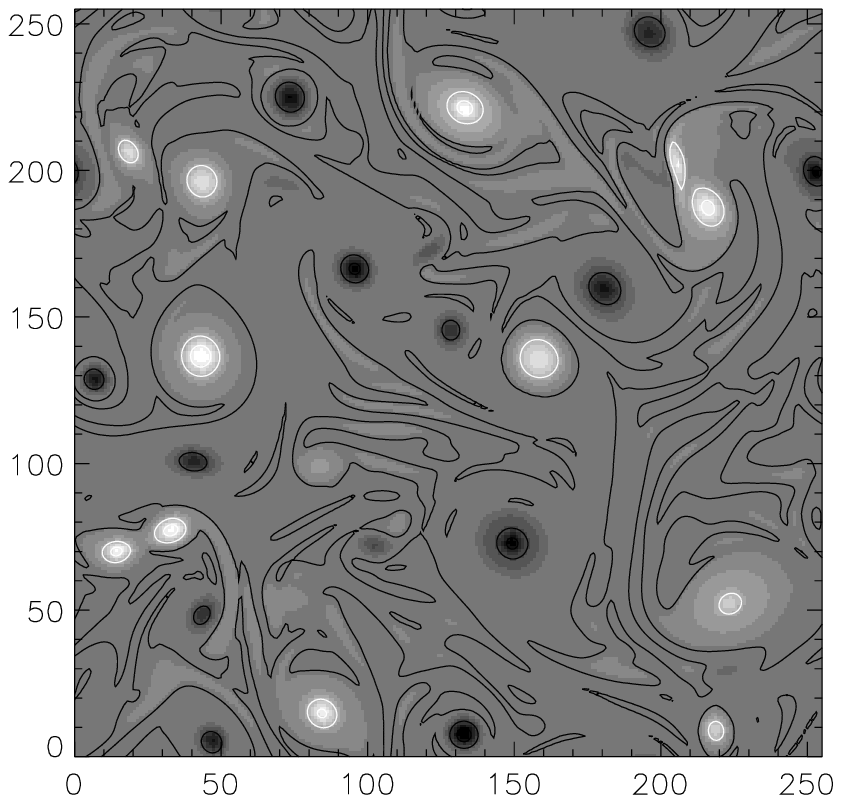,width=6cm,angle=0}}
\centerline{\epsfig{figure=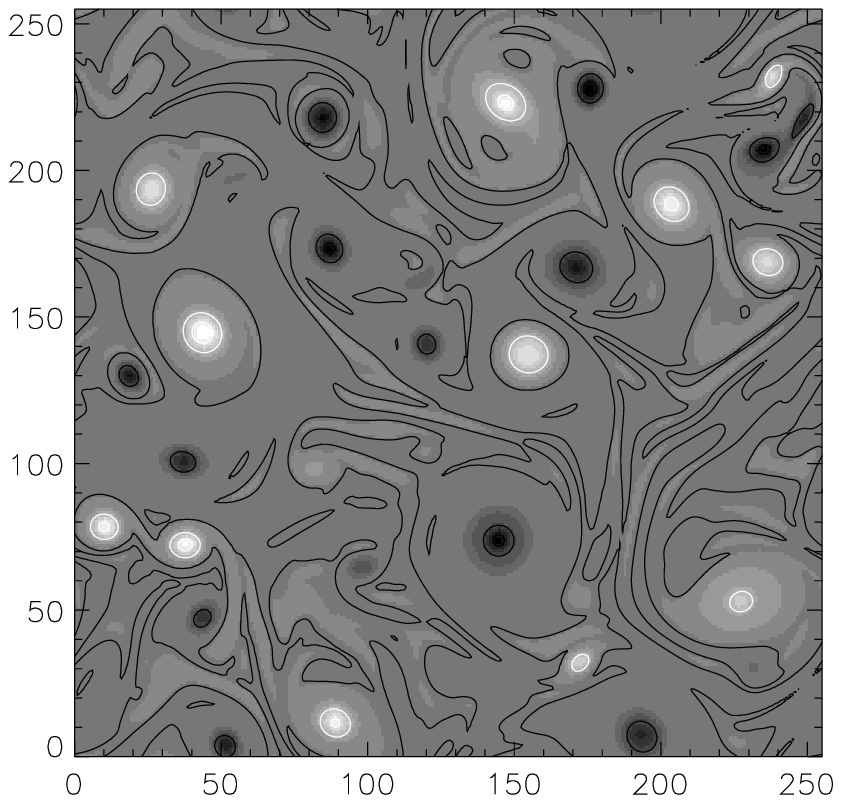,width=6cm,angle=0}}
\centerline{\epsfig{figure=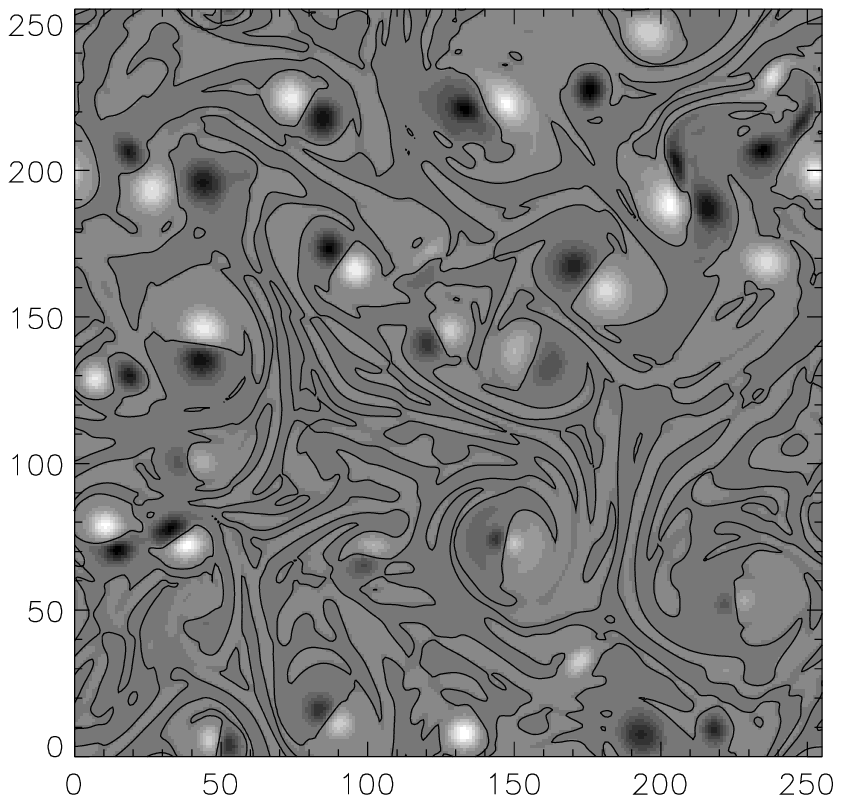,width=6cm,angle=0}}
\caption{Gray scale map of the vorticity fields (obtained by a $256^2$
simulation) at time $T_p=177$. White corresponds to positive vorticity
regions, black to negative ones. (a) Reference field $\omega({\bf x})$
(b) the perturbed one $\omega^{'}({\bf x})$ (c) the error filed
$\delta \omega({\bf x})$ }
\label{fig:3.10-2}
\end{figure}

In Figure~\ref{fig:3.10-2} we plot the vorticity field of the
reference $\omega({\bf x},T_p)$ and perturbed field $\omega'({\bf
x},T_p)$ at the predictability time $T_p$.  Although the two fields
differ, by definition, by 25\% in energy and about 65\% in enstrophy,
they look still remarkably similar for what concerns the distribution
of vortices.  Most of the large coherent structures are 
almost in the same positions.

In Figure~\ref{fig:3.10-2} we also plot the difference field $\delta
\omega({\bf x},T_p)$.  The typical bipolar configuration, usually
observed in simulations \cite{KYO90,O91}, indicates that the error is
concentrated in correspondence of the vortices and that it is
essentially due to the different position of the vortex structures in
the two realizations.

This result suggests that a Lagrangian measure of the error, based on
the vortex positions, would be more suitable for the present system.
For example, to emphasize the limits of the Eulerian measure for the error
(\ref{eq:3.4-3},\ref{eq:3.4-4}), consider the limiting case of singular 
point vortices, where an infinitesimal error in the coordinates 
gives error saturation and hence zero predictability time.
In general, we expect that, in presence
of vortices, an Eulerian-based measure underestimates the
predictability time.  
 
This problem can be overcome by resorting to the natural distance
among vortex centers. We use
a vortex tracking algorithm which recognizes and follows vortices
during the dynamics.  First we need a definition of vortex, the one
here adopted is: a connected region $D_{\alpha}$ in the computational
domain with vorticity maximum $z_{\alpha}$ larger (in absolute value)
than a given threshold and vorticity larger than a fraction (we used
$0.2$) of the vorticity peak.  Given the vortex domains $D_{\alpha}$,
all the physical quantities are computed by integrating inside
the domains.  For example, vortex circulation is defined as
$\Gamma_{\alpha}=\int_{D_{\alpha}} d^2x\, \omega({\bf x})$ and vortex
center ${\bf x}_{\alpha}$ is the center of mass computed from the 
vorticity field.  Finally, vortex trajectories are reconstructed by matching
center positions at different times.
\begin{figure}[hbt]
\centerline{\epsfig{figure=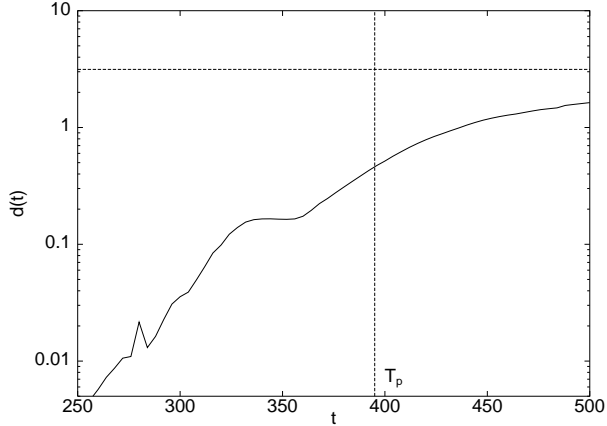,width=8cm,angle=0}}
\vspace{0.4truecm}

\caption{Mean vortex separation $d(t)$ at resolution $512^2$.  At the
classical predictability time $T_p$, the mean vortex separation is
about one-tenth of the saturation level.}
\label{fig:3.10-3}
\end{figure}
A Lagrangian, vortex-based, measure of the error can, e.g., be defined as
\begin{equation}
d^2(t) = {1 \over \sum_{\alpha} |\Gamma_{\alpha}|}\,
\sum_{\alpha} |\Gamma_{\alpha}|\, |{\bf x}'_{\alpha} - {\bf x}_{\alpha}|^2
\label{eq:3.4-7}
\end{equation}
where ${\bf x}_{\alpha}$ and ${\bf x}'_{\alpha}$ are the vortex
positions respectively in the reference and perturbed field.  In
Figure~\ref{fig:3.10-3} we plot $d^2$ obtained from our simulation.
We observe that at the classical predictability time, the mean vortex
separation is $d(T_{\rm p}) \simeq 0.5$, well below the saturation
value ($d_{\rm max} \sim L/2 = \pi$ in the periodic computational
box).  This result is a quantitative confirmation of the observations
drawn from Figure~\ref{fig:3.10-2}, i.e. the existence of an
intermediate regime in which the (finite) error is ruled by the
displacement of the strong coherent structures. If one is interested
in predicting, with some tolerance, positions and intensities of
coherent structures, it is possible to have a much larger
predictability time.

\section{Predictability in fully developed turbulence}
\label{sec:4}
\setcounter{equation}{0}
\subsection{Basic facts of turbulence}
\label{sec:4.1}

Perhaps, {\em fully developed turbulence} \cite{MY75,F95} is the most
important instance of high-dimensional chaotic system.  To give an
example, let us consider a classical experiment in fluid dynamics: in
a wind tunnel, an air mass conveyed by a large fan impinges some
obstacles, which perturb significantly the velocity of fluid
particles. Big and small whirls appear, and the flow evolves
irregularly in time.  One could wonder whether the features of the
flow depend crucially on the physical properties of the fluid, the
size and shape of the obstacle, the mean wind velocity, and so on. It
is easy to understand that, with a given geometry the only relevant
parameter which characterizes the flow is the Reynolds number
$Re=UL/\nu$, where $U$ is the mean wind velocity, $L$ is the typical
size of the obstacle and $\nu$ is the kinematic viscosity of the
fluid. When $Re$ is very large, i.e., of the order of a thousand or
more \cite{B53} turbulence is called fully developed.  The fundamental
physical interest in this regime is motivated by the existence of
universal properties with respect to the details of the experimental
setup \cite{B53,MY75}.  If a velocity probe is placed at some distance
past the obstacle, it is possible to record a temporal series that
gives us statistical information.  If one sits far enough from the
obstacle, there the small-scale properties of the flow do not depend
sensitively on the precise site and orientation of the probe, that is
the turbulence is approximately homogeneous and isotropic.  Since the
flow is swept across the probe at a mean velocity $U$, that largely
exceeds the magnitude of the fluctuations, one can expect that
the time record essentially amounts to a one-dimensional spatial
section of the velocity field. Thus time-scales and length-scales are
interchangeable, this is the essence of the {\it Taylor hypothesis}
\cite{MY75}.  Assuming the above hypothesis, we can safely reinterpret
temporal variations of the velocity, on an interval $\tau$, in a
fixed-point of the space as spatial increments on scale
$\ell=U \tau$, at a fixed-time.

\begin{figure}[t]
\centerline{\epsfig{figure=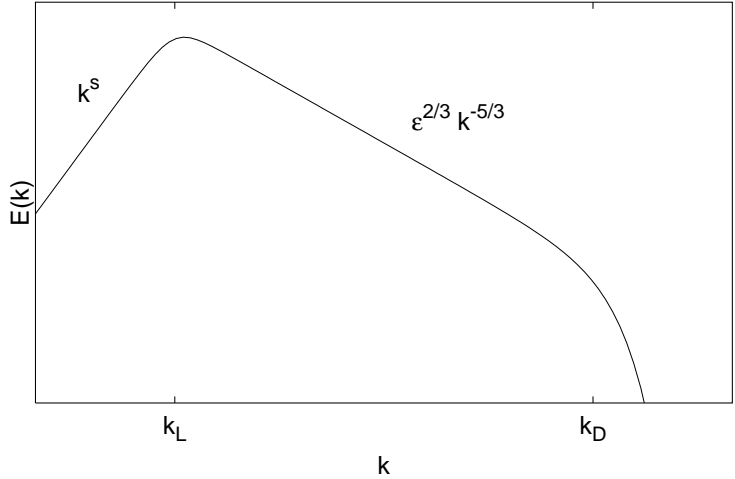,width=10cm, angle=0}}
\caption{ Typical turbulent energy spectrum, $k_{L}^{-1}$ is the
energy containing integral scale and $k_{D}^{-1}$ the dissipative
Kolmogorov scale.}
\label{fig:4.1-1}
\end{figure}

The first important result about the expected universality is the
behavior of  the velocity power spectrum which closely follows a
power law decay $E(k) \propto k^{-5/3}$ on a given range of wave-numbers
\cite{K41,MY75}.  At larger wave-number the spectrum falls off with an
exponential-like behavior, whereas the form at small $k$ (i.e. large
scales) depends on the mechanism of forcing and/or boundary
conditions.  For $k \to 0$ one often observes a self-similar energy
spectrum $E(k) \sim k^{s}$ with scaling exponent $s>0$. In
incompressible decaying turbulence, there are some arguments indicating
that asymptotically $s \le 4$ where the limiting value $s=4$ is
observed if initially the spectrum has $s>4$ \cite{B53}.  A typical
turbulence spectrum is shown in Figure~\ref{fig:4.1-1}.

The two crossovers unveil the presence of two characteristic scales: a
large excitation scale $L \sim k_{L}^{-1}$, associated with the
energy containing eddies, and a small dissipation scale $\ell_{D} \sim
k_{D}^{-1}$, related to the smallest active eddies.  The appearance of
a power law in between these two extremes unveils that no other
characteristic scale is involved.

A simple and elegant explanation of these experimental findings is due
to A.N.~Kolmogorov \cite{MY75}: in a nutshell, it is assumed the
existence of a range of scales where the energy -- injected at the
scale $L$ -- flows down (with a cascade process, as remarked by
Richardson~\cite{R22}) to the dissipative scale $\ell_D$, where it is
dissipated by molecular viscosity. Since, practically, neither injection
nor dissipation takes place in this interval, it is called the
inertial range. In this range the only relevant quantity is the
average energy transfer rate $\bar{\varepsilon}$: dimensional counting
imposes then a power spectrum $E(k)\propto
\bar{\varepsilon}^{2/3}k^{-5/3}$ in agreement with the experimental
observations discussed above.  The scaling for the spectrum is
equivalent to a power law dependence for the second order structure
function (SF)
\begin{equation}
S_2(\ell) = \langle \delta v_{\ell}^2 \rangle =
\langle \left(v(x+\ell)-v(x)\right)^2\rangle \sim \ell^{2/3}\,.
\label{eq:4.1-1}
\end{equation}
The original Kolmogorov theory (K41) assumes self-similarity of the
turbulent flow. As a consequence, the scaling behavior of higher
order structure functions $S_p(\ell)=\langle |v(x+\ell)-v(x)|^p\rangle \sim \ell^{\zeta_p}$ is described
by a single scaling exponent.  The value of the exponent is determined
by the so-called ``4/5 law'', an exact relation derived by Kolmogorov
from the Navier--Stokes equations \cite{K41,F95}, which, under the
assumption of stationarity, homogeneity and isotropy states
\begin{equation}
\langle \delta v^3_{||}(\ell)\rangle = - {4 \over 5} \bar{\varepsilon} \ell\,,
\label{eq:4.1-1b}
\end{equation}
where $\delta v_{||}(\ell)$ is the longitudinal velocity difference between 
two points at distance $\ell$, and
$\bar{\varepsilon}$ is the average rate of energy transfer.  The
structure function exponent $\zeta_p$ is thus predicted by Kolmogorov
similarity theory to be $\zeta_p=p/3$.

Several experimental investigations \cite{AGHA84,F95} have shown that
the Kolmogorov scaling is not exact and $\zeta_p$ is a nonlinear
function (with $\zeta_{3}=1$ as a consequence of the ``4/5 law'').
This means a breakdown of the self--similarity in the turbulent
cascade.  Larger and larger excursions from mean values are observed
as one samples smaller and smaller scales.  This phenomenon goes under
the name of intermittency \cite{F95}.  A complete theoretical
understanding of intermittency in Navier-Stokes turbulence is still
lacking. Nevertheless, there are approaches, as the multifractal model
\cite{PF84}, which are able to characterize at a phenomenological
level the intermittency.

In brief the basic idea of the multifractal model
\cite{PF84,PV87,F95} consists in assuming a local scale-invariance for
the velocity fluctuations, i.e. one has $\delta v_{\ell} \sim \ell^h$,
with a continuous spectrum of (H\"older) exponents $h$, each 
belonging to a given fractal set. In other words, in the inertial
range one has
\begin{equation}
\delta v_{\ell}(x) \sim \ell^h\,,
\label{holder}
\end{equation}
if $x \in S_h$, and $S_h$ is a fractal set with dimension $D(h)$ and
$h \in$ ($h_{min}$, $h_{max}$).  The probability to observe a given
scaling exponent $h$ at the scale $\ell$ is thus $P_{\ell}(h) \sim
\ell^{3-D(h)}$. In this language the Kolmogorov similarity theory
\cite{K41,F95} corresponds to the case of only one singularity
exponent $h=1/3$ with $D(h=1/3)=3$, see also Appendix~\ref{app:mf}.

\subsection{Reduced model of turbulence}
\label{sec:4.2}

In numerical simulations of the Navier-Stokes equations in the regime
of fully developed turbulence, one has to discretize the original PDE
to obtain a set of approximate ODE which must be integrated
numerically. This is the direct numerical simulation approach which,
in its simplest form, is implemented on a regular $3D$ grid of 
$N^3$ points.  Since the dissipative scale (Kolmogorov scale) is
related to the Reynolds number as $\ell_D \sim L Re^{-3/4}$, an
estimate of the number ${\cal N}$ of active spatial degrees of freedom leads to
\begin{equation}
{\cal N} \sim (L/\ell_D)^3 \sim Re^{9/4}\,.
\end{equation}
An obvious consequence of the fast growth of ${\cal N}$ with the
Reynolds number is the unfeasibility of a complete turbulent
simulations at high $Re$. The maximum limit of present computers is
about $N=10^3$ which corresponds to $Re \simeq 10^{4}$.

An alternative approach has been introduced with the so called shell
models by the works of Obukhov, Gledzer and Desnyansky and Novikov
(see \cite{BJPV98} for a detailed discussion).  The basic idea,
originally motivated in the context of closure theory, is to implement
a dynamical cascade with a set of variables $u_n$ ($n=1,..., N$) each
representing the typical magnitude of the velocity fluctuation in a
shell of wave-numbers $k_n < |{\bf k}| < k_{n+1}$.  The representative
wave-numbers are spaced geometrically, $k_n=k_0 2^n$, in this way,
assuming locality in the cascade, interactions are confined to
neighboring shells only.

We will discuss a specific model, known as GOY model (see
\cite{BJPV98} for a review), which makes use of complex velocity
variables $u_n$ and for which the equations of motion are
\begin{eqnarray}
\left({d \over d t} + \nu k_n^2 \right)u_n = 
i k_n \left(u_{n+1} u_{n+2} - {1 \over 4} u_{n-1} u_{n+1} -
{1 \over 8} u_{n-2} u_{n-1} \right)^{*} + f_{n}\,,\nonumber\\
\label{eq:4.1-2}
\end{eqnarray}
where $\nu$ is the viscosity and $f_n$ is a forcing 
term (typically restricted on the first shells).
The coefficients in the nonlinear term (which has the 
same structure of Navier-Stokes equations) are chosen
to conserve energy $E\equiv 1/2\,\sum_{n} |u_n|^2$ in the unforced,
inviscid limit.

Without entering in the details, we recall that shell model
(\ref{eq:4.1-2}) displays energy cascade {\it \`a la} Kolmogorov from
the large scales of forcing to the dissipative scales ($n \sim N$)
with a statistical constant energy flux $\bar{\varepsilon}$. On these
inertial range scales, the moments of velocity show power law scaling
$\langle |u_n|^p \rangle \sim k_{n}^{-\zeta_p}$ with exponents close
to those experimentally observed for fully developed turbulence.

The number of shells $N$ necessary to mimic the cascade mechanism of
 fully developed
turbulence is rather small, due to the geometrical progression
in $k_n$ one has $N \sim \log_{2} Re$. We
have thus a chaotic dynamical system with a reasonably small number of
degrees of freedom where standard methods of deterministic chaos can
be used in order to relate the ``turbulent'' statistical description
in terms of structure functions and intermittency, and  dynamical
properties, such as the spectrum of Lyapunov exponents. The absence of
any stochastic term in (\ref{eq:4.1-2}) makes the shell model a
natural model for investigating the predictability problem in
turbulence.

\subsection{Effects of intermittency on predictability of 
infinitesimal perturbations}
\label{sec:4.3}

The sensitive dependence on initial conditions makes the long term
forecasting in turbulent flow practically impossible.  For instance,
Ruelle \cite{R79} remarked that thermal fluctuations in the atmosphere
produces observable changes on a scale of centimeters after only few
minutes.  As a consequence after one or two weeks, the large scale
atmospheric circulation would be completely unpredictable, even if the
exact evolution equations were known.  This is the so-called {\it
butterfly effect}, in the words of Lorenz: {\it A butterfly moving its
wings in Brazil might cause the formation of a tornado over Texas}.
To support this argument, one can observe that the largest LE of fully
developed turbulence is roughly proportional to the inverse of the
smallest characteristic time of the system, the turn-over time
$\tau_D$ of eddies of the size of the Kolmogorov length $\ell_D$.
From $\ell_D \sim \, L Re^{-3/4}$ one obtains
\begin{equation}
\tau_{D} \sim \ell_{D}/\delta v_{D}\sim \tau_L\, Re^{-1/2}\, ,
\label{single}
\end{equation}
where $\tau_L\approx L/U$ is the eddy turn-over time of the energy
containing scales. As a consequence, as first pointed out by Ruelle
\cite{R79}, the largest LE scales with $Re$ like $\lambda \sim {1 /
\tau_D} \sim Re^{1/2}/\tau_L$. Thus fully developed turbulence
is characterized by a Lyapunov exponent diverging with $Re$. 

Nevertheless a large value of the LE does not prevent the possibility
of long term prediction, at least if one is
 interested in predicting the large scales behavior (which is
related to finite errors), see Sect.~\ref{sec:4.4}.

\begin{figure}[t]
\centerline{\epsfig{figure=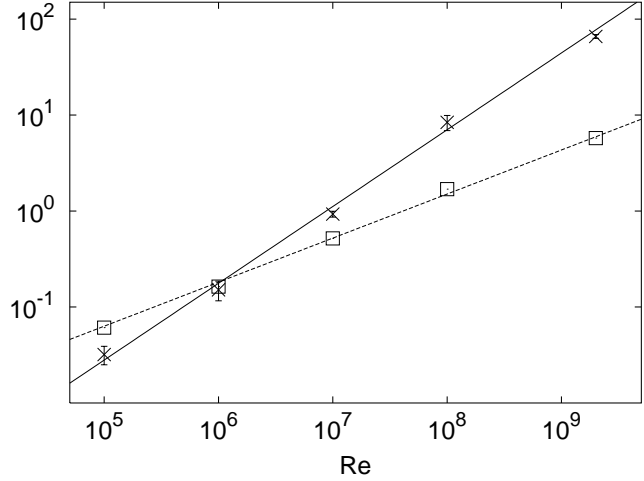,width=10cm, angle=0}}
\caption{ Lyapunov exponent $\lambda$ ($\Box$) and variance $\mu$
($\times$) as a function of the Reynolds number $Re$ for the shell
model (\ref{eq:4.1-2}) with $N=27$ shells. The dashed line is the
multifractal prediction $\lambda \sim Re^{\alpha}$ with
$\alpha=0.459$, with function $D(h)$ obtained by the random beta model
fit of the $\zeta_{p}$ exponents. The full line represents $\mu \sim
Re^{w}$ with $w=0.8$}
\label{fig:4.3-1}
\end{figure}

Remaining in the framework of infinitesimal perturbations, we discuss
 the effects of intermittency on the predictability time.  The
multifractal model \cite{PF84} predicts a spectrum of viscous
cut-offs: each singularity exponent $h$ selects a different damping
scale, $\ell_D(h)\sim L Re^{-1/(1+h)}$, and hence a spectrum of
(dissipative) turn-over times, $\tau_D(h)$, such that (\ref{single})
becomes
\begin{equation}
\tau_{D}(h) \sim \ell_{D}(h)/\delta v_{D} \sim \tau_L Re^{-{1-h \over
1+h}}\,,
\label{eq:multi}
\end{equation}
(see Appendix~\ref{app:mf} for details).
To obtain the largest Lyapunov exponent now we have to integrate $\tau_D(h)^{-1}$,
at the scale $\ell=\ell_D(h)$, over the $h$-distribution
$P_\ell(h) \sim \ell^{3-D(h)}$:
\begin{equation}
\lambda\sim \int  \tau(h)^{-1}\, P_{\ell}(h)\, d h
\sim{1\over \tau_L}\,\int\, 
\left({\ell_D\over L}\right)^{h-D(h)+2}\, d h. 
\label{eq:4.3-1}
\end{equation}
Since the viscous cut-off vanishes in the limit $Re \to \infty$, the
integral can be estimated by the saddle-point method, i.e.
\begin{equation}
\lambda \sim {1\over \tau_L}\,Re^\alpha
\qquad \hbox{\rm with}\
\alpha=\max_h \,\left[\, {D(h)-2-h \over 1+h}\,\right].
\label{eq:4.3-2}
\end{equation}
The value of $\alpha$ depends on the shape of $D(h)$. By using the
function $D(h)$ obtained by fitting the exponents $\zeta_q$ with the
random $\beta$-model \cite{BPPV84} one finds $\alpha=0.459..$,
slightly smaller than the Ruelle prediction $\alpha=0.5$. This result
is confirmed by numerical simulations on the shell model
(\ref{eq:4.1-2}) (see Fig.~\ref{fig:4.3-1}).

We remind that the fluctuations of the effective Lyapunov exponent
$\gamma(t)$ can be characterized by the ratio of $\mu/\lambda$
(Sect.~\ref{sec:2.1}). The variance $\mu$ is
\begin{equation}
\mu = \lim_{t\to \infty} t \left[ \langle \gamma(t)^2 \rangle - 
\langle \gamma(t) \rangle^2 \right] 
\sim t_c \langle (\gamma - \lambda)^2 \rangle
\label{eq:4.3-3}
\end{equation}
where in the last expression we have introduced the integral
correlation time $t_c=\int C(t) dt$ of the effective Lyapunov exponent
\cite{CJPV93,BJPV98}, where $C(t)$ is the normalized correlation
function of the fluctuation of $\gamma(t)$ (i.e. $\gamma(t)-\lambda$).

The quantity $\langle (\gamma - \lambda)^2\rangle$ can be computed by
repeating the argument for $\lambda$:
\begin{equation}
\langle \gamma^2\rangle \sim \langle \tau^{-2} \rangle 
\sim {1 \over \tau_{L}^{2}} Re^{y}\,.
\label{eq:4.3-4}
\end{equation}
An explicit calculation \cite{CJPV93} gives $y=1$ independently of
intermittency. Assuming that the correlation time $t_c$ vanishes as a
power of $Re$
\begin{equation}
t_{c} \sim \tau_L Re^{-z}
\label{eq:4.3-5}
\end{equation}
one ends with the prediction
\begin{equation}
\mu \sim {1 \over \tau_L} Re^{w}
\qquad \hbox{\rm with}\
w=1-z\,.
\label{eq:4.3-6}
\end{equation}

Numerical simulations on the shell model (\ref{eq:4.1-2}) give $w
\simeq 0.8$ (see Figure~\ref{fig:4.3-1}). Because $w>\alpha$ we obtain
that  $\mu/\lambda$ diverges with $Re$. From
Figure~\ref{fig:4.3-1} we see that the strong intermittency regime begins,
for the shell model, at $Re \sim 10^{6}$.  Let us stress that in the
absence of intermittency one would expect that $t_{c} \sim
\lambda^{-1}$ and thus $z=1/2$ and $\mu/\lambda$ constant. The fact
that $z \sim 0.2$ indicates that the presence of quiescent periods in
the turbulent activity is much more relevant for the decay rate of
time correlations than for the Lyapunov exponent.

\begin{figure}[t]
\centerline{\epsfig{figure=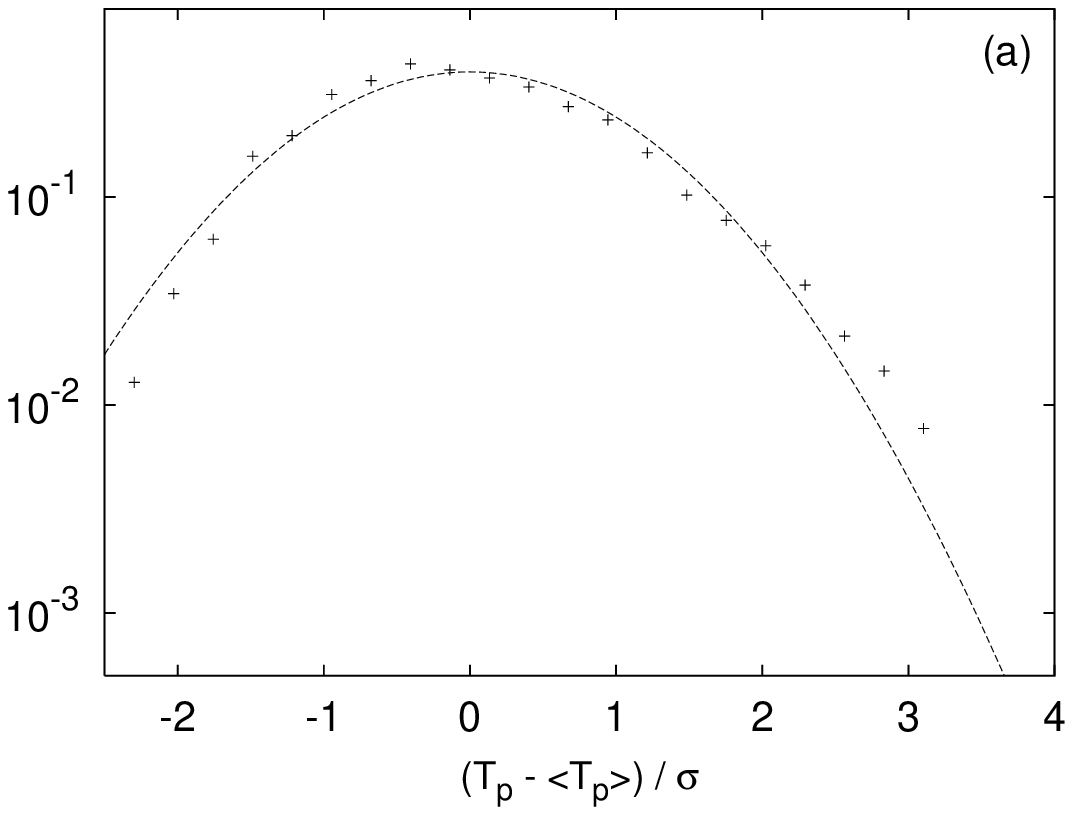,width=9cm, angle=0}}
\centerline{\epsfig{figure=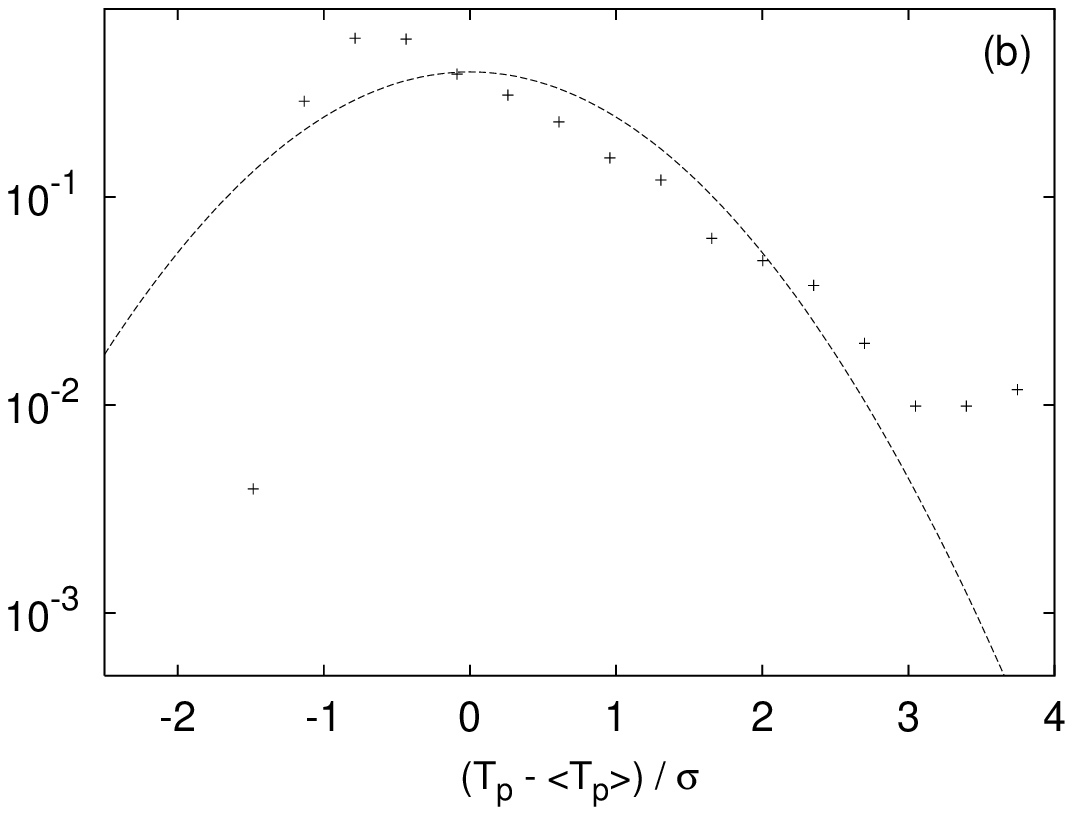,width=9cm, angle=0}}
\caption{Rescaled probability distribution functions of the
predictability time $T_p$ for the shell model (\ref{eq:4.1-2}) 
for (a) $Re=10^{6}$
and (b) $Re=2 \times 10^{9}$. The respective average values are 
$\langle T_p \rangle = 84.0$ and $6.32$ and the standard
deviations are $\sigma(T_p)=22.2$ and $3.16$. The line is the
Gaussian.
}
\label{fig:4.3-2}
\end{figure}

We have seen in Sect.~\ref{sec:2.2} that the fluctuations of the
effective LE affect the distribution of predictability time, and thus
we expect a similar effect in fully developed turbulence.  In the
shell model one can estimate the predictability time by computing the
time $T_p$ at which the difference $\delta u_{m}(t)$ (where $m$
corresponds to the integral scale) among two realizations of the model
becomes larger that the tolerance $\Delta$. The initial difference
$\delta_0$ is restricted to the shell $u_{n^*}$ on the Kolmogorov scale
and $m\ll n^*$.  The predictability time distribution function is
computed at two different Reynolds number. At $Re=10^6$ we are at the
border of the weak intermittent range: the observed PDF 
(Figure~\ref{fig:4.3-2}) is indeed close to a Gaussian with mean value
\begin{equation}
\langle T_p \rangle \simeq {1 \over \lambda} \ln\left({\Delta \over
\delta_0}\right)\,.
\label{eq:4.3-7}
\end{equation}
On the contrary, at $Re=2 \times 10^9$, the PDF exhibits the asymmetric
triangular shape and the mean value is ruled by $\mu$ according to
(\ref{eq:2.2-4}).

\subsection{Growth of non-infinitesimal perturbations}
\label{sec:4.4}

The classical theory of predictability in turbulence has been
developed by Lorenz \cite{L69} (see also \cite{L73}) using physical
arguments, and by Leith and Kraichnan \cite{LK72} on the basis of
closure approximations. The fundamental ingredients of the Lorenz
approach stem from dimensional arguments on the time evolution of a
perturbation in an energy cascade picture.  In this framework, it is
rather natural to assume that the time $\tau_{\ell}$ for a
perturbation at scale $\ell/2$ to induce a complete uncertainty on the
velocity field on the scale $\ell$, is proportional to the typical
eddy turn-over time at scale $\ell$: $\tau_{\ell} \sim {\ell / \delta
v_{\ell}}$ where $\delta v_{\ell}$ is the typical velocity difference
at scale $\ell$. Kolmogorov scaling (\ref{eq:4.1-1}) gives
\begin{equation}
\tau_{\ell} \sim \ell^{2/3}\,.
\label{eq:4.4-1}
\end{equation}
Because of the geometric progression (\ref{eq:4.4-1}), the
predictability time to propagate an uncertainty $O(\delta v_{D})$ from
the Kolmogorov scale $\ell_D$ up to the scale of the energy containing
eddies $L$, is dominated by the longest time
\begin{equation}
T_p \sim \tau_{\ell_d}+\tau_{2\ell_d}+\dots+\tau_{L}  \sim \tau_{L} 
\sim {L \over \delta v_{L}}\,.
\label{eq:4.4-2}
\end{equation}
Closure approximations, where one still uses dimensional arguments,
confirm this result \cite{LK72,L73}.

It is important to stress that, in the Lorenz approach, the
predictability time is independent of the Reynolds number.  This is
only in apparent contradiction with the increase of the Lyapunov
exponent with $Re$ (\ref{eq:4.3-2}).  From the point of view of an
observer interested in forecasting the large scales (i.e. not
infinitesimal perturbations) the Lyapunov exponent is not physically
relevant.  Large scale predictability in turbulence is hence another
example where a large LE coexists with a long predictability time.  We
will see that a coherent description that includes these two
features of predictability is given by the finite size Lyapunov
exponent (\ref{eq:2.3-10}).

It is easy to estimate the scaling behavior of $\lambda(\delta)$ when
the perturbation is in the inertial range $\delta v_{D} \, \le \delta
\le \, \delta v_L$.  Following the phenomenological ideas of Lorenz,
the doubling time of an error of magnitude $\delta$ can be identified
with the turn-over time $\tau_{\ell}$ of an eddy with typical velocity
difference $\delta v_{\ell} \sim \delta$. Using the scaling
(\ref{eq:4.1-1}) one has $\tau_{\ell} \sim \tau_L (\ell/L)^{2/3} \sim
\tau_L (\delta v_{\ell}/\delta v_L)^{-2}$.  In conclusion we obtain 
\cite{ABCPV96}
\begin{equation}
\lambda(\delta) \sim \delta^{-2} \, .
\label{eq:4.4-3}
\end{equation}
In the dissipative range $\delta < \delta v_D$, the error can be
considered infinitesimal, implying $\lambda(\delta) =\lambda$.

Accounting for intermittency, in the framework of the multifractal
approach, one has
\begin{equation}
\lambda(\delta) \sim \tau_L^{-1}\int dh \
(\delta /\delta v_L)^{[3 - D(h)]/h}\, (\delta /\delta v_L)^{1-1/h} \, .
\label{eq:4.4-4}
\end{equation}
From the basic inequality of the multifractal model $D(h)\le 3h + 2$
(see Appendix \ref{app:mf}), we have
\begin{equation}
\frac{2+h-D(h)}{h} \ge -2 \quad \mbox{for all}\ h.
\label{eq:4.4-5}
\end{equation}
As a result of the constancy of the energy flux in the inertial range,
$\bar{\varepsilon}=v^3(\ell)/\ell$, the equality holds for
$h=h^{*}(3)$, and gives $3 h^{*}(3) + 3 - D(h^{*}(3)) = 1$.  Therefore
a saddle point estimation of (\ref{eq:4.4-4}) gives again
(\ref{eq:4.4-3}). The dimensional scaling of the FSLE in fully
developed turbulence $\lambda(\delta) \sim \delta^{-2}$ is thus not
affected by intermittency corrections. This is a direct consequence of
the exact result (\ref{eq:4.1-1b})

\begin{figure}[ht]
\centerline{\epsfig{figure=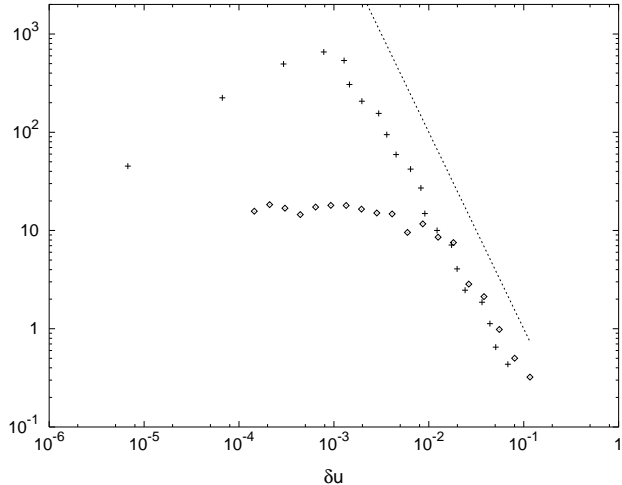,width=9cm, angle=0}}
\vspace{.4cm}

\caption{The inverse of the error doubling times versus $\delta u$ 
(diamond) compared with shell turn-over
times (plus).  Number $N$ of simulated shells is $27$, and Reynolds
number ${\rm Re}= \nu^{-1}= 10^{9}$, $k_0= 0.05$.  The initial
perturbation is randomly uniform over all shells in the inertial
range, with amplitude of order $10^{-6}$.  The first threshold is
$\delta_0=10^{-4}$ and the error growth rate parameter $r$ is
$2^{1/2}$.  The number of error doubling experiments is $400$. The
dashed line has the slope $-2$.  }
\label{fig:4.4-1}
\end{figure}

These findings have been numerically tested on the shell model
(\ref{eq:4.1-2}) for the energy cascade.  Figure~\ref{fig:4.4-1} shows
the scaling of $\langle 1/\tau(\delta v,r)\rangle_t$ as a function of
$\delta v$ in the GOY model, where $\tau(\delta v,r)$ is the
``doubling time'', i.e. the time necessary for a perturbation of size
$\delta v$ to increase by a factor $r$ (see Sect.~\ref{sec:2.3} and
Appendix~\ref{app:fsle}).

For comparison we also plot the eddy turn-over times
$\tau_{\ell}^{-1}= \langle|\delta v_{\ell}|^2\rangle^{1/2}/\ell$.  We
see that below the Kolmogorov scale, the doubling time displays a
constant plateau corresponding to the Lyapunov exponent
(\ref{eq:2.3-11}).  At larger errors we observe a good agreement with
the prediction (\ref{eq:4.4-3}). Let us observe that, even at this
high Reynolds number, the scaling range for the doubling time is
rather small.

\begin{figure}[t]
\centerline{\epsfig{figure=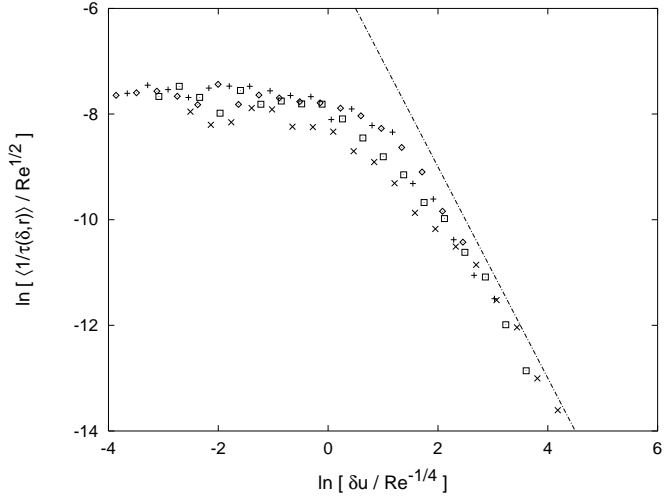,width=9cm, angle=0}}
\vspace{.4cm}
\caption{$\ln\left[\langle 1/\tau(\delta u, r)\rangle/{\rm
Re}^{1/2}\right]$ versus $\ln\left[\delta u/{\rm Re}^{-1/4}\right]$ at
different Reynolds numbers ${\rm Re}=\nu^{-1}$.  ($\Diamond$) $N = 24$
and $Re=10^{8}$; ($+$) $N = 27$ and $Re=10^{9}$; ($\Box$) $N = 32$ and
$Re=10^{10}$; ($\times$) $N = 35$ and $Re=10^{11}$.  The straight line
has slope $-2$.  }
\label{fig:4.4-2}
\end{figure}

It is interesting to look at the doubling time as a function of the
Reynolds number. For small thresholds the inverse of the doubling
time scales as the Lyapunov exponent, i.e. roughly as $Re^{-1/2}$.  We
also observe that the bend away from the infinitesimal growth rate
occurs at smaller scales for larger Reynolds numbers. This
suggests the following scaling ansatz: 
times and errors are scaled with the turn-over time and the
typical scale of fluctuations at the Kolmogorov scale, that is by
$\hbox{Re}^{-1/2}$ and $\hbox{Re}^{-1/4}$, respectively.  
In Figure~\ref{fig:4.4-2} we show the re-scaled data.  The data collapse is
reasonable, allowing to conclude that small-scale predictability, with
small error amplitudes, behaves (apart from intermittency corrections)
as predicted by Ruelle \cite{R79}, whereas large-scale predictability,
characterized by large error amplitudes, is well described by Lorenz
arguments.

\begin{figure}[t]
\centerline{\epsfig{figure=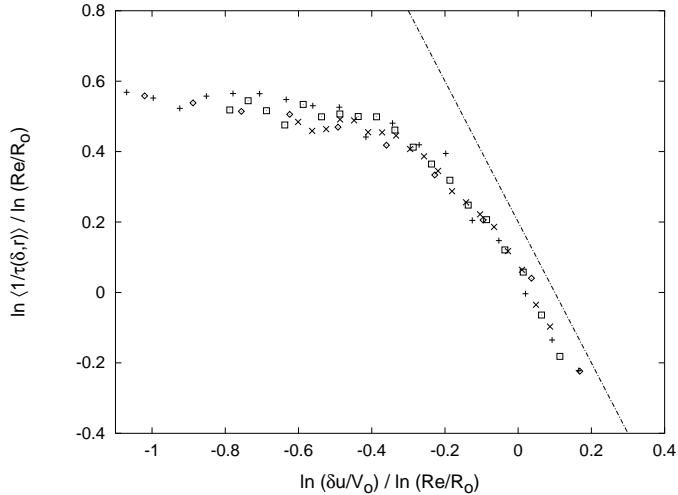,width=9cm, angle=0}}
\vspace{.4cm}
\caption{$\ln\langle 1/\tau(\delta u,r) \rangle / \ln(\mbox{Re}/R_o)$
versus $\ln (\delta u / V_o) / \ln(\mbox{Re}/R_o)$; multiscaling data
collapse at different Reynolds numbers ${\rm Re}=\nu^{-1}$.  The
fitting parameters are $R_o = 6\times 10^{6}$, $V_o= 5\times 10^{-2}$,
and $\mbox{Re}= \nu^{-1}$.}
\label{fig:4.4-3}
\end{figure}

To improve the data collapse, taking into account the multifractal
correction as described in Appendix \ref{app:mf}, one has to make a
multiscaling collapse, i.e. to rescale $\ln \langle 1/\tau(\delta
v,r)\rangle$ and $\ln(\delta v/V_{0})$ with $\ln (Re/Re_{0})$ where
$V_{0}$ and $Re_{0}$ are two parameters to be fixed \cite{ABCPV97}.
The result is shown in Figure~\ref{fig:4.4-3}. The data collapse is
clearly improved.

Finite size predictability has been investigated also in
two-dimensional turbulence, which is relevant for atmospheric flows.
As discussed in Sect.~\ref{sec:3.10}, two-dimensional turbulence in
the inverse energy cascade regime is characterized by a scaling \`a la
Kolmogorov \cite{KM80} with no intermittency \cite{BCV00}. As
discussed above, the scaling exponent in (\ref{eq:4.4-3}) is not
affected by intermittency; however intermittency does reduce the
scaling range because of the intermediate dissipative range (see
Appendix~\ref{app:mf}).  The absence of intermittency corrections in
$2D$ turbulence suggests that the dimensional scaling (\ref{eq:4.4-3})
is observable even in direct numerical simulations at moderate
Reynolds number.

\begin{figure}[t]
\centerline{\epsfig{figure=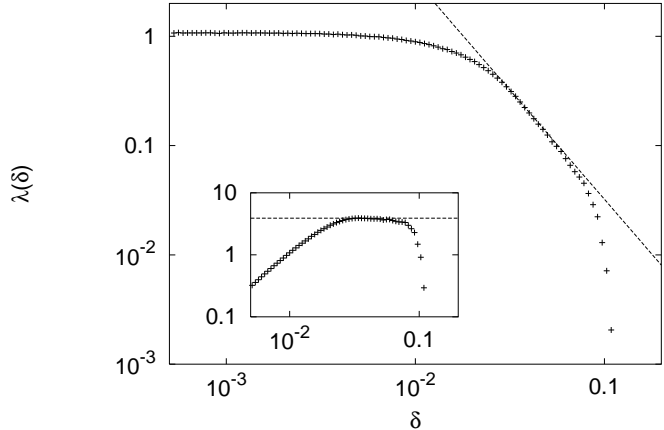,width=9cm, angle=0}}
\caption{Finite size Lyapunov exponent $\lambda(\delta)$ as a function
of velocity uncertainty $\delta$ in a direct numerical simulations with
$1024^2$ grid points of 2D turbulence in the inverse cascade
regime. The asymptotic constant value for $\delta \to 0$ is the
largest Lyapunov exponent of the turbulent flow. The dashed line has
slope $-2$.  In the inset we show the compensated plot
$\lambda(\delta) \delta^{2}/\bar{\varepsilon}$.}
\label{fig:4.4-4}
\end{figure}

Let us consider two realizations of the vorticity field in
(\ref{eq:3.4-1}) starting from very close initial conditions. The
error $\delta$ is defined, following (\ref{eq:3.4-4}), as
$\delta(t)=\sqrt{E_{\delta}(t)}$.  In Figure~\ref{fig:4.4-4} it is
shown the FSLE $\lambda(\delta)$.  It is remarkable the rather wide
scaling range for $\lambda(\delta)\sim \delta^{-2}$ with respect to
the shell model simulations (Fig.~\ref{fig:4.4-1}) obtained at much
larger Re.  As a consequence of the absence of intermittency, also the
 crossover from the infinitesimal regime $\lambda(\delta)=\lambda$
to the inertial range regime (\ref{eq:4.4-3}) is sharp.

\begin{figure}[t]
\centerline{\epsfig{figure=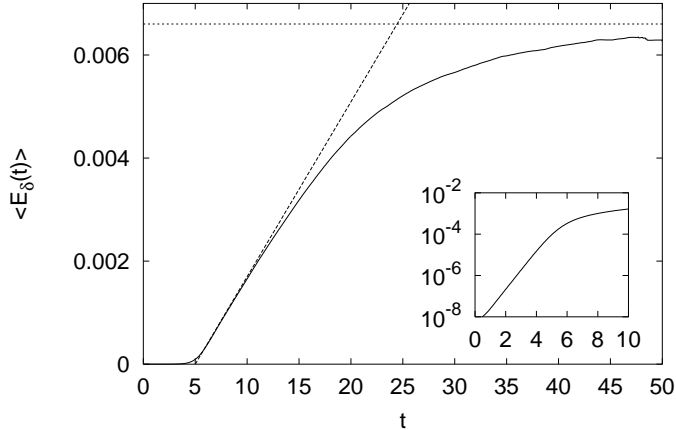,width=9cm, angle=0}}
\caption{Average energy error $\langle E_{\delta}(t) \rangle$ growth.
Dashed line represents linear closure prediction,
dotted line is the saturation value $E$.
The initial exponential growth is emphasized by the lin-log
plot in the inset.}
\label{fig:4.4-5}
\end{figure}

From a general point of view, it is interesting to observe that even
in the absence of intermittency, fixed scale analysis based on the
FSLE overpasses fixed time analysis in the characterization of
predictability. Dimensional considerations and closure approximations
\cite{LK72} predicts a linear growth of the error in the inverse
energy cascade as
\begin{equation}
E_{\delta}(t) = G \bar{\varepsilon} t\,,
\label{eq:4.4-6}
\end{equation}
where $G$ is an adimensional constant. It is easy to realize that
(\ref{eq:4.4-6}) is equivalent to (\ref{eq:4.4-3}), $\lambda(\delta)$
having the dimension of an inverse time and
$\delta=\sqrt{E_{\delta}}$. The result obtained in numerical
simulations is shown in Figure~\ref{fig:4.4-5}, which has to be
compared with Figure~\ref{fig:4.4-4}. The scaling law (\ref{eq:4.4-6})
in Figure~\ref{fig:4.4-5} is barely visible, making the determination
of $G$ difficult. On the contrary, inverting (\ref{eq:4.4-3}) to
(\ref{eq:4.4-6}) one can measure $G$ directly from
Figure~\ref{fig:4.4-4}. The result obtained is in close agreement with
closure computations \cite{BM00}.
 
\subsection{$\epsilon$-entropy for turbulent flows}
\label{sec:4.5}
A complementary way to look at the predictability of turbulent flows
is in terms of its entropy (see Sects.~\ref{sec:1.1.2} and \ref{sec:2.4}).

Unfortunately a direct measurement of the Kolmogorov-Sinai entropy is
practically infeasible. Indeed for $Re\to \infty$ due to the huge
number of active degrees of freedom, the KS-entropy diverges, so that one
needs velocity measurements with an extremely high resolution and
lasting for extremely long times, far beyond the actual experimental
possibilities.  Nevertheless, limiting the analysis to not very high
resolution, one can hope to extract some interesting piece of
information by investigating the behavior of the $\epsilon$-entropy,
$h(\epsilon)$. 
As far as the $\epsilon$-entropy of turbulence is concerned, two
questions can be raised.
\begin{description}
\item{$i)$} Since a direct measurement of the full $3$-dimensional
velocity field is infeasible, one has usually access just to a time
signal measured in one spatial point: which kind of information can we
extract from the $\epsilon$-entropy per unit time of such a signal?
\item{$ii)$} Taking into account $i)$, can we say something about the 
$\epsilon$-entropy of the full $3$-dimensional velocity field?
\end{description}

In $ii)$ we are referring to the $\epsilon$-entropy,
$h^{ST}(\epsilon)$, per unit time and volume (the symbol $^{ST}$ means
space-time).  In other words, we are assuming that 
the total entropy of a turbulent flow
observed for a (very long) time $T$ on a (very large) volume $V$ of
the $3$-dimensional space has the form $H(V,T,\epsilon)\approx
VT h^{ST}(\epsilon)$.  See Ref.~\cite{GW93} for an introduction of
this concept.

Both in $i)$ and $ii)$, as we will see, a crucial role is played by
the sweeping of the large scales of the flow on the small ones,
i.e. the Taylor hypothesis (see Sect.~\ref{sec:4.1}).

\subsubsection{$\epsilon$-entropy for a time signal of turbulence}
\label{sec:4.5.1}
In order to estimate the $\epsilon$-entropy of a given signal one has
to compute the Shannon entropy of the symbolic sequence obtained by
making an $(\epsilon, \tau)$ grid in phase-space
(Sect.~\ref{sec:2.4}).  Unfortunately, this method is rather
inefficient for signals in which many scales are excited
\cite{ABCFVV00a,ABCFVV00b,CFKOV00}, e.g., as in turbulence.
Therefore, here we resort to a recently proposed method
\cite{ABCFVV00a} based on the {\it exit-time} analysis.

In a few words, the idea consists in looking at a sequence of data not
at fixed sampling times but at fixed fluctuation (see
Appendix~\ref{app:eps}), i.e. when the fluctuation of the signal
exceeds a given threshold, $\epsilon$.  In practice we code the signal
$v(t)$ of total duration $T$ in a symbolic sequence
$\Omega^M(\epsilon)=\{t_i(\epsilon),k_i(\epsilon)\}_{i=1}^{M}$, where
$t_i(\epsilon)$ is the first times such that $|v(t_0+\sum_{k=1}^{i-1}
t_k (\epsilon) + t_i(\epsilon) ) - v(t_0 + \sum_{k=1}^{i-1} t_k
(\epsilon)) |\geq \epsilon/2$ (being $t_0$ a reference time) and
$k_i=\pm 1$ tells us in which direction (up or down with respect to
$v(t_0+ \sum_{k=1}^{i-1} t_k (\epsilon))$) the fluctuation has been
realized. $M$ is the total number of exit events, i.e. $\sum_{i=1}^{M}
t_i(\epsilon)=T$. Note that $\Omega^{M}(\epsilon)$ is a faithful
coding of the signal within the required accuracy $\epsilon$. Now the
evaluation of the entropy goes as usual through the evaluation of the
Shannon entropy, $h^{\Omega}(\epsilon)$, of the sequence 
$\Omega^{M}(\epsilon)$.  Finally the $\epsilon$-entropy per unit time
is given by \cite{ABCFVV00a}:
\begin{equation}
h(\epsilon) \approx {h^\Omega(\epsilon,\tau_r) \over \langle
t(\epsilon)\rangle }\,,
\label{eq:4.5-1}
\end{equation}
where a coarse-graining of the possible values assumed by
$t(\epsilon)$ with a resolution time $\tau_r$ has been considered, and
$\langle t(\epsilon)\rangle$ is the average exit time, i.e.  $\langle
t(\epsilon)\rangle= (1/M)\sum_{i=1,M} t_i(\epsilon)$.  The formula
(\ref{eq:4.5-2}) is exact in the limit $\tau_r \to 0$ (in
Appendix~\ref{app:eps} one finds the derivation of (\ref{eq:4.5-1})
and the details of the method).

This procedure allows a noticeable improvement of the computational
possibility to measure the $\epsilon$-entropy. In particular, if one
is interested in the leading scaling behavior of $h(\epsilon)$ with
$\epsilon$, one only needs to estimate the scaling of $\langle
t(\epsilon)\rangle$. Indeed, the correction induced by
$h^{\Omega}(\epsilon,\tau_r)$ can be shown to be sub-leading (in
particular, logarithmic).

Now, we estimate the average exit time for the velocity signal
$v(t)$. This can be done assuming the Taylor hypothesis and the
multifractal model (see Appendix~\ref{app:mf}). In this framework we
can assume that, for $t$ corresponding to scales $R=Ut$ in the
inertial range, the following relation holds $|\delta_t
v|=|v(t_0+t)-v(t)|\sim t^{h}$ and each $h$ is picked with probability
$ P(h)\sim t^{3-D(h)}$.  Since we are interested in the statistics of
the first times necessary to observe a fluctuation $|\delta_t v| \sim
\epsilon$, one can ``invert'' the above relation \cite{BCVV99}:
\begin{equation}
t(\epsilon)\sim \epsilon^{1/h} \;\; {\mbox{with}}\;\; P(h)\sim
\epsilon^{(3-D(h))/h}\,.
\label{eq:4.5-2}
\end{equation}
The exit-time moments \cite{BCVV99}, also called inverse structure
functions \cite{J99}, can be estimated in the multifractal framework
as follows
\begin{equation}  
\langle \langle t^q(\epsilon )\rangle \rangle\ 
\sim \int {\mathrm d}h \, \epsilon^{{q+3-D(h)}\over h} 
                      \sim \epsilon^{\chi(q)}
\end{equation}
where $\chi(q)$ may be obtained with a saddle point estimate in the
limit of small $\epsilon$:
\begin{equation}
\chi(q) = \min_h\left[{q + 3 - D(h) \over h}\right]\, . 
\label{eq:4.5-3} 
\end{equation}
The average $\langle[\dots]\rangle$, obtained by counting the number
of exit-time events $M$, and the average
$\langle\langle[\dots]\rangle\rangle$, computed with the uniform time
sampling are connected by the relation
\begin{equation}
\langle \langle t^q(\epsilon)\rangle \rangle =\lim_{M \rightarrow \infty} 
   \sum_{i=1}^{M} t_i^q {t_i \over {\sum_{j=1}^{M} t_j}} = 
   {{\langle t^{q+1}(\epsilon)\rangle }
 \over {\langle t(\epsilon)\rangle}}\,\,,
\end{equation}
where the term $t_i / {\sum_{j=1}^{M} t_j}$ takes into account the
non-uniformity of the exit-time statistics.  Therefore the quantity we
are looking for, i.e. the mean exit-time, is given by $\langle
t(\epsilon) \rangle= \langle \langle t^{-1}(\epsilon) \rangle
\rangle^{-1} \sim (\epsilon)^{-\chi(-1)}$.  By noting that
\begin{equation}
{-1 + 3 - D(h) \over h}={2-D(h)\over h}\geq -3 \;\;{\mbox {for all}}
\;\;h\,,
\label{eq:4.5-4}
\end{equation}
which is nothing but Eq.~(\ref{eq:4.4-5}), i.e. the $4/5$ law of
turbulence, we finally obtain
\begin{equation}
h(\epsilon)\sim \epsilon^{-3}\,.
\label{eq:4.5-5}
\end{equation} 
\begin{figure}
\centerline{\epsfig{figure=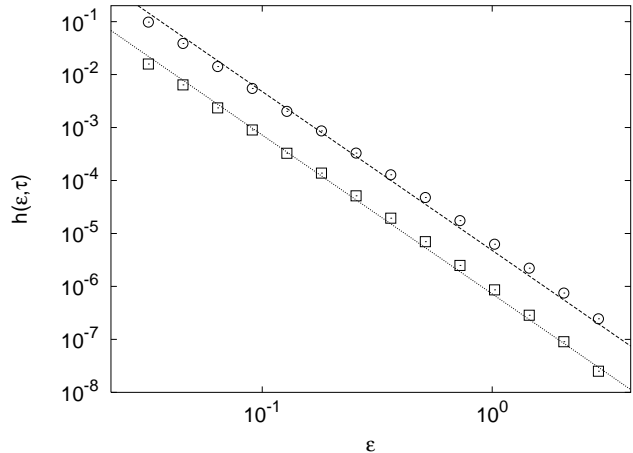,width=9cm}}
\protect\caption{Numerically computed lower bound ($\Box$) and upper
bound ($\circ$), with $\tau=0.1\langle t(\epsilon)\rangle$ for the
$(\epsilon,\tau)$-entropy in the case of a multiaffine signal with
$\zeta(3)=1$.  The signal has been obtained with the method of
Ref.~[29] (see also Appendix~\ref{app:synth}) using a
$D(h)$ which fits experimental data at large Reynolds number.  The two
straight lines show the theoretical scaling $\epsilon^{-3}$.}
\label{fig_eps_multi}
\end{figure}
In Fig.~(\ref{fig_eps_multi}) we report the evaluation of the upper
and lower bound (see Appendix~\ref{app:eps}) of $h(\epsilon)$ for a
synthetic signal, $v(t)$, constructed in such a way to reproduce the
statistical properties of turbulence \cite{BBCCV98}.

Let us now compare the above results with a previous study of the
$\epsilon$-entropy in turbulence \cite{WG92}, where it was argued that:
\begin{equation}
h(\epsilon) \sim \epsilon^{-2}\,,
\label{eq:wrong}
\end{equation}
a behavior that differs from the prediction (\ref{eq:4.5-5}).  The behavior
(\ref{eq:wrong}) has been obtained by assuming that $h(\epsilon)$, at
scale $\epsilon$, is proportional to the inverse of the typical eddy
turnover time at that scale: since the typical eddy turnover time for
velocity fluctuations of order $\delta v \sim \epsilon$ is
$\tau(\epsilon) \sim \epsilon^2$, Eq.~(\ref{eq:wrong}) follows. Indeed
this is the argument used to derive (\ref{eq:4.4-3}) for the FSLE.
The difference between (\ref{eq:wrong}) and (\ref{eq:4.5-5}) can be
understood by considering that even if $\lambda(\delta)$ and
$h(\epsilon)$ are two complementary concepts (the fact that for both
the estimate of the scaling behavior reduces to the ``$4/5$ law'' is
not a coincidence), in the latter case one has to consider the sweeping
induced by the large scales. On the contrary, since the former is
related to the distance of two realizations which differ in the small
scales ($<\delta$) but not on the large scales ($>\delta$), the
sweeping of the large scales is not effective.

\subsubsection{$\epsilon$-entropy of turbulence and the Taylor Hypothesis}
\label{sec:4.5.2}
Now we study the $\epsilon$-entropy per unit time and volume for the
velocity field of turbulent flows in $3+1$ dimensions,
$h^{ST}(\epsilon)$.  We will show that, by assuming the usually
accepted Taylor hypothesis, one has a spatial correlation which can be
quantitatively characterized by an ``entropy'' dimension ${\cal
D}=8/3$. As already remarked, $h^{ST}(\epsilon)$ cannot be directly
measured so we will discuss its estimation in a theoretical framework
by introducing a multi-affine field.  For the sake of simplicity, we
neglect intermittency by assuming a pure self-affine field with a
unique H\"older exponent $h=1/3$.

Let us first introduce a multi-affine field with the proper spatial
and temporal scaling \cite{ABCFVV00b}.  The idea consists in defining
the signal as a dyadic three-dimensional superposition of wavelet-like
functions $\varphi(({\bf x} - {\bf x}_{n,{\bf k}}(t))/\ell_n)$ whose
centers move due to the sweeping.  The coefficients of the
decomposition $a_{n,{\bf k}}(t)$ are stochastic functions chosen with
suitable self-affine scaling properties both in time and in space.  A
field with spatial H\"older exponent $h$ in $d$-dimensions is (see
Appendix~\ref{app:synth}):
\begin{equation}
\label{aff-proc}
v({\bf x},t)= \sum_{n=1}^M \sum_{k=1}^{2^{d(n-1)}} a_{n,k}(t)\,
\varphi \! \left(\frac{{\bf x} - {\bf x}_{n,k}(t)}{\ell_n}\right)\,\,,
\end{equation}
where ${\bf x}_{n,k}$ is the center of the $k^{th}$ wavelets at the
level $n$, i.e. for eddies with size $\ell_n \sim 2^{-n}$.  According
to the Richardson-Kolmogorov cascade picture, one assumes that
sweeping is present, i.e., ${\bf x}_{n+1,k}={\bf x}_{n,k^\prime}+{\bf
r}_{n+1,k}$ where $(n,k^\prime)$ labels the ``mother'' of the
$(n+1,k)$-eddy and ${\bf r}_{n+1,k}$ is a stochastic vector which
depends on ${\bf r}_{n,k^\prime}$ and evolves with characteristic time
$\tau_n \propto (\ell_n)^{1-h} $.

If the coefficients $\{a_{n,k}\}$ and $\{ {\bf r}_{n,k}\}$ have
characteristic time $\tau_n \sim (\ell_n)^{1-h}$ and $\{a_{n,k}\} \sim
(\ell_n)^h$, it is possible to show (see Appendix~\ref{app:synth} for
details) that the field (\ref{aff-proc}) has the correct
spatio-temporal statistics, i.e.
\begin{eqnarray}
|v({\bf x}+{\bf R},t_0) - v({\bf x},t_0)| &\sim& |{\bf R}|^{h}
\label{eq:aggiunta1}\\ | v({\bf x},t_0+t) - v({\bf x},t_0)| &\sim&
t^{\,h} \,.
\label{eq:aggiunta2}
\end{eqnarray}
In addition the proper Lagrangian sweeping is satisfied. Now we are
ready for the $\epsilon$-entropy analysis of the field
(\ref{aff-proc}).  If one wants to look at the field $v$ with a
resolution $\epsilon$, one has to take $n$ in (\ref{aff-proc}) up to
$N$ given by:
\begin{equation}
(\ell_N)^{h} \sim \epsilon\;,
\end{equation}
in this way one is sure to consider velocity fluctuations of order
$\epsilon$.  Then the number of terms contributing to (\ref{aff-proc})
is
\begin{equation}
\#(\epsilon) \sim (2^d)^N \sim \epsilon^{-d/h}\;.
\label{eq:numero}
\end{equation}
By using a result of Shannon \cite{S48} one estimates the
$\epsilon$-entropy of the single process $a_{n,k}(t)$ (and also of
${\bf r}_{n,j}$) as:
\begin{equation}
h_n(\epsilon) \sim \frac{1}{\tau_n} \ln \left ({1 \over \epsilon}
\right)\;,
\end{equation}
where the above relation is rigorous if the processes $a_{n,k}(t)$ are
Gaussian and with a power spectrum different from zero on a band of
frequency $\sim 1/\tau_n$.  The terms which give the main contribution
are those with $n \sim N$ with $\tau_N \sim (\ell_N)^{1-h} \sim
\epsilon^{(\frac{1-h}{h})}$.  Their number is given by
(\ref{eq:numero}) so that, collecting the above results, one finds
\begin{equation}
\label{scaling-H}
h^{ST}(\epsilon) \sim
 {\#(\epsilon) \over \tau_N} \sim \epsilon^{-\frac{d-h+1}{h}}\;.
\end{equation}
For the physical case $d=3$, $h=1/3$, one obtains
\begin{equation}
h^{ST}(\epsilon) \sim \epsilon^{-11}\;.
\label{32}
\end{equation}
By denoting with $v_{\eta}$ the typical velocity at the Kolmogorov scale
$\eta$, one has that Eq.~(\ref{32}) holds in the inertial range, i.e.,
$\epsilon \ge v_{\eta} \sim Re^{-1/4}$, while for $\epsilon \le v_{\eta}$,
$h^{ST}(\epsilon)=$ constant $\sim Re^{11/4}$.

Let us now consider an alternative way to compute the
$\epsilon$-entropy of the field $v({\bf x},t)$: divide the $d$-volume
in boxes of edge length $\ell(\epsilon)\sim \epsilon^{1/h}$ and look
at the signals $v({\bf x}_\alpha,t)$, where the ${\bf x}_\alpha$ are
the centers of the boxes.  Denoting with $h^{(\alpha)}(\epsilon)$ the
$\epsilon$-entropy of the temporal sequence of the velocity field
measured in ${\bf x_{\alpha}}$, we have
\begin{equation}
h^{(\alpha)}(\epsilon) \sim \epsilon^{-1/h}\;
\label{entrotime}
\end{equation}
because of the scaling (\ref{eq:aggiunta2}). 
Therefore, $h^{ST}(\epsilon)$ is obtained summing up
all the ``independent'' contributions (\ref{entrotime}), i.e.
\begin{equation}
h^{ST}(\epsilon) \sim {\cal N}(\epsilon) h^{(\alpha)}(\epsilon)
 \sim {\cal N}(\epsilon) \epsilon^{-1/h}\;,
\end{equation}
where ${\cal N}(\epsilon)$ is the number of  independent cells.  It
is easy to understand that the simplest assumption ${\cal
N}(\epsilon)\sim l(\epsilon)^d\sim \epsilon^{d/h}$ gives a wrong
result, indeed one obtains
\begin{equation}
h^{ST}(\epsilon) \sim \epsilon^{-\frac{d+1}{h}}\;,
\end{equation}
which is not in agreement with (\ref{scaling-H}).  In order to obtain
the correct result (\ref{32}) it is necessary to assume
\begin{equation}
{\cal N}(\epsilon)\sim l(\epsilon)^{{\cal D}}\;,
\end{equation}
with ${\cal D}=d-h$.  In other words, one has that the sweeping
implies a nontrivial spatial correlation, quantitatively measured by
the exponent ${\cal D}$, which can be considered as a sort of
``entropy'' dimension.  Incidentally, we note that ${\cal D}$ has the
same numerical value of the fractal dimensions of the velocity
iso-surfaces \cite{M75,Voss89}. From this observation, at first
glance, one could conclude that the above result is somehow trivial
since it is simply related to a geometrical fact.  However, a closer
inspection reveals that this is not true.  Indeed, one can construct a
self-affine field with spatial scaling $h$ and thus with the fractal
dimension of the velocity iso-surfaces given by $d-h$ for geometrical
reasons, while ${\cal D}=d$.  Such a process can be simply obtained by
eliminating the sweeping, i.e.,
\begin{equation}
v({\bf x},t) = \sum_{n=1}^M\sum_{k=1}^{2^{d(n-1)}} a_{n,k}(t)\,
 \varphi \! \left({{\bf x}-{\bf x}_{n,k} \over \ell_n}\right)\,\,,
\label{lagrangian_field}
\end{equation}
where now the ${\bf x}_{n,k}$ are fixed and no longer time-dependent,
while $a_{n,k} \sim (\ell_n)^{h}$ but $\tau_n \sim \ell_n$.

We conclude by noting that it is possible to obtain (see \cite{GW93})
the scaling~(\ref{scaling-H}) using equation~(\ref{lagrangian_field}),
i.e. ignoring the sweeping, assuming $\tau_n \sim (\ell_n)^{1-h}$ and
$a_{n,k} \sim (\ell_n)^h$; this corresponds to take separately the
proper temporal and spatial spectra.  However, this is not
satisfactory since one has not the proper scaling in one fixed point
(see Eq.~(\ref{entrotime}) the only way to obtain this is through the
sweeping).

\section{Uncertainty in the evolution equations}
\label{sec:5}
\setcounter{equation}{0}
The study of a large class of problems in science (physics, chemistry,
biology,...) is reduced to the investigation of evolution laws, which
describe some aspects of the system. The assumption 
that natural processes can be described by mathematical models is 
at the foundation of this approach \cite{W60,DH84}.  The purpose of this
Section is to discuss how the unavoidable uncertainty in the equation
of motion puts limits on the long time forecasting.

To be more concrete, let us consider a system described by a
differential equation:
\begin{equation}
{d \over d t} \, {\bf x}(t)={\bf f}({\bf x},t) \, , \; \; \; \;
{\bf x, f} \in {\mathrm I\!R}^{n}  \, .
\label{eq:5.0-1}
\end{equation}
As a matter of fact, we do not know exactly the equations, so we have
to devise a model which is different from the true dynamics. In
practice this means that we study
\begin{equation}
{d \over dt} \, {\bf x}(t)={\bf f}_{\epsilon}({\bf x},t)
\; \; \; \; \mbox{where} \; \; \;
{\bf f}_{\epsilon}({\bf x},t)={\bf f}({\bf x},t)+\epsilon
\delta{\bf f}({\bf x},t) \, .
\label{eq:5.0-2}
\end{equation}
Therefore, it is natural to wonder about the relation between the true
evolution ({\em reference} or {\em true} trajectory ${\bf x}_{T}(t)$)
given by (\ref{eq:5.0-1}) and the one effectively computed ({\em
perturbed} or {\em model} trajectory ${\bf x}_{M}(t)$) given by
(\ref{eq:5.0-2}).  A typical example is 
 the relation between the true dynamics of the
physical system and the one obtained by a computer simulation. This
issue is of particular relevance for the study of weather forecast
where it is referred to as {\it predictability of the second kind}
\cite{P95}.

In this context it is particularly relevant the {\it shadowing lemma}
\cite{B70} which implies that, for Anosov systems, a computer may not 
calculate the true orbit but what it does find is nevertheless an approximation
of the true one. As a consequence, the statistical properties are
well reproduced by an accurate numerical integration \cite{G86}.

A central point in the discussion of the second kind predictability
problem is the issue of {\it structural stability} \cite{GH86}: since
the evolution laws are known only with finite precision it is highly
desirable that at least certain properties were not too sensitive to
the details of the equations of motion.  For example, in a system with
a strange attractor, small generic changes in the evolution laws
should not change drastically the statistical properties of the
dynamics \cite{ER85,HLB96}.

In order to see that a non generic perturbation, although very
``small'' in some sense, can produce dramatic changes in the
statistical properties of the dynamics, following
Refs.~\cite{B94,HLB96}, we consider the one-dimensional chaotic map
$x(t+1)=f(x(t))$ with $f(x)=4 x$ mod $1$, and a perturbed version of
it:
\begin{equation}
f_{\epsilon}(x)=\left\{ 
\begin{array} {ll}

8x-\frac{9}{2} & \;\;\; x \in \left[\frac{5}{8},\frac{247}{384}\right]  \\ 
\\
\frac{1}{2}x+\frac{1}{3} & \;\;\;x \in \left[\frac{247}{384},
\frac{265}{384}\right]\\ 
\\
8x-\frac{29}{6} &  \;\;\; x \in \left[\frac{265}{384},\frac{17}{24}\right] \\
\\
4x\;{\mbox {mod}}\;1 & \;\;\;{\mbox{otherwise}}\,.
\end{array}
\right.    
\label{eq:ap0}
\end{equation}
The perturbed map is identical to the original outside the interval
$[5/8,17/24]$, and the perturbation is small in $L_{2}$ norm.
Nevertheless, the fixed point $x=2/3$, which is unstable in the
original dynamics, becomes stable in the perturbed one, and it is a
{\it global attractor} for $f_{\epsilon}(x)$, i.e. almost every point
in $[0,1]$ asymptotically approaches $x=2/3$.

\begin{figure}[htb]
\centerline{\epsfig{figure=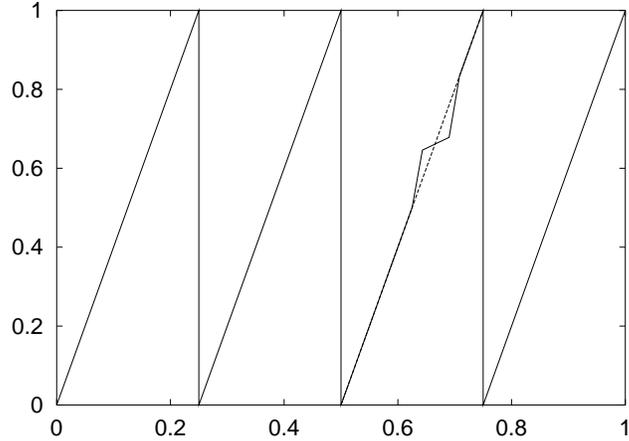,width=9cm, angle=0}}
\caption{The map $f_{\epsilon}$ of equation (\ref{eq:ap0}) (solid
line) and the original chaotic map $f$ (dashed line).}
\label{fig:5.0-1}
\end{figure}

If one compares the trajectories obtained iterating $f(x)$ or 
$f_{\epsilon}(x)$ it is not difficult to understand that they may 
remain identical for a
certain time but unavoidably differ utterly in the long time
behavior.  The transient chaotic behavior of the perturbed orbits
can be rendered arbitrarily long by reducing the interval in which the
two dynamics differ \cite{B94}.

As for the problem of predictability with respect to perturbations on
the initial conditions, the problem of second kind predictability in
the limit of infinitesimal perturbations is essentially understood in
terms of the Lyapunov exponents. Indeed, it is possible to show (see
below) that a small uncertainty on the evolution laws of chaotic
systems has the same effects of an error of the same order of
magnitude on the initial conditions.  However, also in the case of
second kind predictability one has often to deal with errors which are
far from being infinitesimal.  Moreover, in real systems the size of
an uncertainty on the evolution equations is determinable only {\it a
posteriori}, based on the ability of the model to reproduce some of
the features of the phenomenon.  Typical examples are systems
described by partial differential equations (e.g. turbulence,
atmospheric flows). The numerical study of these systems is performed
by using a model with unavoidable severe approximations, the most
relevant due to the necessity to cut some degrees of freedom off
(i.e. the small scale variables).  A relevant problem in this case is
to quantify the effect of the unresolved scales on the predictability
of the resolved ones.

From a general point of view, in the second kind predictability
problem we can distinguish three main cases depending on the original
dynamics.  In particular, Eq.~(\ref{eq:5.0-1}) may display:

(i) trivial attractors:
asymptotically stable fixed points or attracting periodic
orbits;

(ii) marginally stable fixed points or periodic/quasi-periodic orbits
 as in integrable Hamiltonian systems;

(iii) chaotic behavior.

In case (i) small changes in the equations of motion do not modify the
qualitative features of the dynamics.  Case (ii) is not generic and
the outcome strongly depends on the specific perturbation $\delta {\bf
f}$, i.e. it is not structurally stable (see \cite{CFV89} for a
discussion on this point).  In the chaotic case (iii) one expects that
the perturbed dynamics is still chaotic. In the following we will
consider only this latter case.

In chaotic systems, the effects of a small uncertainty on the evolution
law is, for many aspects, similar to those due to imperfect knowledge
of initial conditions. As an example let us consider the Lorenz system
(\ref{eq:2.2-5}).  In order to mimic an indetermination in the
evolution law we assume a small error $\epsilon$ on the parameter $r$:
$r \rightarrow r+\epsilon$. Let us consider the difference $\delta
{\bf x}(t)={\bf x}_{M}(t)-{\bf x}_{T}(t)$,  for simplicity,
$\delta {\bf x}(0)=0$, i.e. we assume a perfect knowledge of the
initial condition.  For small $\epsilon$ one has, with obvious
notation:
\begin{equation}
{d \delta {\bf x} \over d t}={\bf f}_{\epsilon}({\bf x}_{M})-
{\bf f}({\bf x}_{T}) \simeq
\frac{\partial {\bf f}}{\partial {\bf x}} \, \delta {\bf x}
+\frac{\partial {\bf f}_{\epsilon}}{\partial r} \, \epsilon \, .
\label{eq:5.0-4}
\end{equation}
Since at time $t=0$ one has $|\delta {\bf x}(0)|=0$, $|\delta {\bf
x}(t)|$ initially grows under the effect of the second term in
(\ref{eq:5.0-4}).  At later times, when $|\delta {\bf x}(t)|\approx
O(\epsilon)$ the first term becomes the leading one, and we recover
the first kind predictability for an initial uncertainty $\delta_0
\sim \epsilon$.  Therefore, apart from an initial growth, which
depends on the specific perturbation, for small enough $\epsilon$ the
evolution of $\langle \ln(|\delta {\bf x}(t)|) \rangle$ follows the
usual linear growth with the slope given by the largest LE.  Typically
the value of the LE computed by using the model dynamics differs from
the true one by a small amount of order $\epsilon$,
i.e. $\lambda_{M}=\lambda_{T}+O(\epsilon)$ \cite{CFV89}.

\begin{figure}[htb]
\centerline{\epsfig{figure=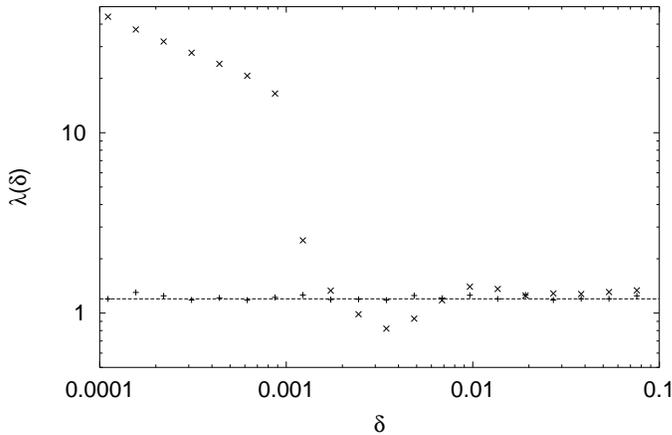,width=9cm, angle=0}}
\caption{Finite Size Lyapunov Exponents $\lambda_{TT}(\delta)$ ($+$)
and $\lambda_{TM}(\delta)$ ($\times$) versus $\delta$ for the Lorenz
model (\ref{eq:2.2-5}) with $\sigma=c=10$, $b=8/3$, $r=45$ and
$\epsilon=0.001$. The dashed line represents the largest Lyapunov
exponent for the unperturbed system ($\lambda_{T} \approx 1.2$).  The
statistics is over $10^4$ realizations.}
\label{fig:5.0-2}
\end{figure}

A picture of the error growth, valid also for finite errors, can be
obtained by considering the Finite Size Lyapunov Exponent.  In
addition to the FSLE of the model, $\lambda_{MM}(\delta)$, we
introduce the FSLE for the true dynamics (\ref{eq:5.0-1})
$\lambda_{TT}(\delta)$ (which cannot be measured in real situations)
and $\lambda_{TM}(\delta)$, the FSLE computed following the distance
between one true trajectory and one model trajectory starting at the
same point. In the case of a perfect model
$\lambda_{MM}(\delta)=\lambda_{TT}(\delta)$.  The results of the
computation for the Lorenz model (\ref{eq:2.2-5}) are shown in
Figure~\ref{fig:5.0-2}.  $\lambda_{TT}(\delta)$ displays the chaotic
plateau with $\lambda \simeq 1.2$.  As discussed above, for $\delta >
\epsilon$ the second term in (\ref{eq:5.0-4}) becomes negligible and
we observe $\lambda_{TM}(\delta) \simeq \lambda_{TT}(\delta) \simeq
\lambda$.  In this range of errors the model system recovers the
intrinsic predictability of the true system.  For very small errors,
$\lambda_{TM}$ is dominated by the second term in (\ref{eq:5.0-4}) and
deviates from $\lambda_{TT}$.

\subsection{Uncertainty introduced by numerical computations}
\label{sec:5.1}
In numerical computations, an unavoidable source of errors is due to
the representation of numbers on the computer, as computers work
with integers. This has two main consequences: the phase space of the
simulated system is necessarily discrete (and finite); and the
computation introduces a sort of noise due to the round-off.

A direct consequence of the discreteness in phase space is that any
numerical trajectory is periodic.  At first sight, this seems a very
serious problem, especially when integrating chaotic systems which have
non periodic behavior. However, as discussed in \cite{CFV89}, apart
from cases in which one uses very low precision, and very low
dimensional systems, the period is usually extremely large and one
works with an effective continuous phase space dynamical system (see
Sect.~\ref{sec:6.1}).

The round-off produces on (\ref{eq:5.0-1})-(\ref{eq:5.0-2}) a
perturbation $\delta{\bf f}({\bf x},t)$ of order $\epsilon \sim
10^{-\alpha}$ ($\alpha=$number of digits in floating point
representation) which depends on ${\bf f}$ and on the software
\cite{K69}.  In general, the round-off error is very small and may
have a positive role in selecting the physical probability measure,
the so-called {\it natural measure}, from the set of the admissible
invariant ones \cite{ER85}.

\begin{figure}[htb]
\centerline{\epsfig{figure=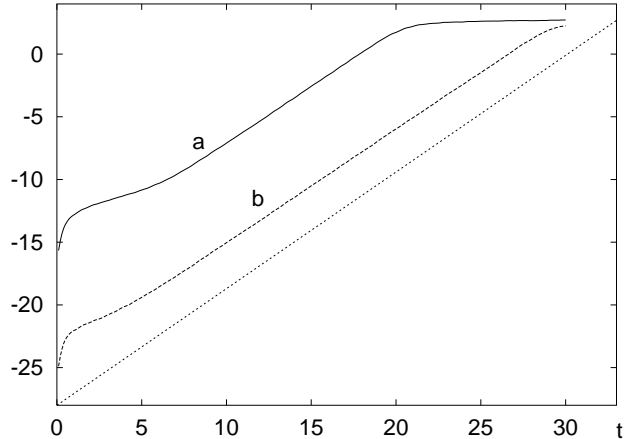,width=9cm, angle=0}}
\caption{$\langle \ln|\delta {\bf x}(t)\rangle|$ versus $t$, where
$|\delta {\bf x}(t)|$ is the Euclidean distance between two
trajectories of the Lorenz model (\ref{eq:2.2-5}) for $r=28$.  Curve
(a) refers to the comparison between trajectories obtained using a
fourth order Runge-Kutta algorithm with $\Delta t=4\times 10^{-3}$ and
$\Delta t=4\times 10^{-5}$.  Curve (b) shows the same quantity
obtained with $\Delta t=4\times 10^{-4}$ and $\Delta t=4\times
10^{-5}$. The dotted line with slope $\lambda\approx 0.9$ is shown for
comparison.}
\label{fig:5.1-1}
\end{figure}

In order to show the effect of the numerical precision on the
predictability, let us consider again the Lorenz model
(\ref{eq:2.2-5}).  At variance with the previous Section, here we
assume to have a perfect knowledge of the model (i.e. of the parameter
$r$), and the error is introduced only by the numerical integration,
e.g. by different time step $\Delta t$.  The most precise integration
with smallest $\Delta t$ is taken as the reference trajectory and the
other is the perturbed one.  The result is shown in
Figure~\ref{fig:5.1-1} for two different values of $\Delta t$ for the
perturbed integration. In both cases, for small values of the error,
the exponential growth rate is given by the largest LE $\lambda$. The
same behavior is observed by introducing the numerical error in other
ways, e.g.  by using different precision (single or double) or
different integration algorithms \cite{CFV89}.

\subsection{Finite resolution effects and  parameterization of 
unresolved scales}
\label{sec:5.2}
Let us now consider more complex situations, in which many interacting
degrees of freedom and different characteristic times are involved
\cite{BCCLV00b}.  We will consider the particular examples of an
extremely simplified model of global circulation \cite{L96,LE98} and
the shell model (Sect.~\ref{sec:4.2}).

For systems with many different scales usually one is able to
represent only the large scale variables.  A typical situation is the
discretization of partial differential equations. The small scale
modes, below the computational grid, are unresolved and are typically
parameterized according to some phenomenological prescription
(e.g. the eddy viscosity parameterization of the small scales
\cite{L90,F95}).  So we consider systems of the following form
\begin{equation}
\displaystyle{ \frac{d{\bf x}}{dt} }= {\bf f}({\bf x},{\bf y}) \qquad
\displaystyle{ \frac{d{\bf y}}{dt} }= {\bf g}({\bf x},{\bf y})
\label{eq:5.2-1}
\end{equation}
where ${\bf x}\in {\mathrm I\!R}^{n}$ represent the large (and
typically slow) variables while ${\bf y}\in {\mathrm I\!R}^{m}$
represent the small (and fast) ones.  As explained above, in many
practical situations the small variables cannot be explicitly
resolved. In this framework, a natural question is: how must we
parameterize the unresolved modes in order to predict the resolved
ones?  In this respect, the optimal parameterization is that
one for which the predictability on the resolved modes is not worse
than the intrinsic predictability of the same variables in the
complete system, i.e. in our notation $\lambda_{TM}=\lambda_{TT}$.

An example in which it is relatively simple to develop a model for the
small scale modes is represented by skew systems, i.e., ${\bf g}$
depends only on the fast variables ${\bf y}$.  In this case, simply
neglecting the fast variables or parameterizing them with a suitable
stochastic process does not drastically affect the prediction of the
slow variables \cite{BPV96}.

On the other hand, in typical cases ${\bf y}$ feels some feedback from
${\bf x}$, and, therefore, we cannot simply neglect the unresolved
modes (see Ref.~\cite{BCCLV00b} for details).  In practice one has to
construct an effective equation for the resolved variables:
\begin{equation}
{d {\bf x} \over d t}= {\bf f}_{M}({\bf x},{\bf y({\bf x})}) \,,
\end{equation}
where the functional form of ${\bf y({\bf x})}$ and ${\bf f}_{M}$
is built  by phenomenological arguments and/or by numerical studies
of the full dynamics (if available).

Let us now discuss a simplified model for atmosphere circulation
\cite{L96,LE98} which includes large scales $x_{k}$ (synoptic scales)
and small scales $y_{j,k}$ (convective scales):
\begin{equation}
\begin{array}{lll}
\displaystyle{ {d x_{k} \over dt} }& = & - x_{k-1}
\left(x_{k-2}-x_{k+1}\right) - \nu x_{k} + F - \sum_{j=1}^{J} y_{j,k}
\vspace{5pt}\\ \displaystyle{ {d y_{j,k} \over dt} } & = & - c b
y_{j+1,k} \left(y_{j+2,k}-y_{j-1,k}\right) - c \nu y_{j,k} + x_{k}
\,,\\
\end{array}
\label{eq:5.2-3}
\end{equation}
where $k=1,...,K$ and $j=1,...,J$. As in \cite{L96} we assume periodic
boundary conditions on $k$ ($x_{K+k}=x_{k}$, $y_{j,K+k}=y_{j,k}$)
while for $j$ we impose $y_{J+j,k}=y_{j,k+1}$. The variables $x_k$
represent some large scale atmospheric quantities in $K$ sectors
extending on a latitude circle, while the $y_{j,k}$ represent
quantities on smaller scales in $J \cdot K$ sectors.  The parameter
$c$ is the ratio between fast and slow characteristic times and $b$
measures the relative amplitude (both larger than unity).  Model
(\ref{eq:5.2-3}), even if rather crude, contains some non trivial
aspects of the general circulation problem, namely the coupling among
variables with very different characteristic times.

\begin{figure}[htb]
\centerline{\epsfig{figure=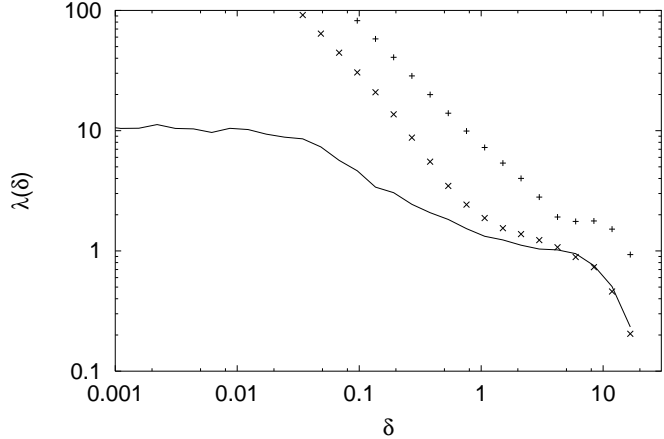,width=9cm, angle=0}}
\caption{Finite Size Lyapunov Exponents for the Lorenz model
(\ref{eq:5.2-3}) $\lambda_{TT}(\delta)$ (solid line) and
$\lambda_{TM}(\delta)$ versus $\delta$ obtained by dropping the fast
modes ($+$) and with eddy viscosity parameterization ($\times$).  
The parameters are
$F=10$, $K=36\,,\;J=10$, $\nu=1$ and $c=b=10$, implying that the
typical $y$ variable is $10$ times faster and smaller than the $x$
variable.  The value of the parameter $\nu_e=4$ is chosen after a
numerical integration of the complete equations as discussed in
Ref.~[36]. The statistics is over $10^4$ realizations.}
\label{fig:5.2-1}
\end{figure}

Being interested in forecasting the large scale behavior of the
atmosphere by using only the slow variables, a natural choice for the
model equations is:
\begin{equation}
{d x_{k} \over dt} = - x_{k-1} \left(x_{k-2}-x_{k+1}\right)
- \nu x_{k} + F - G_{k}({\bf x})\,,
\label{eq:5.2-4}
\end{equation}
where $G_{k}({\bf x})$ represents the parameterization of the fast
components in (\ref{eq:5.2-3}).  Following the approach discussed in
Ref.~\cite{BCCLV00b}, a physical reasonable parameterization is
\begin{equation}
G_{k}({\bf x}) = \nu_e x_k\,,
\label{eq:5.2-5}
\end{equation}
where $\nu_e$ is a numerically determined parameter.

In Figure~\ref{fig:5.2-1} we plot $\lambda_{TM}(\delta)$ obtained from
different choices of $G_{k}$.  The simplest possibility is to neglect
the fast variable, i.e. $G_{k}=0$. Also for large errors we have
$\lambda_{TM}(\delta)>\lambda_{TT}(\delta)$ because this crude
approximation is not able to capture the intrinsic predictability of
the original system.  More refined parameterizations in terms of
stochastic processes with the correct probability distribution
function and correlation time do not improve the forecasting ability.
On the contrary Eq.~(\ref{eq:5.2-5}) gives the result shown in
Figure~\ref{fig:5.2-1}.  At small scales we still observe deviations
from $\lambda_{TT}$ but, at variance with the previous case, we
recover intrinsic predictability for error of the size of the resolved
scale. 

As a more complex example, let us consider a version of the shell
model discussed in Sect.~\ref{sec:4.2}, more precisely we study
\cite{LPPPV98}:
\begin{eqnarray}
{d u_n \over d t} = i \left(k_{n+1} u^*_{n+1} u_{n+2} -
{1 \over 2} k_{n} u^*_{n-1} u_{n+1}  \right. &+&\left.
{1 \over 2} k_{n-1} u_{n-2} u_{n-1} \right)\nonumber\\
&-&\nu k_{n}^{2} u_{n} + f_{n}\,,
\label{eq:sabra}
\end{eqnarray}
with $n=1,\dots,N$.
\begin{figure}[htb]
\centerline{\epsfig{figure=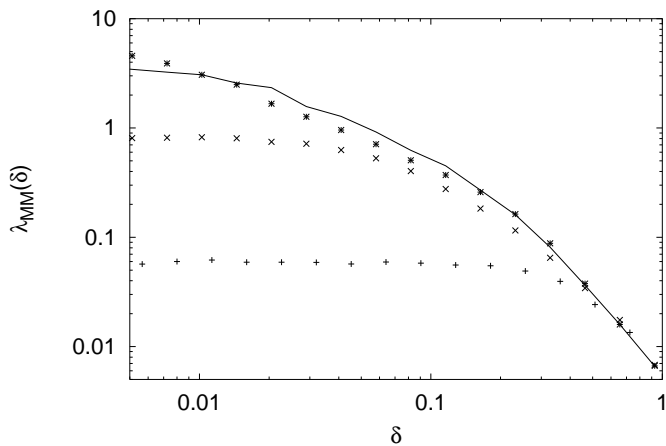,width=9cm, angle=0}}
\caption{The FSLE for the eddy-viscosity shell model
(\ref{eq:5.2-6}-\ref{eq:5.2-7}) $\lambda_{MM}(\delta)$ at
various resolutions $N_M=9(+),15(\times),20(\ast)$.
For comparison it is drawn  $\lambda_{TT}(\delta)$ (continuous line). 
Here $\kappa=0.4$, $k_0=0.05$.}
\label{fig:5.2-2}
\end{figure}

At variance with the previous example, here we have a set of scales
$\ell_{n} \simeq 1/k_{n}$ with characteristic times $\tau_{n} \sim
k_n^{-2/3}$ (see Sect.~\ref{sec:4.4}).  In order to simulate a finite
resolution in the model, we consider a model of
(\ref{eq:sabra}) in terms of an eddy viscosity \cite{BBST98}:
\begin{eqnarray}
{d u_n \over d t} = i \left(k_{n+1} u^*_{n+1} u_{n+2}- 
{1 \over 2} k_{n} u^*_{n-1} u_{n+1} \right.&+&\left.
{1 \over 2} k_{n-1} u_{n-2} u_{n-1} \right) \nonumber\\
&-& \nu^{(e)}_{n} k_{n}^{2} u_{n} + f_{n}\,,
\label{eq:5.2-6}
\end{eqnarray}
where now $n=1,...,N_M<N$ and the eddy viscosity, restricted to the
last two shells, has the form
\begin{equation}
\nu^{(e)}_{n} = \kappa {|u_n| \over k_n} \left(\delta_{n,N_{M}-1}+
\delta_{n,N_M}\right)\,,
\label{eq:5.2-7}
\end{equation}
where $\kappa$ is a constant of order $1$ \cite{BBST98}.  In the model
equation $N_M<N$ the molecular viscosity term is much smaller than the
eddy viscosity term and can be simply neglected. Model equations
(\ref{eq:5.2-6}-\ref{eq:5.2-7}) are essentially the large eddy
simulation for the shell model. Thus, although shell models are not
realistic models for large scale geophysical flows (being nevertheless
a good model for small scale turbulent fluctuations), the study of the
effect of truncation in term of eddy viscosity is of general interest.

\begin{figure}[htb]
\centerline{\epsfig{figure=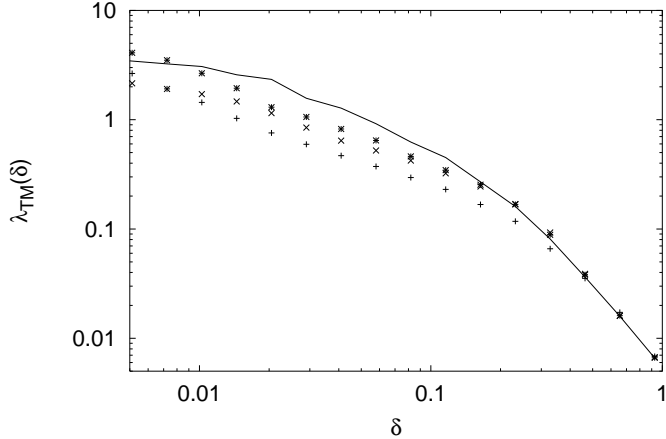,width=9cm, angle=0}}
\caption{The FSLE between the eddy-viscosity shell model and the full
shell model $\lambda_{TM}(\delta)$, at various resolutions
$N_M=9(+),15(\times),20(\ast)$. For comparison it is drawn
$\lambda_{TT}(\delta)$ (continuous line).  The total number of shell
for the complete model is $N=24$, with $k_0=0.05$, $\nu=10^{-7}$.}
\label{fig:5.2-3}
\end{figure}

In Figure~\ref{fig:5.2-2} we show $\lambda_{MM}(\delta)$, i.e.  the
FSLE computed for the model equations (\ref{eq:5.2-6}) with $N=24$ at
different resolutions $N_{M}=9,15,20$.  A plateau is detected for
small amplitudes of the error $\delta$, corresponding to the LE, which
increases with the resolution according to $\lambda \sim
k_{N_M}^{2/3}$.  At larger $\delta$, the curves collapse onto the
$\lambda_{TT}(\delta)$, showing that large-scale statistics of the
model is not affected by small-scales resolution.

The ability of the model to predict satisfactorily the features of the
``true'' dynamics is not anyway determined by $\lambda_{MM}(\delta)$
but by $\lambda_{TM}(\delta)$, which is shown in
Figure~\ref{fig:5.2-3}.  Increasing the resolution $N_M=9,15,20$
towards the fully resolved case $N=24$ the model improves, in
agreement with the expectation that $\lambda_{TM}$ approaches
$\lambda_{TT}$ for a perfect model.  At large $\delta$ the curves
practically coincide, showing that the predictability time for large
error sizes (associated with large scales) is independent of the
details of small-scale modeling.

\subsection{Lyapunov exponents and complexity in dynamical systems
with noise}
\label{sec:5.3}

We saw how in deterministic dynamical systems there exist well established ways
to define the complexity of a temporal evolution, either in terms of
the Lyapunov exponents and the Kolmogorov-Sinai entropy, or by means
of their generalization to non infinitesimal perturbations, like
$FSLE$ and $\epsilon$-entropy.  The situation is much more ambiguous
with random perturbations, which are always present in physical
systems as a consequence of thermal fluctuations or hidden changes of
control parameters, and, in numerical experiments, because of the
roundoff errors \cite{MM89}.
 
The combined effect of the noise and the deterministic part of the
evolution law can produce highly non-trivial behaviors
\cite{BJS90,C90,DF92,HES86,HP87,MT83}.  Let us mention stochastic
resonance, where there is a synchronization of the jumps between two
stable points \cite{BSV81,BPSV82,N82} (for a recent review see
\cite{GHJM98}), and the phenomena of the so called noise-induced order
\cite{MT83} and noise-induced instability \cite{BJS90,C90}.
 
When facing  systems with noise, the simplest possibility is to
treat the random term as a time-dependent term, that is to consider
the separation of two close trajectories with the same realization of
noise. In this way one computes the largest LE, $\lambda_{\sigma}$,
associated with the separation rate of two nearby trajectories with
the same realization of the stochastic term (where $\sigma$ indicates
the noise strength).  Although $\lambda_{\sigma}$ is a well defined
quantity, i.e. the Oseledec theorem \cite{O68} holds, it is not the
most useful characterization of complexity.  In addition, a moment of
reflection shows that it is practically impossible to extract
$\lambda_{\sigma}$ from experimental data.
 
We will show how, for noisy and random systems, a more natural
indicator of complexity can be obtained by computing the separation
rate of nearby trajectories evolving with different noise 
realizations. This measure of complexity, defined in \cite{PSV95,LPSV95},
and inspired by the contributions of Shannon \cite{S48} and Kolmogorov
\cite{K65}, is related to the mean number of bits per unit time
necessary to specify the sequence generated by a random evolution law.
\subsubsection{The naive approach:  noise treated as a standard 
 function of time}
\label{sec:5.3.1}
The approach in which one treats the random term as an usual
time-dependent external force can lead to  misleading results, 
as illustrated in the following example.
  
Let us consider a one-dimensional Langevin equation \begin{equation} {
{d x} \over {d t} }= - { {\partial V(x)} \over {\partial x}} +
\sqrt{2\sigma}\, \eta \, , \end{equation} where $\eta(t)$ is a white
noise and $V(x)$ diverges for $\mid x\mid \to \infty$, like, e.g., the
usual double well potential $V=-x^2/2+x^4/4$.
 
The Lyapunov exponent $\lambda_{\sigma}$, associated with the
separation rate of two nearby trajectories with the same realization
of $\eta(t)$, is defined as \begin{equation}
\lambda_{\sigma}=\lim_{t\to \infty} { 1 \over t} \ln |z(t)|
\label{deflyap} \end{equation} where the evolution of the tangent
vector is given by: \begin{equation} { {d z} \over {d t} }= - {
{\partial^2 V(x(t))} \over {\partial x^2}}z(t). \end{equation} Since
the system is ergodic with invariant probability distribution $P(x)=C
e^{-V(x)/\sigma}$, one has: \begin{equation} \begin{array}{ll}
\lambda_{\sigma}&=\lim_{t\to \infty} { 1 \over t} \ln |z(t)|=
-\lim_{t\to \infty} { 1 \over t} \int_0^t \partial^2_{xx} V(x(t'))
dt'= \\ &\\ &-C \int \partial^2_{xx} V(x) e^{-V(x)/\sigma} \,\, dx=
-{C \over \sigma } \int (\partial_x V(x))^2 e^{-V(x)/\sigma} \,\, dx <
0 \, .  \end{array} \end{equation} This result has a rather intuitive
meaning: the trajectory $x(t)$ spends most of the time in one of the
``valleys'' where $-\partial^2_{xx} V(x) < 0$ and only  short intervals
on the ``hills'' where $-\partial^2_{xx} V(x) > 0$, so that the distance
between two trajectories evolving with the same noise realization
decreases on average.  Notice that in Ref.~\cite{VN93}, supported by a
wrong argument, an opposite conclusion has been claimed.
 
A negative value of $\lambda_{\sigma}$ implies a fully predictable process only if the
realization of the noise is known.  In the case of two initially close
trajectories evolving under two different noise realizations, after a
certain time $T_{\sigma}$, the two trajectories can be very distant,
because they can be in two different valleys. For $\sigma \to 0$, due
to the Kramers formula \cite{C43}, one has $T_\sigma \sim \exp \Delta
V/\sigma$, where $\Delta V$ is the difference between the values of
$V$ on the top of the hill and at the bottom of the valley. The result
obtained for the one dimensional Langevin equation can easily be
generalized to any dimension for gradient systems if the noise is
small enough \cite{LPSV95}.

Another example showing the limitations of this approach is
provided by the case of stochastic resonance in chaotic systems.  In
this case, in fact, one can find the same qualitative behavior both
for a positive and a negative LE.  We refer to \cite{LPSV95} for more
details.

\subsubsection{An information theory approach} 
\label{sec:5.3.2} 
The main difficulties in defining the notion of ``complexity'' of an
evolution law with a random perturbation already appears in 1D maps.
The generalization to $N$-dimensional maps or to ordinary differential
equations is straightforward.

Therefore, we consider the model
\begin{equation} x(t+1)=f[x(t),t]+\sigma w(t), 
\end{equation}
where $t$ is an integer and $w(t)$ is an uncorrelated random process,
e.g. $w$ are independent random variables uniformly distributed in
$[-1/2,1/2]$. For the largest LE $\lambda_{\sigma}$, as defined in
(\ref{deflyap}), now one has to study the equation
\begin{equation} z(t+1)=f'[x(t),t] \,z(t), 
\end{equation} where $f'=df/dx$.
 
Following the approach of Sect.~\ref{sec:1.3}, 
let $x(t)$ be the trajectory starting at
$x(0)$ and $x'(t)$ be the trajectory starting from $x'(0)=x(0)+\delta
x (0)$.  Let $\delta_0 \equiv |\delta x(0)|$ and indicate by $\tau_1$
the minimum time such that $|x'(\tau_1)-x(\tau_1)|\ge \Delta$. Then,
we put $x'(\tau_1)=x(\tau_1)+\delta x(0)$ and define $\tau_2$ as the
time such that $|x'(\tau_1+\tau_2)-x(\tau_1+\tau_2)|>\Delta$ for the
first time, and so on. In this way the Lyapunov exponent 
can be defined  as
\begin{equation} 
\lambda= 
{1 \over \overline{\tau} } \,
\ln \left( { \Delta \over \delta_0} \right)\,
\label{eq:lll}
\end{equation}
being $\overline{\tau}=\sum \tau_i$ (see also
Appendix~\ref{app:fsle}). If the above procedure is applied 
by considering the same noise realization for both trajectories, 
$\lambda$ in (\ref{eq:lll}) coincides with $\lambda_{\sigma}$ 
(if $\lambda_{\sigma}>0$). Differently, by considering two
different realizations of the noise for the two trajectories, we have a 
new quantity 
\begin{equation} 
K_{\sigma}= \, {1
\over \overline {\tau} } \, \ln \left( { \Delta \over \delta_0}\right)\,,
 \label{7} 
\end{equation} 
which  naturally arises in the framework of
information theory \cite{AY81} and algorithmic complexity theory.  The
times $\tau_1, \tau_2, \dots $ are nothing but the intervals at which
it is necessary to repeat the transmission of $x(t)$, with a precision
$\delta_0$, and $K_{\sigma}/\ln 2$ is the number of bits {\it per}
unit time one has to specify in order to transmit the sequence. If the
fluctuations of the effective Lyapunov exponent $\gamma(t)$ are very
small (i.e. weak intermittency) one has:
\begin{equation} 
K_{\sigma} \, = \, \lambda +O(\sigma/\Delta)\,.
\end{equation} 
The interesting situation happens for strong intermittency when there
are alternations of positive and negative $\gamma$ during long time
intervals: this induces a dramatic change for the value of
$K_{\sigma}$.  This becomes particularly clear when we consider the
limiting case of positive $\gamma^{(1)}$ in an interval $T_1>>
1/\gamma^{(1)}$ followed by a negative $\gamma^{(2)}$ in an interval
$T_2>>1/|\gamma^{(2)}|$, and again a positive effective LE and so on.
During the intervals with positive effective LE the transmission has
to be repeated rather often with $\simeq T_1/(\gamma^{(1)} \ln 2)$
bits at each time, while during the ones with negative effective LE no
information has to be sent.  Nevertheless, at the end of the
contracting intervals one has $|\delta x| =O(\sigma)$, so that, at
variance with the noiseless case, it is impossible to use them to
compensate the expanding ones.  This implies that in the limit of very
large $T_i$ only the expanding intervals contribute to the evolution
of the error $\delta x(t)$ and $K_{\sigma}$  is given by an
average of the positive effective Lyapunov exponents:
\begin{equation} 
K_{\sigma} \simeq \langle   \gamma \, \theta(\gamma)\rangle \, . 
\label{12} 
\end{equation} 
Note that it may happen that $K_{\sigma} > 0$ with
$\lambda_{\sigma}<0$.  We stress again that (\ref{12})
holds only for strong intermittency, while for uniformly expanding
systems or rapid alternations of contracting and expanding behaviors
$K_{\sigma} \simeq \lambda_{\sigma}$.
 
Note that $K_{\sigma}$ is a sort of $\epsilon$-entropy (see
Sect.~\ref{sec:2.4}), indeed, the complexity we consider is defined
for $\delta_0$ not too small ($\delta_0 \gg \sigma$). If $\delta_0$
and $\Delta$ are small enough, but still much larger than $\sigma$,
$K_{\sigma}$ is essentially independent of their values.

The relation $K_{\sigma} \simeq \langle \gamma \,
\theta(\gamma)\rangle$ is the time analogous of the Pesin relation
(\ref{eq:1-12}) $h_{KS}\leq \sum_i \lambda_i \, \theta(\lambda_i)$.
The latter relation expresses the fact that negative Lyapunov exponents do not
decrease the value of $h_{KS}$, because the contraction along the
corresponding directions cannot be observed for any finite space
partition.  In the same way the contracting time intervals, if long
enough, do not decrease $K_{\sigma}$.  Another important remark is
that in the usual treatment of the experimental data, where noise is
usually present, one practically computes $K_{\sigma}$ and the result
can be completely different from $\lambda_{\sigma}$.

\begin{figure}[ht]
\centerline{\epsfig{figure=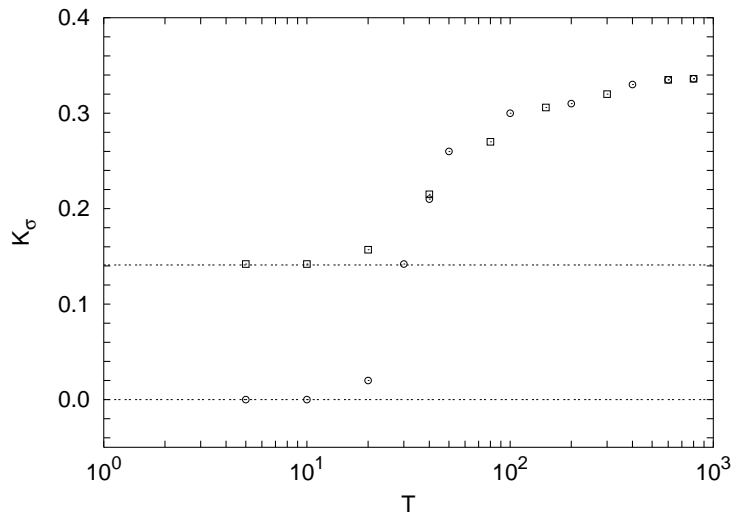,width=10cm, angle=0}}
\caption{ $K_{\sigma}$ versus $T$ with $\sigma = 10^{-7}$ 
for the map (\ref{periodic}). The parameters of the map are 
$a=2$ and $b=2/3$ (squares) or $b=1/4$ (circles). The dashed 
lines are the noiseless limit of $K_{\sigma}$.}
\label{fig:5.3-1}
\end{figure}

\begin{figure}[ht]
\centerline{\epsfig{figure=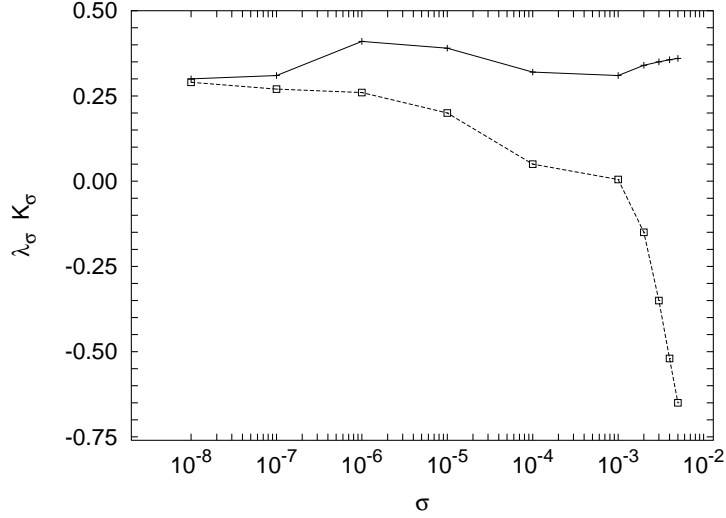,width=10cm, angle=0}}
\caption{$\lambda_{\sigma}$ (squares) and $K_{\sigma}$ 
(crosses) versus $\sigma$ for the map (\ref{beluzov}).}
\label{fig:5.3-2}
\end{figure}

Let us now briefly discuss some numerical results obtained with two
different systems (Fig.~\ref{fig:5.3-1} and Fig.~\ref{fig:5.3-2}).
The first example consists in a periodic alternation of two piecewise
linear maps of the interval $[0,1]$ into itself:
\begin{equation} 
f[x,t]=\cases { a \, x \qquad {\rm mod} \, 1 \qquad {\rm if} \
(2n-1)T\le t < 2nT ; \cr b \, x \qquad {\rm mod} \, 1 \qquad {\rm if}
\ 2nT \le t < (2n+1)T \cr}
\label{periodic} 
\end{equation} 
where $a>1$ and $b<1$.  Note that in the limit of small $T$,
$K_{\sigma} \to \max [\lambda_{\sigma}, 0]$, because it is a
non-negative quantity as shown in Fig.~\ref{fig:5.3-1}.
  
The second example (Fig.~\ref{fig:5.3-2}), strongly intermittent
without external forcing, is the Beluzov-Zhabotinsky map
\cite{HP87,MT83}, introduced for describing the famous chemical
reaction: 
\begin{equation} 
f(x)= \left\{ 
\begin{array}{l} 
\left[(1/8-x)^{1/3}+a \right] e^{-x}+b \qquad {\rm if} \ 0 \le x < 1/8  \\
\\
\left[ (x-1/8)^{1/3}+a \right] e^{-x}+b \qquad {\rm if} \ 1/8 \le x <3/10  \\ 
\\
c(10\, x \, e^{-10 x/3})^{19} +b \;\; \qquad \qquad {\rm if} \
3/10 \le x \end{array} 
\right.  
\label{beluzov} 
\end{equation} 
with $a=0.50607357, b=0.0232885279, c=0.121205692$.  The map exhibits
a chaotic alternation of expanding and contracting time intervals.  In
Figure~\ref{fig:5.3-2}, one sees that while $\lambda_{\sigma}$ passes
from negative to positive values at decreasing $\sigma$, $K_{\sigma}$
is not sensitive to this transition \cite{MT83}.  Considering the
system with a given realization of noise as the ``true'' evolution law,
one has that $\lambda_{\sigma}$ corresponds to $\lambda_{TT}$ while
$K_{\sigma}$ corresponds to $\lambda_{TM}$.

The previous results show that the same system can be regarded either
as regular (i.e. $\lambda_{\sigma}<0$), when the same noise
realization is considered for two nearby trajectories, or as chaotic
(i.e. $K_{\sigma}>0$), when two different noise realizations are
considered.
\subsection{Random dynamical systems}
\label{sec:5.3.3}
We discuss now dynamical systems where the randomness is not
simply given by an additive noise. This kind of systems has been the
subject of interest in the last few years in relation to the problems
involving disorder \cite{KF93}, such as the characterization of
the so-called {\em on-off intermittency} \cite{PST93} and to
model transport problems in turbulent flows
\cite{YOC93,YOC93b,GV94}.  In these systems, in general, the random
part represents an ensemble of hidden variables believed to be
implicated in the dynamics. Random maps exhibit very interesting
features ranging from stable or quasi-stable behaviors, to chaotic
behaviors and intermittency.  In particular {\em on-off
intermittency} is an aperiodic switching between static, or laminar,
behavior and chaotic bursts of oscillation. It can be generated by
systems having an unstable invariant manifold, within which it is
possible to find a suitable attractor (i.e. a fixed point).  For
further details we refer to \cite{PST93}.

A random map can be defined in the following way.  Denoting with ${\bf
x}(t)$ the state of the system at discrete time $t$, the evolution law
is given by \begin{equation} {\bf x}(t+1)= {\bf f}({\bf x}(t),J(t)),
\label{rap} \end{equation} where $J(t)$ is a random variable.
 
As for the case of additive noise examined in the previous Section,
the simplest approach is the introduction of the LE $\lambda_J$
computed considering the separation of two nearby trajectories
evolving with the same realization of the random process
$J(t)=i_1,i_2,...,i_t$.  The Lyapunov exponent $\lambda_J$ generalizes
$\lambda_{\sigma}$ of Sect.~\ref{sec:5.3.1} and can be computed from
the tangent vector evolution:
\begin{equation} 
\lambda_J=\lim_{N \to \infty} {1\over N} \ln |{\bf z}(N)|
\end{equation}
where
\begin{equation} 
z_{m}(t+1)=\sum_{n} {\partial f_m ({\bf x}(t),i_t) \over \partial x_n}
z_{n}(t) \, .
\end{equation} 

On the other hand, also for these systems, as in the case of additive
noise, it is possible to introduce a measure of complexity, $K$, which
better accounts for their chaotic properties \cite{PSV95,LPSV95}
\begin{equation} 
K \simeq h_{Sh}+\lambda_J \theta(\lambda_J)\, , 
\label{compl} 
\end{equation}
where $h_{Sh}$ is the Shannon entropy of the random sequence $J(t)$.
The meaning of $K$ is rather clear: $K/\ln 2$ is the mean number of
bits, for each iteration, necessary to specify the sequence
$x_1,...,x_t$ with a certain tolerance $\Delta$. Note that there are
two different contributions to the complexity: (a) one has to specify
the sequence $J(1),J(2),...,J(t)$ which implies $h_{Sh}/\ln 2$ bits
per iteration; (b) if $\lambda_J$ is positive, one has to specify the
initial condition ${x}(0)$ with a precision $\Delta \exp^{-\lambda_J
T}$, where $T$ is the time length of the evolution. This requires
$\lambda_J/\ln 2$ bits per iteration; if $\lambda_J$ is negative the
initial condition can be specified using a number of bits independent
of $T$.

\subsubsection{A toy model: one dimensional random maps}

\begin{figure}[htb]
\centerline{\epsfig{figure=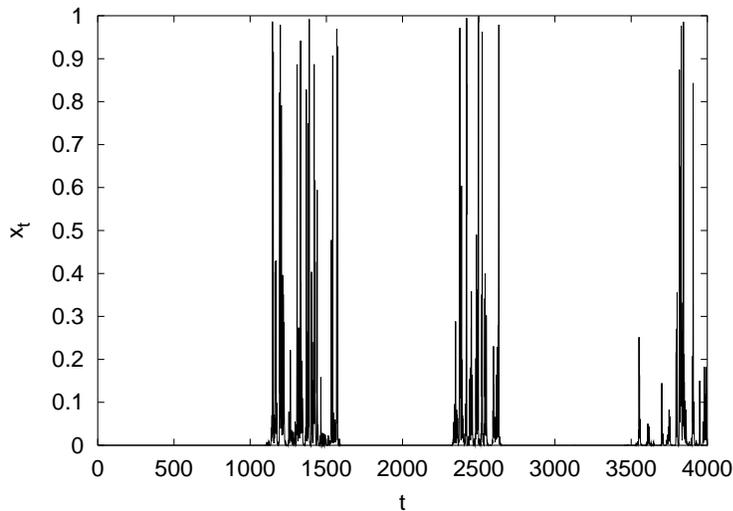,width=10cm, angle=0}}
\caption{$x(t)$ versus $t$ for the random map 
(\ref{rap2} -- \ref{rap3}), with $p=0.35$. }
\label{fig:5.3-3}
\end{figure}

Let us discuss a random map which, in spite of its simplicity,
captures some basic features of this kind of systems
\cite{PST93,HPH94}:
\begin{equation} 
x(t+1)=a_t x(t) (1-x(t)) 
\label{rap2} \,,
\end{equation}
where $a_t$ is a random dichotomous variable given by 
\begin{equation}  
a_t= \left\{ 
\begin{array}{l}
4 \phantom{aaaaa} \mbox{with probability $p$}\\
1/2 \phantom{aaa} \mbox{with probability  $1-p$} \,.
\end{array} \right. 
\label{rap3} 
\end{equation}
For $x(t)$ close to zero, we can neglect the non linear term to obtain
\begin{equation} 
x(t)=  \prod_{j=0}^{t-1} a_j x(0) \,;  
\end{equation}
from the law of large numbers one has that the typical behavior is
\begin{equation} x(t) \sim x(0) e^{<\ln a> t}.  \end{equation} Since
$\langle \ln a \rangle =p \ln 4+(1-p) \ln 1/2=(3p-1) \ln 2$ one has
that, for $p < p_{c} =1/3$, $x(t) \rightarrow 0$ for $t \rightarrow
\infty$. On the contrary for $p > p_c$ after a certain time $x(t)$
escapes from the fixed point zero and the non-linear term becomes
relevant. Figure~\ref{fig:5.3-3} shows a typical {\em on-off
intermittency} behavior for $p$ slightly larger than $p_c$.  
Note that, in spite of this irregular behavior, numerical
computations show that the LE $\lambda_J$ is negative for $p < \tilde
p_c \simeq 0.5$: this is essentially due to the non-linear terms.
 
By introducing a finite threshold $\epsilon$, in order to discriminate
laminar and intermittent phases, we can define a complexity
$K(\epsilon)$.  We denote with $l_L$ and $l_J$ the average life times
respectively of the laminar and of the intermittent phases for $p$
close to $p_c$ ($l_J << l_L$).  The mean number of bits, per
iteration, one has to specify in order to transmit the sequence is
\cite{LPPV96}
\begin{equation} 
{K(\epsilon)\over {\ln 2}} \simeq
\frac{l_J h_{Sh}}{(l_J+l_L) \ln 2} \simeq \frac{l_J}{l_L}
\frac{h_{Sh}}{\ln2}.  
\label{compl2} 
\end{equation}
 
To obtain (\ref{compl2}) first notice that on an interval $T$ one has
approximatively $T/{(l_J+l_L)}$ intermittent bursts and the same number of
laminar phases.  Then notice that, during a laminar phase, there is
not an exponential growth of the distance between two trajectories
initially close and computed with the same sequence of $a_t$. Since
during a laminar phase one has to send a number of bits which does not
depend on its duration, one can send all the necessary information
simply by giving the sequence of $a_t$ during the intermittent bursts.
Eq.(\ref{compl2}) has an intuitive interpretation: in systems with a
sort of ``catastrophic'' events, the most important feature is the mean
time between two subsequent events.

\subsubsection{Sandpile models as random maps} 

Another example of a system which can be treated in the framework of
random maps is represented by the so-called sandpile models
\cite{BTW88}. These models are a paradigmatic example of the
 Self-Organized Criticality (SOC) \cite{BTW87}.  This term
refers to the tendency of some large dynamical systems to evolve {\em
spontaneously} toward a critical state characterized by spatial and
temporal self-similarity.  The original Sandpile Models are
probabilistic cellular automata inspired to the dynamics of avalanches
in a pile of sand.  Dropping sand slowly, grain by grain on a limited
base, one reaches a situation where the pile is critical, i.e. it has
a critical slope. This means that a further addition of sand will
produce sliding of sand (avalanches) that can be small or cover the
entire size of the system. In this case the critical state is
characterized by scale-invariant distributions for the size and the
lifetime and it is reached without tuning of any critical parameter.
 
We will refer in particular to the Zhang model \cite{Z87}, a
continuous version of the original sandpile model \cite{BTW88},
defined on a $d$-dimensional lattice.  The variable on each site
$x_{i}$ (interpretable as energy, sand, heat, mechanical stress etc.)
can vary continuously in the range $\left[ 0,1 \right]$ with the
threshold fixed to $x_{c}=1$.  The dynamics is the following:
 
\begin{description}
\item{(a)} one chooses at random a site and adds to it an energy
$\delta e$,
\item{(b)} if at a certain time $t$ the energy in a site, say $i$, exceeds the
threshold $x_{c}$ a relaxation process is triggered defined as:
\begin{equation}
\left\{ 
\begin{array}{l}
x_{i+nn} \rightarrow x_{i+nn} + x_{i} /2d \\
x_{i} \rightarrow 0\,,
\end{array}
\right.
\end{equation}
where ${\it nn}$ indicates the $2d$ nearest neighbors of the site $i$;
\item{(c)} one repeats point (b) until all the sites are relaxed;
\item{(d)} one goes back to point (a).
\end{description}

Let us now discuss the problem of predictability in sandpile models on
the basis of the rigorous results \cite{CL96}, which clarify the role
of the LE for this class of systems.
 
In Ref.~\cite{CL96} it has been proved that the LE $\lambda_J$ is
negative.  In fact the dynamics of a little difference between two
configurations follows the same rules (a)-(d), i.e., the ``error'' is
redistributed to the nearest neighbors site, so that one has
\begin{equation} 
\lambda _J \le - {{\rm const} \over R^2}
\end{equation} 
where $R$ is the diameter of the system.
 
As for other examples already discussed, the existence of a negative
LE does not mean a perfect predictability.  This can be understood by
looking at the growth of the distance, $\delta (t)$, between two
initially close trajectories computed with two different realizations
of randomness, i.e., by adding sand in different sites.  Let us
consider the case of the ``minimal error'':  in the reference
realization one adds sand on a site $i$ chosen at random. In the
perturbed realization, instead, one adds a sand grain at one of the
nearest sites of $i$. In such a case $\delta (t)$ increases up to a
maximal distance in few avalanches \cite{LPPV96}. Practically, one has
the same kind of phenomenon, already discussed, of the Langevin
equation with two noise realizations.

Let us now estimate the complexity $K$ of this system.  An upper bound
can be given by using (\ref{compl}) $K= h_{Sh} +\lambda_{J}
\theta(\lambda_{J})$, where $h_{Sh}$ is the entropy of the random
sequence of addition of energy.  In sandpile models, since each site
has the same probability to be selected, one has $h_{Sh}=\ln V$, where
$V$ is the number of sites of the system.  Since the Lyapunov exponent
is negative, the complexity is just determined by $h_{Sh}$.

\section{Irregular behavior in discrete dynamical systems}
\setcounter{equation}{0}
\label{sec:6}
For the sake of completeness we include in this review a discussion on
the characterization of irregular behaviors in systems whose states
are discrete.  Such systems include Cellular Automata (CA), which have
been intensively studied both for their intrinsic interest \cite{W86}
and for applications as, e.g., to simulate hydrodynamic equations
\cite{FHP86} or to study various forms of chemical turbulence
\cite{OK85,OY87,BDKL96}. Other interesting systems with discrete
states are the neural networks used for modeling some brain functions
\cite{Amit92}. It is also relevant to note that in every simulation
with a computer, because of the finite number of digits it can use,
one deals with a system with discrete states (see 
Sect.~\ref{sec:5.1}). In addition, the general problem of dynamics 
of systems with discrete states is important in the debated 
issue of quantum chaos. Indeed quantum mechanics can be regarded 
as a discretized version of the classical one, acting on a suitable 
lattice in phase space, where the number of the possible states is 
proportional to the inverse of the Planck constant 
\cite{ford91,cachi95,chiri96}. 

\begin{figure}[htb]
\centerline{\epsfig{figure=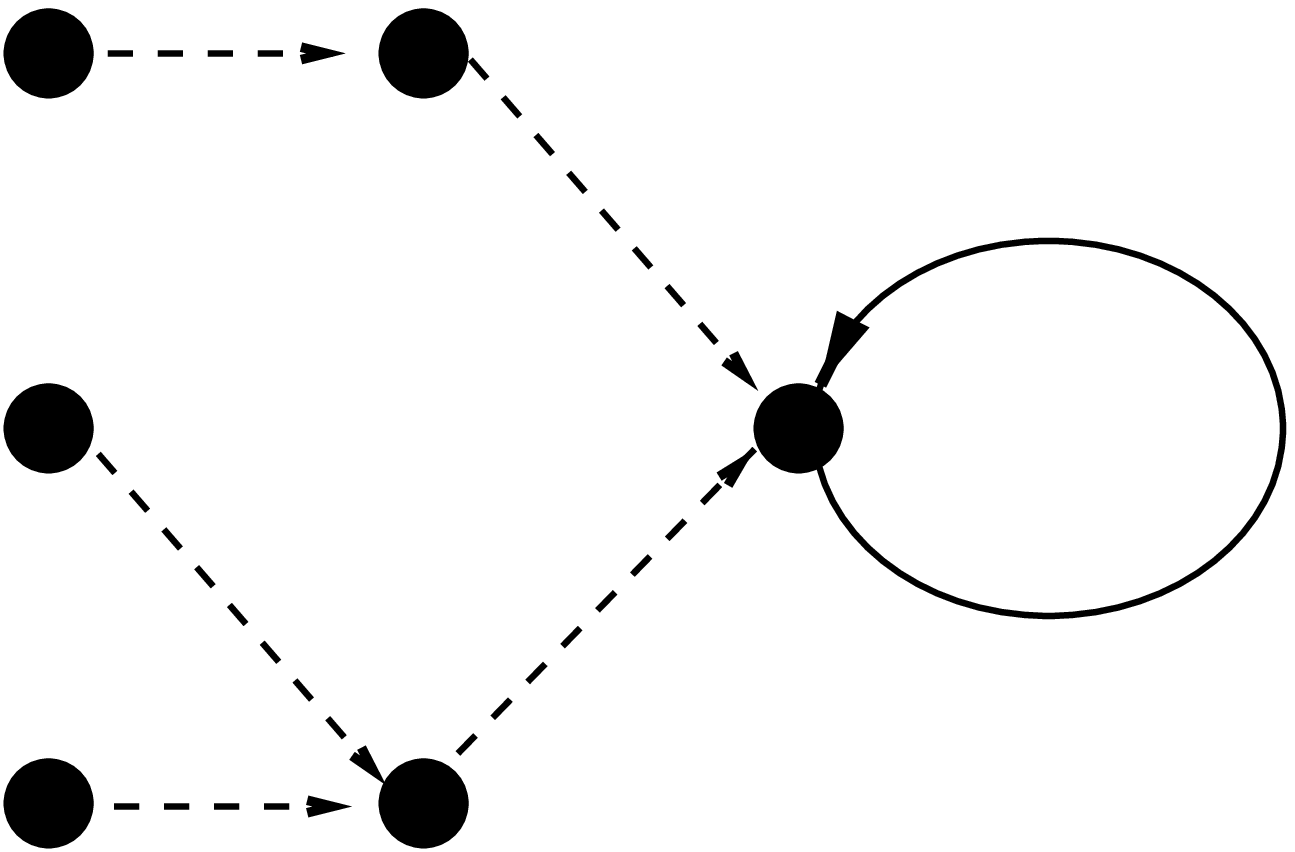, width=6cm, angle=0}
\hfill \epsfig{figure=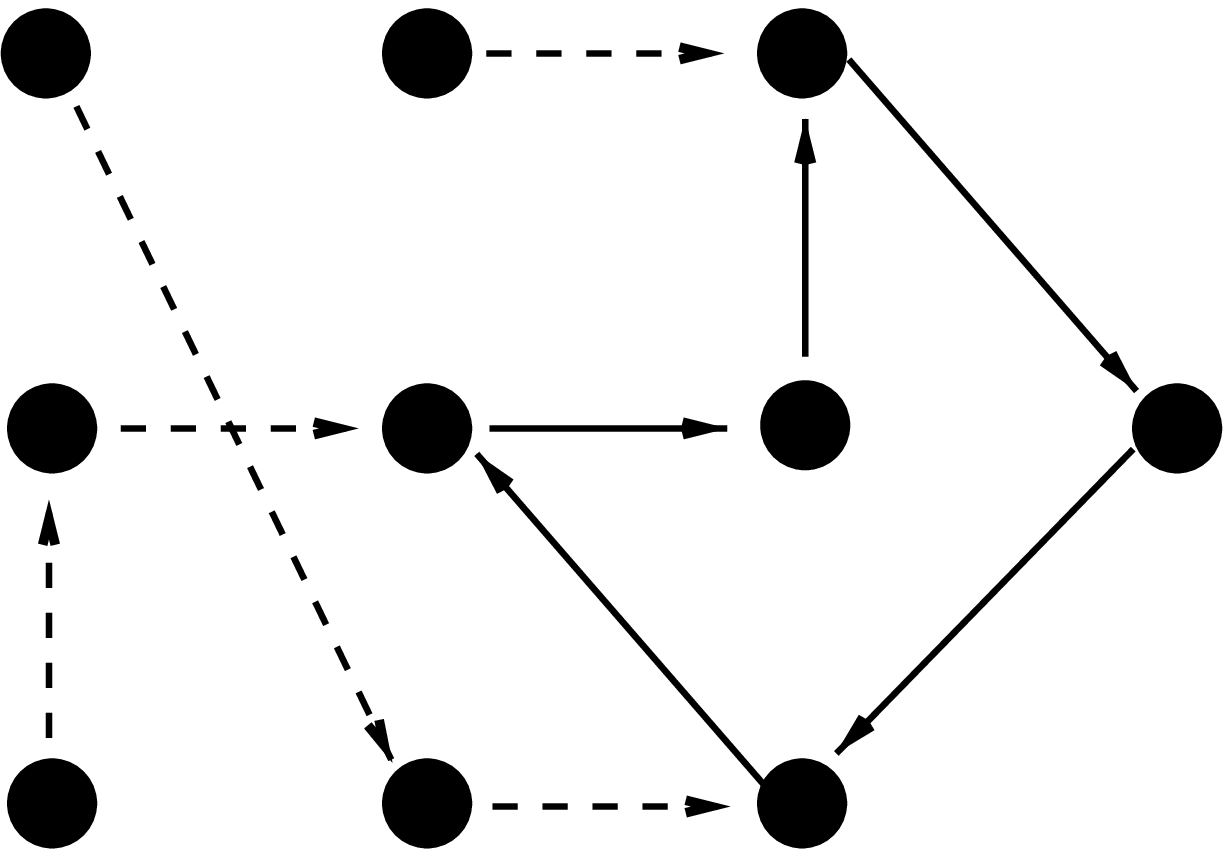, width=6cm, angle=0}}
\vspace{0.4truecm}
\caption{\label{fig:6.2-1} Schematic representation of the evolution
of a deterministic rule with a finite number of states: (a) with a
fixed point , (b) with a periodic cycle.}
\end{figure}

If a system consists of $N$ elements and each element can assume an
integer number $k$ of distinct values, ${\cal N}=k ^N$ is the
number of states. When these states evolve
with a deterministic rule, the dynamics can be depicted in terms of
oriented graphs: a set of points, representing the states, are
connected by arrows, indicating the time evolution. Of course, each
point has one, and only one, outgoing arrow; but different arrows can
end at the same point. For any finite system each initial condition
evolves to a definite attractor, which can be either a fixed point (as
in Fig.~\ref{fig:6.2-1}a), or a periodic orbit
(Fig.~\ref{fig:6.2-1}b).

In systems of this kind, obviously, it is not possible to use the
previously introduced indicators of chaos, e.g. the Lyapunov exponents
or the Kolmogorov-Sinai entropy, whose definitions rely on the
continuous character of the system states. Moreover, the asymptotic
periodic behavior seems to force the conclusion that discrete states
systems are trivial, from an entropic or algorithmic complexity point
of view.

The above conclusions, although mathematically correct, are rather
unsatisfactory from the physical point of view, indeed from this side
the following questions deserve some interest:
\begin{enumerate}
\item What is the ``typical'' period, ${\cal T}$, in a system
with $N$ elements, each assuming $k$ distinct values ?
\item When ${\cal T}$ is very large, how can we characterize the
(possible) irregular behavior of the trajectories, on times that are
large enough but still much smaller than ${\cal T}$ ?
\item What does it happen in the transition from discrete
to continuous states, i.e. in the limit $k \to \infty$ ?
\end{enumerate}

In the next subsections we will deal to the above questions.

\subsection{Dependence of the period on the number of the states}
\label{sec:6.1}
For deterministic discrete state systems the dependence of the period
of the attractor on the number of the states, may be addressed with a
statistical approach in terms of random maps \cite{CH86}. We recall
that this problem is important for computer simulations of chaotic
systems (see Sect.~\ref{sec:5.1}).  If ${\cal N} = k^N \gg 1$ is the
number of states of the system, the basic result for the average
period, ${\cal T}$, is
\begin{equation}
{\cal T}({\cal N}) \sim \sqrt{{\cal N}}.
\label{period}  
\end{equation}
In the following we give a simple argument, by Coste and H\'enon
\cite{CH86}.
 
For simplicity of notation, we consider the case with $k=2$, so that
the state of the system is a string of $N$ bits.  A deterministic
evolution of such a system is given by a map which is one among the
possible functions connecting the $2^{N}$
states. Let us now assume that all the possible functions can
be extracted with the same probability. Denoting with $I(t)$ the state
of the system, for a certain map we have a periodic attractor of
period $m$ if $I(p+m)=I(p)$ and $I(p+j) \neq I(p)$, for $j<m$.  The
probability, $\omega (m)$, of this periodic orbit is obtained by
specifying that the $(p+m-1)^{th}$ first successive images of the map
are distinct from all the previous ones and the $(p+m)^{th}$ iterates
coincides with the $p^{th}$ one. Since one has $I(p+1) \neq I(p+m)$, with
probability $\left( 1 - 1 / {\cal N} \right)$; $I(p+2) \neq I(p+m)$,
with probability $\left( 1 - 2 / {\cal N} \right)$; $\dots \dots ;
I(p+m-1) \neq I(p+m)$, with probability $\left( 1 - (m-1) / {\cal N}
\right)$; and, finally, $I(p+m)=I(p)$ with probability $\left( 1 /
{\cal N} \right)$,  one obtains
\begin{equation}
\omega (m) = 
   \left( 1-\frac{1}{{\cal N}}\right)
  \left( 1-\frac{2}{{\cal N}}\right) \cdots 
  \left( 1-\frac{(m-1)}{{\cal N}}\right) \frac{1}{{\cal N}} .
\label{probperiod}  
\end{equation}
The average number, $M(m)$, of cycles of period $m$ is 
\begin{equation}
M (m) = \frac{{\cal N}}{m} \, \omega (m)\quad  \stackrel { ({\cal N} \gg 1)}
{\approx} \quad  \frac{e^{- m^2/2 {\cal N} }}{m} \,,
\label{average}  
\end{equation}
from which one obtains 
${\cal T} \sim \sqrt{{\cal N}}$
for the average period. 

It is here appropriate to comment on the relevance of
Eq.~(\ref{period}) for computer-generated orbits of chaotic dynamical
systems.  Because of the finite number, $n$, of digits used in the
floating point representation, when one iterates a dynamical system,
one basically deals with a discrete system with a finite number ${\cal N}$ of
 states. If $d_2$ indicates the correlation dimension of the
system \cite{GP83,GP83a}, one can reasonably assume that ${\cal N} \sim 10
^{n d_2}$, so that, from Eq.~(\ref{period}) one has:
\begin{equation}
{\cal T} \sim 10 ^{n d_2 /2} .  
\end{equation}
This estimation gives an upper limit for the typical number of
meaningful iterations of a map on a computer.  Note that this number,
apart from the cases of one or two dimensional maps with few digits,
is very large for almost all  practical purposes.

\subsection{The transition from discrete to continuous states} 
\label{sec:6.2}
Following the basic ideas of Ford \cite{F83,F86}, as discussed in
Sect.~\ref{sec:1.3}, and the results of Sect.~\ref{sec:5} -- on the
predictability in systems whose evolution law is not completely known
-- we describe now a way to introduce a practical definition of chaos
for systems with discrete states. In addition, we deal with the
problem of the transition from discrete to continuous states.

Given a system with ${\cal N}$ possible states, denoting with ${\bf I}
(t)$ its state at time $t$ we can write its evolution law as:
\begin{equation}
{\bf I} (t+1) = {\bf F} [{\bf I} (t)] \,. 
\label{evolution}  
\end{equation}
A single state ${\bf I}$ is a sequence of (at most) $\ln _2 {\cal N}$
bits, and its time evolution for $M$ steps can be surely translated in
a binary sequence $\Sigma$ of length $ \ell _{\Sigma} (M,{\cal N})
\leq M \ln _2 {\cal N}$.

Relying one the definition of algorithmic complexity
(Sect.~\ref{sec:1.2.3}) we can make the following classification: we
call regular (compressible) those sequences that can be encoded by a
computer program whose length $ \ell _{\Sigma} (M,{\cal N})$ increases
less than linearly in $M$, when $M$ ranges over a physically
significant interval, at fixed values of ${\cal N}$. Otherwise
the system will be called chaotic or incompressible. Let us call $\ell
_F$ the binary length of the algorithm for one step: $\ell _F \leq 2
{\cal N} \ln _2 {\cal N} $. The sequence $\Sigma$ can be expressed by
the record composed by the initial state ${\bf I} (0)$ (specified by
$\ln _2 {\cal N}$ bits), the number of steps $M$ (specified by $\ln _2
M$ bits) and the rule ${\bf F}$ for one step (specified by $\ell _F$
bits). Therefore
\begin{equation}
 \ell _{\Sigma} (M,{\cal N})\leq  (2 {\cal N}+1) \ln _2 {\cal N}
 +  \ln _2  M + O(1) .
\label{length}  
\end{equation}
Let us note that from the above equation one has that -- when $M$
grows indefinitely and ${\cal N}$ is constant -- $\ell _{\Sigma}$ is
logarithmically bounded and hence the sequence appears to be
compressible. This is somewhat trivial since, because of the discrete
nature of the states, the motion at $M > {\cal N}$ (in practice $M >
{\cal T} \sim \sqrt{{\cal N}}$) is periodic. Therefore it is
interesting to consider only $1 \ll M < {\cal T} \ll {\cal
N}$. Although the evolution law (\ref{evolution}) can be carried out,
in principle, in exact arithmetic, in practice in real computations
one has unavoidable errors due to truncations and approximations.  Let
us now regard the evolution law (\ref{evolution}) as a computer
program with input ${\bf I} (0)$ and a set ${\bf C}$ of parameters,
with $C$ components, needed to define the algorithm ${\bf F}$. If
these parameters are all known within precision $O(2^{-q})$, the
binary length of the coding of ${\bf C}$ is $O(qC)$.

Consider the following problem: given two identical initial conditions
${\bf I}^{(1)} (0) ={\bf I}^{(2)} (0)={\bf I} (0) $, and two different
realizations ${\bf C}^{(1)}$ and ${\bf C}^{(2)}$ of the set of
coefficients ${\bf C}$ (with difference $O(2^{-q})$), what is the
dependence on $\epsilon = 2^{-q}$ of the first-error time
$\widetilde{M}$ (i.e. the first time for which ${\bf I}^{(1)} (t) \neq
{\bf I}^{(2)} (t) $)?  Of course, the answer depends on the
realizations of the components of ${\bf C}$ and on the initial
conditions ${\bf I} (0)$. Let us consider ${\bf C}^{(2)}$ as an
$\epsilon$-perturbation of ${\bf C}^{(1)}$, i.e. we pose, for each
component of the parameter vector:
\begin{equation}
C_i ^{(2)} = C_i ^{(1)} +\epsilon _i \, ,
\end{equation}
where the random variables $\epsilon _i$ are uniformly distributed in
 $[-2^{(-q-1)}, 2^{(-q-1)}]$. Let us note that the coding
length $O(C q) + \ln _2 {\cal N}$ is enough to define the sequence
$\Sigma$ up to the first error time $\widetilde{M}$. Performing an
average on the $ \epsilon _i$ and on the initial conditions ${\bf I}
(0)$, one can compute an average first-error time $\langle
\widetilde{M} (\epsilon) \rangle$, and a typical first-error time $
\widetilde{M}_{typ} (\epsilon) = \exp \langle \ln \widetilde{M}
(\epsilon)\rangle $. For $\langle \widetilde{M} \rangle$ and $
\widetilde{M}_{typ} $ the dependence on $\epsilon$ can be of the
following type:
\begin{enumerate}
\item[(a)]  $\langle \widetilde{M} \rangle \sim \epsilon ^{-\beta} 
\sim 2^{q \beta }$ ,
\item[(b)]   $\langle \widetilde{M} \rangle \sim \ln _2 (1/\epsilon)
\sim q$.
\end{enumerate}
In the case $(a)$ we say that the system is compressible, while if
$(b)$ holds one has a chaotic (incompressible) case. The above
classification is rather obvious: in case $(a)$ a trajectory of length
$\widetilde{M}$ can be coded by a program of length $O(\ln _2 {\cal
N}) + O(\ln \widetilde{M})$, while in case $(b)$ one has a length
$O(\ln _2 {\cal N}) + O( \widetilde{M})$. For a detailed discussion
see Ref.~\cite{CFMV94}.

Let us now discuss in an explicit example the previous approach and
the problems in the transition to the continuous state case.  We
consider a discretized standard map, as obtained by considering
lattice points in the torus $[0,2\pi]^2$ of the form $(x,y)$ with
$x=2\pi Q/L$ and $y=2\pi P/L$, where $P$ and $Q$ are integers between
$1$ and $L$. The action of the map is
\begin{equation}
 \left\{
\begin{array}{ll}
Q (t+1)= & \bigl[\, Q (t) +\alpha \,\frac{L}{2\pi} \;\sin (P (t)
\frac{2\pi}{L}) \, \bigr] \quad {\rm mod ~L}

\\

\\

  P(t+1)= & 
(P (t) + Q(t+1) ) \quad {\rm mod ~L} \\  
\end{array}
\right.  
\label{standard}
\end{equation}
where $\alpha$ is the control parameter and $[\cdot]$ means integer
part.  From the results of Sect.~\ref{sec:6.1} one has that the
typical period for the map (\ref{standard}) is ${\cal T}_L \sim L$; so
if $L$ is large the periodic motion will be seen only for sufficiently
large times.  In the system (\ref{standard}) one has just one
parameter, i.e. the ``kick strength'' $\alpha$.  Numerical evidence
supports the following picture: at fixed $L$, the first-error time is
roughly constant for large values of the error $\epsilon$, while it
goes as $\epsilon^{-1}$ for small errors $\epsilon$.  The transition
between the two regimes occurs at a critical value $\epsilon_{{\rm
c}}(L)$ which scales as $\epsilon_{{\rm c}}\sim 1/L $. In formulae:
\begin{equation}
\label{eq:nr4} 
\langle \widetilde{M}(\epsilon)\rangle 
\sim \left\{ 
\begin{array}{ll}
O(1) & \mbox{for $\epsilon > \epsilon_{\rm c}(L)$} \\  
1/\epsilon & \mbox{for $\epsilon < \epsilon_{\rm c}(L)$} . 
\end{array}
\right.  
\end{equation}
It is rather easy to give analytical estimates supporting the
numerical evidence \cite{CFMV94}
\begin{equation}
\label{eq:err4}
\langle \widetilde{M} (\epsilon ) \rangle  =\left\{ 
\begin{array}{ll}
\left( {\pi l \over \epsilon} \right) & \mbox{for $\epsilon < l  $} \\  
\Big( {\epsilon \over \pi l} (1-\cos \theta )
+1 - {2\theta \over \pi} + {l \over \pi \epsilon }
\ln \tan ({\frac \theta 2}) \Big)^{-1} & \mbox{for $\epsilon > l $.} \\
\end{array}
\right. 
\end{equation}
where the angle $\theta $ is defined via $\sin \theta =l/\epsilon $,
and $l=2\pi / L$. Numerical simulations show that the behavior
proposed in Eq.~(\ref{eq:nr4}) holds.  To have a comparison with 
the usual standard map, we have computed the
average time $\langle M_\Delta (\epsilon) \rangle$ required for two
trajectories to reach lattice points farther than a fixed distance
$\Delta $ in the discrete phase space of Eq.~(\ref{standard}).  We
found:
\begin{equation}
\label{eq:nr71}
\langle M_\Delta (\epsilon) \rangle \sim \left\{ 
\begin{array}{ll}
(1 / \lambda ) \ln (\Delta /\epsilon ) & \mbox{for $\epsilon >
\epsilon_{\rm c}(L)$} \\ 1/ \epsilon & \mbox{for $\epsilon <
\epsilon_{\rm c}(L)$} .
\end{array}
\right. 
\end{equation}
We remark that when $\epsilon <\epsilon _{{\rm c}}(L)$, $M_\Delta
(\epsilon )$ is weakly dependent, i.e. logarithmically, on $\Delta $.
This is just another verification of the similarity of the effect of a
small disturbance on the equations of motion and of a small error in
the initial conditions for a dynamical evolution (see
Sect.~\ref{sec:5}).

These results unveil the nature of the dynamics of this discrete
system: its trajectories are incompressible and therefore chaotic only
for large values of $\epsilon$, the cutoff value decreasing as
$1/L$. This helps us also to understand the extent to which the
dynamics of the discrete standard map Eq.~(\ref{standard}) is
equivalent to its continuum counterpart.  When $\alpha$ is large, and
$\epsilon > \epsilon_{\rm c}$, the two systems possess chaotic
trajectories.   Simple
calculations show that, to the leading order, $\widetilde{M} \sim
\log_2(L)$.  After this time,
 the discrete system appears ``regular'', i.e. compressible
and predictable. Therefore, continuous and discrete systems are
similar (as far as chaos is concerned) only over logarithmically short
times.  It is important to stress that the system appears ``regular''
on time scales much smaller than the typical period ${\cal T}_L \sim
L^{d_2/2}$ (being $d_2$ the correlation dimension of the attractor
\cite{GP83,GP83a}).

Recently Mantica \cite{mantica2000} studied the algorithmic complexity
in classical polygonal billiards with $L$ sides. The system, for any
finite value of $L$, is regular; on the other hand, as $L \to \infty$,
it tends to a curved billiard, which can be chaotic. This system is
very similar to the discrete dynamical system (\ref{standard}) and may
be used to study the transition from quantum to classical mechanics
and the principle of correspondence.  The average complexity of
symbolic trajectories in the polygonal billiards has the same scaling
behavior (as function of $L$ and of the precision $\epsilon$) of that
one of the system (\ref{standard}), i.e. a compressible (regular)
regime for $\epsilon < \epsilon_c \sim 1/L$ and an incompressible
(chaotic) one for $\epsilon < \epsilon_c $.

It is interesting to note that a similar feature is characteristic of
quantum dynamics of systems whose classical behavior is chaotic.
Roughly speaking, a quantum system behaves as a system with discrete
states whose number is proportional to $\hbar^{-1}$. A semi-classical
wave function follows a dynamics which is approximately classical up
to a time $t_c \sim (1/\lambda) \ln (I/\hbar)$, where $\lambda$ is the
Lyapunov exponent of the classical motion, and $I$ is a typical action
of the motion. Over this time, the quantum system has the same
complexity of its classical counterpart, while for larger times its
quantal (quasi-periodic) nature appears \cite{ford91,cachi95,chiri96}.

\subsection{Entropies and Lyapunov exponents in Cellular Automata} 
\label{sec:6.3}
Cellular automata (CA) consist of discrete valued state variables,
$\sigma_i(t)$, defined on a discrete lattice, and updated
synchronously at discrete time according to a local rule.  They
can be defined in any dimensions and for any finite number of possible
values for $\sigma_i(t)$. For the sake of simplicity we consider
Boolean CA, i.e. $\sigma_i(t)=\{0,1\}$, in a 1-dimensional lattice.
An evolution rule can be written as:
\begin{equation}
\sigma_i(t+1)=F[\sigma_{i-r}(t), \dots,\sigma_i(t),\dots,
\sigma_{i+r}(t)]\,,\qquad
i=1,\dots,N\,,
\label{eq:6.3-1}
\end{equation}
where $r$ defines the range of the coupling, i.e. the variable in a  site
depends on the variables in the 
$2r$ neighbor sites.  If $F$ in (\ref{eq:6.3-1}) only
depends on the sum of the state variables, one speaks of ``totalistic''
CA. Another usual requirement is to have symmetric rules.  For further
details we refer to Ref.~\cite{W86}, where the standard scheme for the
classification of the possible rules and a number of examples of
CA-behavior are discussed.

In the following we refer to $1d$ Boolean cellular automata with local
symmetric rules -- as those ones systematically studied by Wolfram
\cite{W86}.
\subsubsection{Classification of Cellular Automata according to 
the transient times}
\label{sec:6.3.1}
For finite lattices with $N$ sites the number of possible states of CA
is finite and equal to ${\cal N}=2^N$. As already discussed, this
means that, strictly speaking, from an entropic (or algorithmic) point
of view CA are trivial.  Therefore the problem of the
characterization of irregular behaviors in CA has, in principle, some
meaning only in the limit $N \to \infty$. In more physical terms, for
finite $N$ one expects the characterization in terms of entropy to be
possible for times shorter than the typical period ${\cal T}(N)$ or
the typical transient time $\widetilde{{\cal T}} (N)$, provided ${\cal
T}(N)$ and $\widetilde{{\cal T}}(N)$ are long enough.

Actually, cellular automata behaviors can be classified according to
the dependence of ${\cal T}(N)$ and $\widetilde{{\cal T}}(N)$ on
$N$. One has three possible classes of behavior.

\begin{figure}[hbt]
\centerline{\epsfig{figure=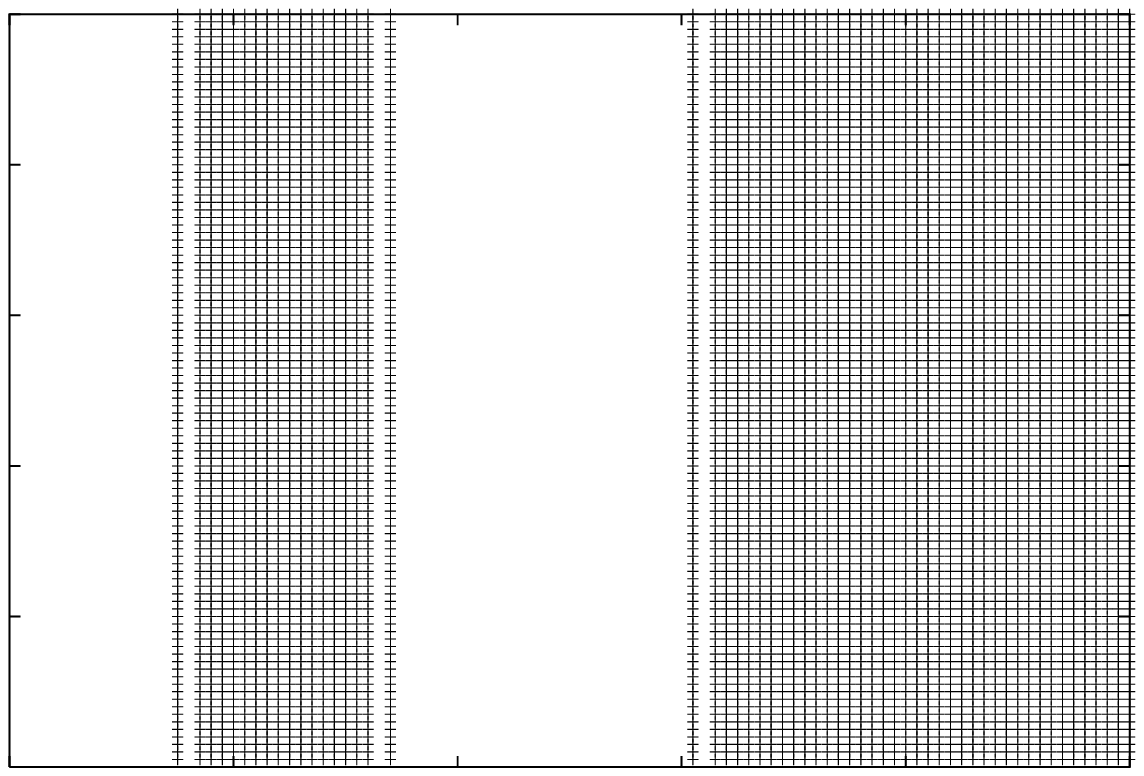, width=4cm, height=6cm, angle=0} 
\hfill \epsfig{figure=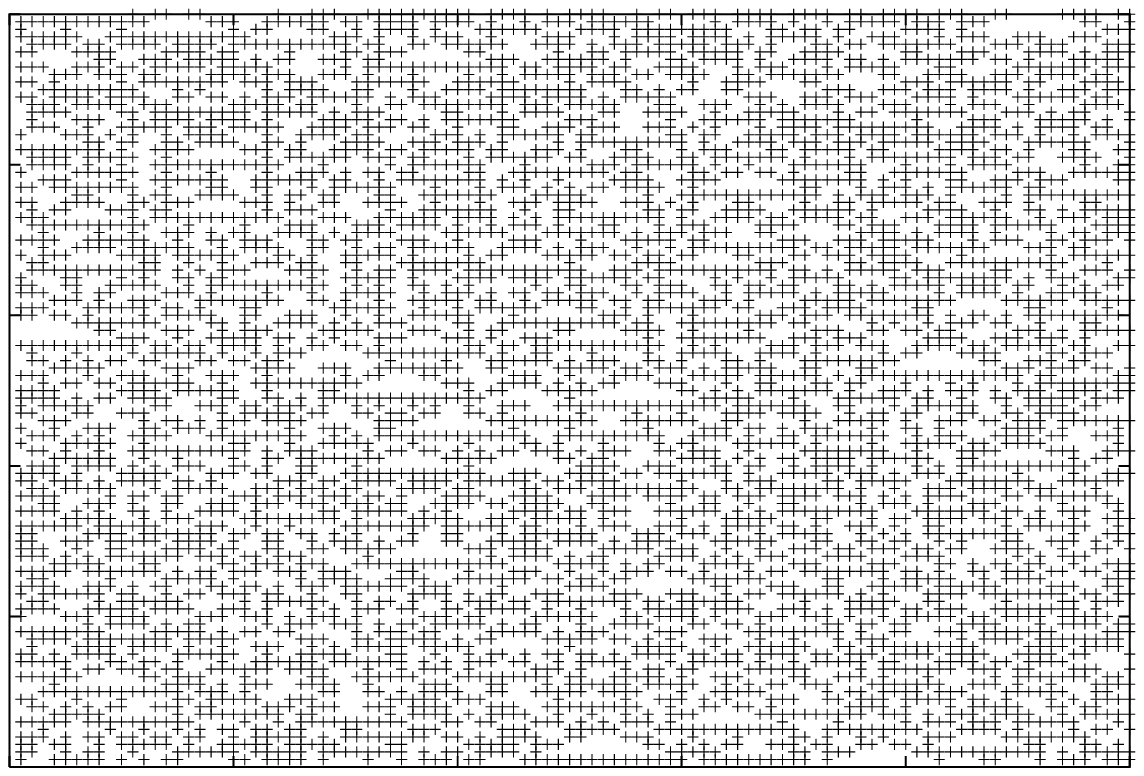 ,width=4cm, height=6cm,angle=0}
\hfill \epsfig{figure=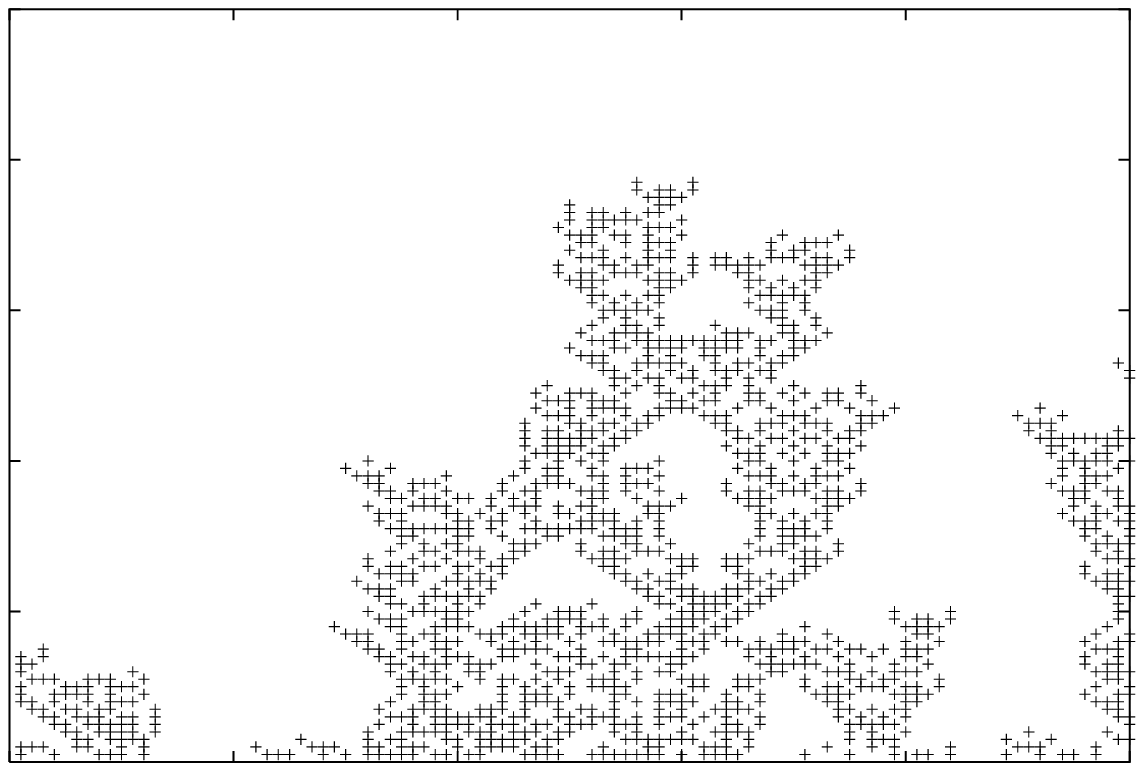,width=4cm, height=6cm,angle=0}}
\vspace{0.4truecm}
\caption{\label{fig:rules} Typical behavior of (a) regular $[Rule\;
52]$, (b) chaotic $[Rule\; 22]$, (c) complex $[Rule\; 20]$. We used
totalistic $r=2$ cellular automata.  Time flows from below to above}
\end{figure}

{\it Regular} cellular automata (class 1 and 2 in Wolfram's
classification \cite{W86}) evolve either on homogeneous states both in
time and space (the analogous of fixed point in dynamical systems)
either to a set of simple stable periodic structures (analogous to
limit cycles) which, in general, depend on the initial configuration
(Fig.~\ref{fig:rules}a). In these CA, ${\cal T}(N)$ and
$\widetilde{{\cal T}}(N)$ can range from being almost independent of
$N$ to be, at maximum, proportional to $N$.

{\it Chaotic} cellular automata (class 3 in \cite{W86}) yield
disordered patterns (Fig.~\ref{fig:rules}b).  For any finite $N$ these
CA reach periodic states, but there are rather clear numerical
evidences that the transient time $\widetilde{{\cal T}}$ increases
exponentially with the system size:
\begin{equation}
\widetilde{{\cal T}}(N) \sim \exp(c N).
\label{transient}  
\end{equation}
Moreover, also the cycle period shows in most of the cases a similar
dependence on $N$, this is a reminiscence of what we discussed in
Sect.~\ref{sec:6.1}.

{\it Complex} cellular automata (class 4 in \cite{W86},
Fig.~\ref{fig:rules}c) usually evolve toward complex localized
structures (gliders) which interact in a complicate way.  For these CA
numerical simulations \cite{G91} have shown that both the transient
time and the cycle period display a non trivial $N$-dependence
(i.e. the average, the typical values or the median depend in a
different way on $N$).  The unpredictability of these system manifests
itself in the distribution of these times. In particular, the large
variability of these times in dependence of the initial conditions and
the lattice size inhibits any forecasting of the duration of the
transient.

In the following we limit the discussion to chaotic rules, i.e. class
3 in the Wolfram classification. A detailed characterization of {\it
complex} CA would require the introduction of concepts and tools that
are beyond the aim of this review, for further details  see
Refs.~\cite{BP97,G86,G91,W86}.

\subsubsection{Sensitive dependence on initial conditions}
\label{sec:6.3.2}
A first intuitively reasonable characterization of irregular
behaviors is in terms of sensitive dependence on initial conditions,
but in CA it is not possible to have arbitrary
small distances between two states. Nevertheless, for large $N$, when
considering two states with only one different element, one can say
that, in some sense (i.e. in an appropriate norm), 
the difference is small. Denoting with $R_t$ the
number of different elements at time $t$, we can define the damage
propagation speed as \cite{W86}
\begin{equation}
v =  \lim _{t \to \infty} \frac {R_t}{2 t} .
\label{speed}  
\end{equation}

It is not difficult to see that $v$ is, in a proper space, a Lyapunov
exponent (i.e.  it measures the rate of divergence of two
configurations) \cite{URE97}. Consider two initial bi-infinite configurations
${\mbox{\boldmath{$\sigma$}}} (0) = \left( \cdots , \sigma_{-2} (0),
\sigma_{-1} (0), \sigma_{1} (0), \sigma_{2} (0), \cdots \right)$ and
${\mbox{\boldmath{$\sigma^{'}$}}} (0) = \left( \cdots ,
\sigma^{'}_{-2} (0), \sigma^{'}_{-1} (0), \sigma{'}_{1} (0),
\sigma^{'}_{2} (0), \cdots \right) $, with $\sigma_{i} (0)=
\sigma^{'}_{i} (0)$ for $|i| < N_0$, and their evolutions
${\mbox{\boldmath{$\sigma$}}} (t)$ and
${\mbox{\boldmath{$\sigma$}}}^{'} (t)$. One can define a distance, $
|| \delta {\mbox{\boldmath{$\sigma$}}} (t) || $, between
${\mbox{\boldmath{$\sigma$}}} (t)$ and
${\mbox{\boldmath{$\sigma$}}}^{'} (t)$, as follows:
\begin{equation}
|| \delta {\mbox{\boldmath{$\sigma$}}} (t) || = \sum _{n=1}^{\infty} 
  \frac{|\delta \sigma_n (t)| +|\delta \sigma_{-n} (t)|}{2^n} \,
\label{diff}  
\end{equation}
where $\delta \sigma_n=\sigma_n^{'}-\sigma_n$. With the above norm
two systems can be arbitrarily close: one only needs $N_0 \gg 1$. At
this point it is possible to define the Lyapunov exponent as
\begin{equation}
 \lambda = \lim _{t \to \infty} \; \; 
\lim _{|| \delta {\mbox{\boldmath{$\sigma$}}} (0) || \to 0 } \; \; 
 \frac {1}{t} \ln \frac {|| \delta {\mbox{\boldmath{$\sigma$}}} (t) || }
{|| \delta {\mbox{\boldmath{$\sigma$}}} (0) || }\,  .
\label{lyap}  
\end{equation}
Note that in (\ref{lyap}) it has been implicitly taken the limit $N\to
\infty$.
\begin{figure}[hbt]
\centerline{\epsfig{figure=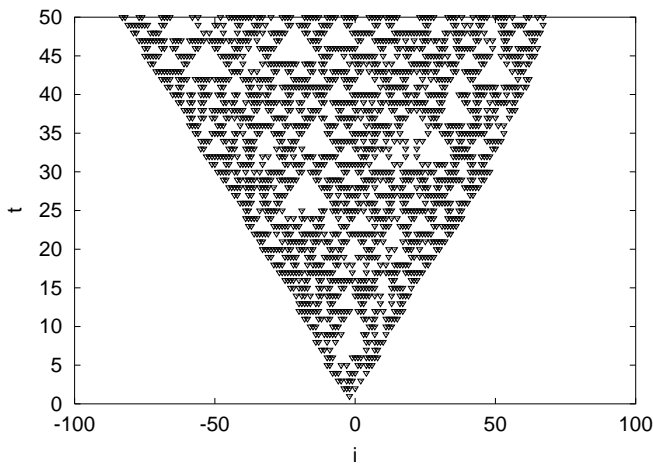,width=9cm,angle=0}}
\caption{ Damage spreading analysis performed on a totalistic $[Rule\;
10]$ r=2 cellular automaton with $N=200$. At time $t=0$ a replica is
initialized by flipping the value at the center of the lattice.}
\label{fig:6.3.2-1}
\end{figure}

Noting that $\delta \sigma_n (t) =0$ for $| n - N_0 | > R_t /2 \simeq
v t$, while $|\delta \sigma_n (t)| =1$ for $|n-N_0| < R_t /2 \simeq v
t$, from the definition (\ref{diff}) one has:
\begin{equation}
|| \delta {\mbox{\boldmath{$\sigma$}}} (t) || \sim 2 ^{-N_0 + v t} , 
\end{equation}
and therefore 
\begin{equation}
\lambda = v \ln 2 .
\label{eq:damage}
\end{equation}
In other words, the linear damage spreading in the physical space
corresponds to an exponential growth in the norm (\ref{diff}). Oono
and Yeung \cite{OY87} stressed a conceptual (and practical) difficulty
with the above approach. In systems with continuous states it is clear
that by performing an infinitesimal change on a typical configuration one
does not destroy the ``typicality'', i.e. the new initial condition
will generate a trajectory belonging to the same attractor. On the
contrary, it is not obvious that for a, however large, system with
discrete states in a typical configuration a change of only one
element gives another typical state. For instance, this seemingly
innocent change can induce a jump among basins of attraction, so that
the perturbed trajectory goes to a different attractor
\cite{W86,BRR92}.  However, taking into account the above criticism,
numerically one finds, for most the initial conditions, $v> 0$ for
chaotic CA, and $v=0$ for regular CA.

We conclude this subsection mentioning a proposal, by Bagnoli {\it
et. al.} \cite{BRR92}, to introduce a Lyapunov exponent for cellular
automata, defining it in analogy with continuous states dynamical
systems.

In this approach, the equivalent of an infinitesimal perturbation (as
for the damage spreading analysis) is the difference between the
system and one of its replicas in which one site has been flipped at
time $t=0$. Then one formally introduces the Boolean derivatives, a
sort of Jacobian of the rule, $F^{'}_{i,j}$, the elements of which are
$0$ or $1$.  Here, for simplicity, we consider a generic nearest
neighbor ($r=1$) rule so that $F^{'}_{i,j}=0$ for $|i-j|\geq 2$ and
$$
F^{'}_{i,i-1}= {\partial \sigma_i(t+1) \over \partial \sigma_{i-1}(t)} \equiv
F[\sigma_{i-1},\sigma_{i},\sigma_{i+1}]\; {\rm \bf \scriptstyle  XOR}\;
F[\sigma_{i-1}\,{\rm \bf \scriptstyle XOR}\, 1,\sigma_{i},\sigma_{i+1}]\,
$$
where the other nonzero terms are obtained by shifting the ${\rm \bf
\scriptstyle XOR}$ operation to $i$ and $i+1$ (respectively). We
recall that ${\rm \bf \scriptstyle XOR}$ is the Boolean exclusive
operation (i.e. $0\, {\rm \bf \scriptstyle XOR}\, 0= 0$, $1\, {\rm \bf
\scriptstyle XOR}\, 1= 0$, $0\, {\rm \bf \scriptstyle XOR}\, 1=1$ and
$1\, {\rm \bf \scriptstyle XOR}\, 0=1$). Of course as time goes on
the initial perturbation spreads, i.e. new defects appear.  As in
continuous systems, one needs to maintain the perturbation
``infinitesimal''. One introduces a vector ${\bf N}$ (whose components,
$N_i$, take integer values) which plays the role of the tangent
vector. In order to mimic an infinitesimal perturbation at the initial
time one assumes $N_i(0)=\delta_{ij}$, i.e. only one defect on the
site $j$. The dynamics of $N_i$ is ruled by the Boolean derivative
i.e.
\begin{equation}
N_i(t+1)=\sum_{j}F^{'}_{ij}(t)N_j(t)\, .
\end{equation}
Finally, putting $|{\bf N}(t)|=\sum_j N_j(t)$, one can define the
``Lyapunov exponent'', $\lambda_B$, of the cellular automaton as:
\begin{equation}
\lambda_B=\lim_{T\to \infty} {1 \over T}\ln (|{\bf N}(T)|)\,.
\end{equation} 
Now, in analogy with continuous systems, $\lambda_B<0$ indicates an
exponential decrease of the perturbation, while for $\lambda_B>0$ the
damage spreads. Just to give an example, if one considers the rule
$150$ of Wolfram classification, i.e.
($F[0,0,1]=F[0,1,0]=F[1,0,0]=F[1,1,1]=1$ and $0$ otherwise) it is easy
to see that $F^{'}_{ij}$ is a tridiagonal matrix with all the elements
equal to $1$ so that $\lambda=\ln(3)$. For a generic rule one has to
compute a suitable average over a long trajectory or on many initial
configurations.

The Lyapunov exponent, $\lambda_B$, has been demonstrated to
be relevant in the synchronization problem \cite{BR99} and allows for
a qualitative characterization of the cellular automata in agreement
with the classification proposed by Wolfram \cite{BRR92,BR99}.

\subsubsection{Entropies}
\label{sec:6.3.3}
For cellular automata one can define a
spatial/temporal entropy density by looking at the evolution of the
elements in a subset ${\cal L}_L$, of size $L$, of the
system. Denoting with $C(L,T)$ a ``word'' of spatial size $L$ and time
length $T$ appearing in the time evolution of the elements in ${\cal
L}_L$, one defines the entropy of the subset ${\cal L}_L$,
\begin{equation}
h(L) =   \lim_{T\to \infty} -\frac{1}{T}
 \sum _{C(L,T)} P(C(L,T)) \ln P(C(L,T)) \,,
\label{hl}
\end{equation}
and then the spatio-temporal entropy density as
\begin{equation}
h^{ST} = \lim _{L \to \infty} {1 \over L } h(L)\,.
\label{hst}
\end{equation} 
This entropy cannot be practically computed. 
A more accessible quantity is the temporal entropy:
\begin{equation}
h^T=h(1)= \lim_{T \to \infty} - \frac{1}{T}
 \sum _{C(1,T)} P(C(1,T)) \ln P(C(1,T))\,,
\end{equation}
i.e. the Shannon entropy of the time sequence of one element
 $(\sigma_n (0), \sigma_n (1), \cdots)$. In principle, $h^T$ can
 depend on the site $n$ and one can classify as nontrivial a system
 for which the majority of the elements have $h^T>0$ \cite{OK85}. An
 average measure of the ``temporal disorder'' is given by the spatial
 average $\langle h^T \rangle$.  A systematic study of $h(1), h(2),
 h(3), \dots$ -- although very difficult in practice -- could give, in
 principle, relevant information on the spatial/temporal behavior. A
 characterization of the spatial properties can be obtained studying,
 at a given time $t$, the spatial sequences. In practice, one studies
 $C(L,1)$ at increasing $L$:
\begin{equation}
h^S =\lim _{L \to \infty} - \frac{1}{L}
 \sum _{C(L,1)} P(C(L,1)) \ln P(C(L,1)) .
\label{hs}
\end{equation}
One can associate to $h^S$ a sort of ``effective'' dimension
$d=h^S/\ln 2$ \cite{W86}. In a completely disordered cellular
automaton configuration one has $d=1$, as expected, while a
homogeneous (or spatially periodic) configuration gives $d=0$.

From the definition of cellular automata (\ref{eq:6.3-1}) one easily
sees that the value of $\sigma_i(t)$ depends on sites at maximum
distance $r$ from $i$ at the previous time step. This means that after
$T$ time steps, the value $\sigma_i(t)$ can depend (at maximum) on
sites at distance $rT$ on both direction, so that the maximum speed
for information propagation is $r$ (i.e. the range of
interaction). However, for many CA the actual velocity of information
propagation, $v_p$, is less than $r$, i.e.  $\sigma_i(T)$ depends only
on $v_pT<rT$ sites.
\begin{figure}[hbt]
\centerline{\epsfig{figure=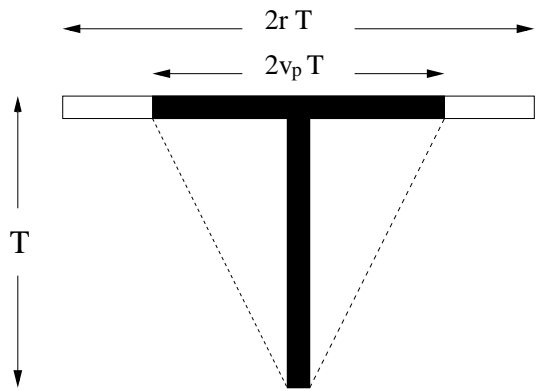,width=7cm,angle=0}}
\caption{Sketch of the dependence of temporal sequences on spatial ones.} 
\label{fig:sT}
\end{figure}
By considering a simple construction (see Fig.~\ref{fig:sT}) one can
understand that the spatial and temporal entropies are related
to each other by the inequality \cite{W86}:
\begin{equation}
h^T\leq 2v_p h^S\,,
\label{ineq}
\end{equation}
where a good estimate of $v_p$ can be given in terms of the damage spreading
velocity (\ref{eq:damage}) \cite{W86}. 

The possible scenario arising from (\ref{ineq}) can be summarized as
follows.  One can have ``spatial chaos'' ($h^S>0$) in absence of
``temporal chaos'' ($h^T=0$), while the existence of ``temporal chaos''
requires not only a non zero spatial entropy but also the existence of
a finite propagation velocity.  This confirms somehow that the
classifications of a CA as chaotic in terms of damage spreading
velocity and entropy are related to each others.

However, as stressed by Oono and Kohomoto \cite{OK85}, the seemingly
natural assumption of calling ``turbulent'' a cellular automaton for
which one has $h^S>0$ and $ \langle h^T \rangle >0$ is not correct in
general.  This is particularly clear by considering a single direction
shift imposed on a ``frozen'' disordered background.  Nevertheless, in
spite if this specific counterexample, the attempts based on entropic
concepts, for the characterization of the irregular spatial and/or
temporal behavior of systems with discrete states, in our opinion,
are the most promising ones.
In this context Casartelli and coworkers \cite{Ca90,CZ00} 
introduced the concept of 
rational partitions in order to define a complexity measure for systems
which can be reduced to $1d$ CA.

Let us conclude this Section with a brief discussion and comparison
between the unpredictability which characterizes cellular automata
evolution with respect to the one encountered in the context of
continuous states dynamics, e.g.  in coupled map lattices (see
Sect.~\ref{sec:3}).  The latter indeed seems to be the natural
candidate for such a comparison. We limit the discussion to
1-dimensional lattices with $r=1$, i.e. CML and CA
with nearest neighbor coupling.
\begin{figure}[hbt]
\centerline{\epsfig{figure=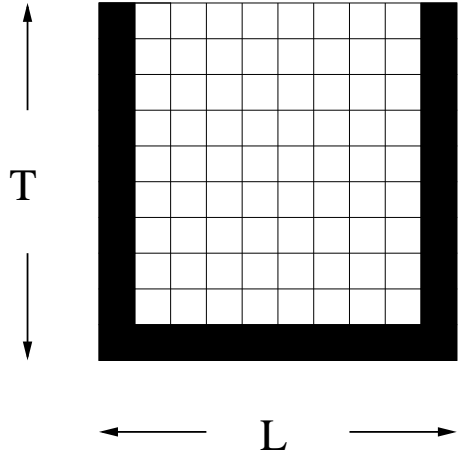,width=6cm,angle=0}}
\caption{The values of the sites in black together with the
specification of the rule completely specify the values of the sites
in white.}
\label{fig:box}
\end{figure}

Let us now ask the amount of information we have
to specify for knowing all the $LT$ sites  of spatial size $L(<N)$ and
temporal length $T$, as shown in Fig.~\ref{fig:box}.  Since
both CA and CML are ruled by a local deterministic dynamics one needs
to specify the rule of evolution and the values of the $L+2(T-1)$
states at the border, in black in Figure~\ref{fig:box}. Basically, one
has to specify the initial conditions on the $L$ sites and the
``boundaries'' $\sigma_{1}(t)$ and $\sigma_{L}(t)$ for $1< t \leq T$.
But while for CA this specification unambiguously determines the $LT$
values, for a chaotic CML this is not enough. Indeed, one has to
specify the precision, $\epsilon$, with which he wants to know the
$LT$ values. Once specified $\epsilon$, one knows the necessary
initial precision, $\epsilon_0$, on the $L+2(T-1)$ sites in black.  A
conservative estimate gives $\epsilon_0 \approx \epsilon
\exp(-LT{\cal H}_{KS})$, where ${\cal H}_{KS}$ is the entropy density
defined in Eq.~(\ref{eq:3.7}).  This very simple argument suggests that
the main difference between CA and continuous systems is the absence
of ``local'' production of information, i.e. in CA the complexity only
arises by the spatial propagation of information \cite{G89}.
Nevertheless there exist counterexamples in which starting from simple initial
configuration complex pattern are generated \cite{W85}.

From this point of view it is interesting to consider the behavior of
certain CMLs which, in spite of their continuous nature, seem to be
rather similar to ``chaotic'' cellular automata.  Indeed, it has been
found that a class of stable (i.e. $\lambda<0$) CMLs
\cite{CK88,PLOK93} displays an unpredictable dynamics on times
exponentially large with the system size. So that in the limit of
infinite lattices they are completely unpredictable.  Moreover, these
CMLs have a finite velocity of propagation for initially localized
disturbances (provided that the value of the disturbance was $O(1)$)
\cite{PLOK93,PT94}. Recalling the discussion of Sect.~\ref{sec:3.8}, we
know that this cannot be predicted in terms of the comoving Lyapunov
exponents, it is a fully non linear phenomenon.  The strong analogies
with ``chaotic'' CA have been furtherly explored in Ref.~\cite{PLOK93},
where it has been proposed to classify these CML as large memory
cellular automata according to the behavior of their spatial and
temporal entropy.

\section{The characterization of the Complexity 
and system modeling}
\label{sec:7}
\setcounter{equation}{0}
In the previous Sections we discussed the characterization of
dynamical behaviors when the evolution laws are known either exactly
or with an uncertainty. On the other
hand in experimental investigations only time records of
some observable are  typically available, 
and  the equation of motions are not known. 
For the predictability problem, 
this latter case, at least from a conceptual point of view, can
be treated in the same framework of when the evolution laws are
known. Indeed, in principle, with the embedding technique one can
reconstruct the phase space \cite{T80,A93,A96,KS97}. Nevertheless
there are rather severe limitations in high dimensional systems
\cite{G89} and even in low dimensional ones non trivial features
appear in presence of noise \cite{KS97}.

In this Section we show that an entropic analysis at different
resolution scales allows us for a pragmatic classification of a signal
and gives suggestions for modeling of systems.  In particular we
illustrate, using some examples, how quantities such as the
$\epsilon$-entropy or the FSLE can display a subtle transition from
the large to the small scales.  A negative consequence of this is the
difficulty in distinguishing, only from data analysis, a genuine
deterministic chaotic system from one with intrinsic randomness
\cite{CFKOV00}.  On the other hand, the way  the
$\epsilon$-entropy or Finite Size Lyapunov Exponent depends on the
(resolution) scale, allows us for a pragmatic classification of the
stochastic or chaotic character of the signal, and  this gives some 
freedom in modeling the system.

\subsection{How random is a random number generator?}
\label{sec:7.2.1}

It is rather natural to wonder about the ``true character'' of the
number sequence $(x_1, x_2, \dots)$ obtained with a (pseudo) random
number generator (PRNG) on a computer. One would like to have a
sequence with a random character; on the other hand, one is forced to
use deterministic algorithms to generate $(x_1, x_2, \dots)$. This
subsection is mainly based on the paper \cite{KO00}.  A simple and
popular PRNG is the multiplicative congruent one \cite{PTVF86}:
\begin{equation}
\begin{array}{ll}
z_{n+1} = & N_1 z_n  \:\: \mbox{mod} \:\: N_2  \nonumber \\
x_{n+1} = & z_{n+1}/N_2 \, , 
\end{array}
\label{prng}
\end{equation}
with an integer multiplier $N_1$ and modulus $N_2$. The $\{z_n\}$ are
integer numbers and one hopes to generate a sequence of random
variables $\{ x_n \}$, which are uncorrelated and uniformly
distributed in the unit interval.  A first problem one has to face is
the periodic nature of (\ref{prng}), because of its discrete character
(see Sect.~\ref{sec:6}). In practice one wants to fix $N_1$ and $N_2$
in such a way to maximize this period. Note that the rule (\ref{prng})
can be interpreted as a deterministic dynamical system, i.e.
\begin{equation}
x_{n+1} =  N_1 x_n  \:\: \mbox{mod} \:\: 1 \, ,
\label{xdyn}
\end{equation}
which has a uniform invariant measure and a KS entropy $h_{KS}=\lambda
= \ln N_1$.  When imposing the integer arithmetics of Eq.~(\ref{prng})
onto this system, we are, in the language of dynamical systems,
considering an unstable periodic orbit of Eq.~(\ref{xdyn}), with the
particular constraint that, in order to achieve the period $N_2-1$
(i.e.\ all integers $< N_2$ should belong to the orbit of
Eq.~(\ref{prng})) it has to contain all values $k/N_2$, with
$k=1,2,\cdots, N_2-1$. Since the natural invariant measure of
Eq.~(\ref{xdyn}) is uniform, such an orbit represents the measure of a
chaotic solution in an optimal way.  Every sequence of a PRNG is
characterized by two quantities: its period ${\cal T}$ and its
positive Lyapunov exponent $\lambda$, which is identical to the
entropy of a chaotic orbit of the equivalent dynamical system. Of
course a good random number generator has a very large period, and as
large as possible entropy.

It is natural to ask how this apparent randomness can be reconciled
with the facts that (a) the PRNG is a deterministic dynamical systems
(b) it is a discrete state system.

If the period is long enough on shorter times one has to face only
 point (a). In the following we discuss this point in terms of
the behavior of the $\epsilon$-entropy, $h(\epsilon)$ (see
Sect.~\ref{sec:2.4}). It seems rather reasonable to think that at a
high resolution, i.e. $\epsilon \leq 1/N_1$, one should realize the
true deterministic 
chaotic nature of the system and, therefore, $h(\epsilon) \simeq
h_{KS}= \ln N_1$. On the other hand for $\epsilon \geq 1/N_1$ one
expects to observe the ``apparent random'' behavior of the system,
i.e. $h(\epsilon) \sim \ln (1/ \epsilon)$.
\begin{figure}[htb]
\centerline{\psfig{file=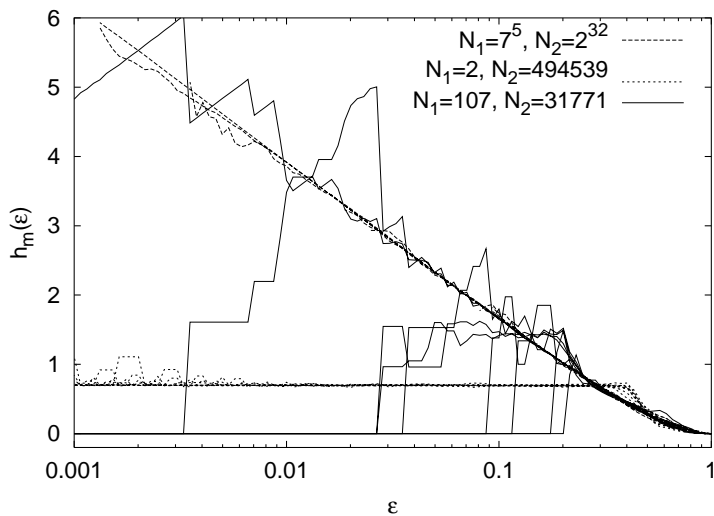,width=10cm,angle=0}}
\vspace{.5cm}
\caption{The $\epsilon$-entropies, $h_m(\epsilon)$, at varying 
the embedding dimension $m$ for the multiplicative congruential random number
generator Eq.~\ref{prng} for different choices of $N_1$ and $N_2$.
This figure has been taken from Ref.~[116]. }
\label{fig:rng_entropy}
\end{figure}
When the spatial resolution is high enough so that
every point of this periodic orbit is characterized by its own symbol,
then, for arbitrary
block length $m$, one has a finite number of $m$-words whose
probabilities are different from $0$. Therefore,
 the block entropy $H_m$ (\ref{eq:2-4}) is $m$-independent and $h_m=0$.

In Fig.~\ref{fig:rng_entropy} it is shown the behavior of
$h_m(\epsilon)$, computed on sequences of length 60000 of the PRNG
with three different pairs $(N_1,N_2)$ chosen to be ($7^5$,$2^{32}$),
(2,494539), and (107,31771).  The first one is optimal and no
deviation from the stochastic behavior is visible.  The second one
has a small pseudo-entropy, and this is seen by the saturation of all
$h_m(\epsilon)$ at $\ln N_1=\ln 2$, and the last one has large entropy
but a rather short period, so that all $h_m(\epsilon)$ drop to zero
for some $\epsilon_m$, where $\epsilon_m$ becomes dramatically larger
for increasing $m$ (strong fluctuations arise from the fact that data
are confined to a grid of spacing 1/31771).

\subsection{High dimensional systems}
\label{sec:7.2.2}
\begin{figure}[th]
\centerline{\psfig{file=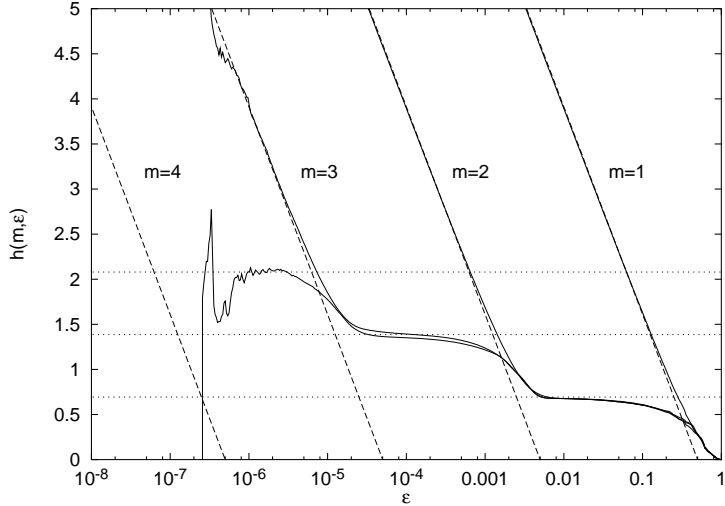,width=7cm,angle=270}}
\vspace{.5cm}
\caption{$h_m(\epsilon)$ for the system (\ref{eq:dresden} where
$f(x)=2|1/2-|x-1/2||$ is the tent map and $\sigma=0.01$. The
horizontal lines indicate the entropy steps which appears at
decreasing $\epsilon$.
The $\epsilon$-entropy is computed with the
Grassberger-Procaccia method [92]. For further details see Ref.~[165]}
\label{fig:olbrich}
\end{figure}

Now we discuss high-dimensional systems that show non-trivial
behavior at varying the resolution scales.  Olbrich et
al. \cite{OHK98} analyzed an open flow system described by
unidirectionally coupled map lattice:
\begin{equation}
x_j(t+1)=(1-\sigma) f(x_{j+1}(t))+\sigma x_j(t)
\label{eq:dresden}
\end{equation}
where $j=1,\dots,N$ denotes the site of a lattice of size $N$, $t$ the
discrete time and $\sigma$ the coupling strength.  A detailed
numerical study (also supported by analytical arguments) of the
$\epsilon$-entropy $h_m(\epsilon)$ at different $\epsilon$, in the
limit of small coupling, gives the following scale-dependent scenario:
for $1 \geq \epsilon \geq \sigma$ there is a plateau
$h(\epsilon)\simeq\lambda_s$ where $\lambda_s$ is the Lyapunov exponent of
the single map $x(t+1)=f(x(t))$. For $\sigma \geq \epsilon \geq
\sigma^2$ another plateau appears at $h(\epsilon) \simeq 2
\lambda_s$, and so on: for $\sigma^{n-1} \geq \epsilon \geq
\sigma^{n}$ one has $h(\epsilon) \simeq n \lambda_s$ (see
Fig.~\ref{fig:olbrich}).
Similar results hold for the correlation dimension which increases
step by step as the resolution increases, showing that the
high-dimensionality of the system becomes evident only as $\epsilon
\to 0$.  Therefore one understands that the dynamics at different
scales is basically ruled by a hierarchy of low-dimensional systems
whose ``effective'' dimension $n_{eff}(\epsilon)$ increases as
$\epsilon$ decreases \cite{OHK98}:
\begin{equation}
n_{eff} (\epsilon) \sim \left[ {\ln(1/\epsilon) \over
\ln(1/\sigma)}\right]\,,
\end{equation}
where $[\dots]$ indicates the integer part.  In addition, for a given
resolution  $\epsilon$, it is possible to find a suitable
low-dimensional noisy system (depending on $\epsilon$) which is able
to mimic $x_1(t)$ given by Eq.~(\ref{eq:dresden}).  It is interesting
to note that, on an extended range of values of $\epsilon$ ($\epsilon
\geq \sigma^{N}$), $h(\epsilon)$ can be roughly approximated as
log-periodic fluctuations around
\begin{equation}
 h(\epsilon) \sim \ln{1\over \epsilon} 
\label{eq:behaviour}
\end{equation}
i.e. the typical behavior of a stochastic process.  Of course for
$\epsilon \leq \sigma^{N}$ one has to realize that the system is
deterministic and $h(\epsilon) = O( N \lambda_s)$.

Let us now briefly reconsider the issue of the macroscopic chaos,
discussed in Sect.~\ref{sec:3.9}.  The main result can be summarized
as follows:
\begin{itemize}
\item at small $\epsilon\,\,(\ll 1/\sqrt{N})$, where $N$ is the number
of elements, one recovers the ``microscopic'' Lyapunov
exponent\footnote{From hereafter we use the same symbol $\epsilon$
both for the FSLE and the $\epsilon$-entropy in order to make a direct
comparison between the two quantities}, i.e. $\lambda(\epsilon)\approx
\lambda_{micro}$
\item at large $\epsilon\,\,(\gg 1/\sqrt{N})$ one observes another
plateau $\lambda(\epsilon) \approx \lambda_{macro}$ which can be much
smaller than the microscopic one.
\end{itemize}
The emerging scenario is that at a coarse-grained level,
i.e. $\epsilon \gg 1/\sqrt{N}$, the system can be described by an
``effective'' hydro-dynamical equation (which in some cases can be
low-dimensional), while the ``true'' high-dimensional character
appears only at very high resolution, i.e.
$$
\epsilon \leq \epsilon_c = O\left({1 \over \sqrt{N}}\right).
$$

\subsection{Diffusion in deterministic  systems and Brownian motion}
\label{sec:7.2.3}

Consider the following map which generates a diffusive behavior on
the large scales \cite{SFK82}:
\begin{equation}
\label{eq:3-1}
x_{t+1} = \lbrack x_{t} \rbrack + F\left(x_{t} - \lbrack x_{t} \rbrack
 \right) ,
\end{equation} 
where $\lbrack x_{t} \rbrack$ indicates the integer part of $x_{t}$
and $F(y)$ is given by:
\begin{equation}
F(y)=\left\{\begin{array}{ll}
(2+\alpha) y & \:\: \mbox{if}\:\: y\, \in [0,1/2[ \nonumber \\
(2+\alpha) y-(1+\alpha) &\:\: \mbox{if}\:\: y\, \in \,[1/2,1]\,.
\end{array}\right.
\label{eq:3-1.1}
\end{equation}
The largest Lyapunov exponent $\lambda$ can be obtained immediately:
$\lambda = \ln |F'|$, with $F'=dF/dy=\!2\!+\!\alpha$. 
One expects the following scenario for $h(\epsilon)$:
\begin{equation}
\label{eq:3-2}
h(\epsilon) \approx \lambda \qquad {\rm for} \quad \epsilon < 1 ,
\end{equation} 
\begin{equation}
\label{eq:3-3}
h(\epsilon) \propto {D\over \epsilon ^2} \qquad {\rm for} 
\quad \epsilon > 1 , 
\end{equation} 
where $D$ is the diffusion coefficient, i.e. 
\begin{equation}
\label{eq:3-4}
\langle \left( x_t -x_0 \right)^2 \rangle \approx 2 \,\, D \,\, t
\qquad {\rm for} \quad {\rm large} \quad t .
\end{equation}
\begin{figure}[bh]
\centerline{\psfig{file=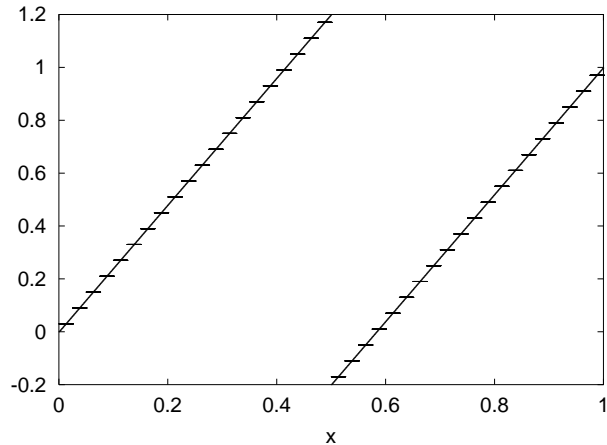,width=9cm,angle=0}}
\vspace{.5cm}
\caption{The map $F(x)$ (\ref{eq:3-1.1}) for $\alpha=0.4$ is shown
with superimposed the approximating (regular) map $G(x)$
(\ref{eq:3-5}) obtained by using $40$ intervals of slope $0$. }
\label{map}
\end{figure}
Consider now a stochastic system, namely a noisy map
\begin{equation}
\label{eq:3-5}
x_{t+1} = \lbrack x_{t} \rbrack + G \left(x_{t} - \lbrack x_{t} \rbrack
 \right) + \sigma \eta _{t},
\end{equation} 
where $G(y)$, as shown in Fig.~\ref{map}, is a piece wise linear map
which approximates the map $F(y)$, and $\eta _{t}$ is a stochastic
process uniformly distributed in the interval $\lbrack -1, 1 \rbrack$,
and no correlation in time.  When $|dG/dy| < 1$, as is the case we
consider, the map (\ref{eq:3-5}), in the absence of noise, gives a
non-chaotic time evolution.

\begin{figure}[bh]
\centerline{\psfig{file=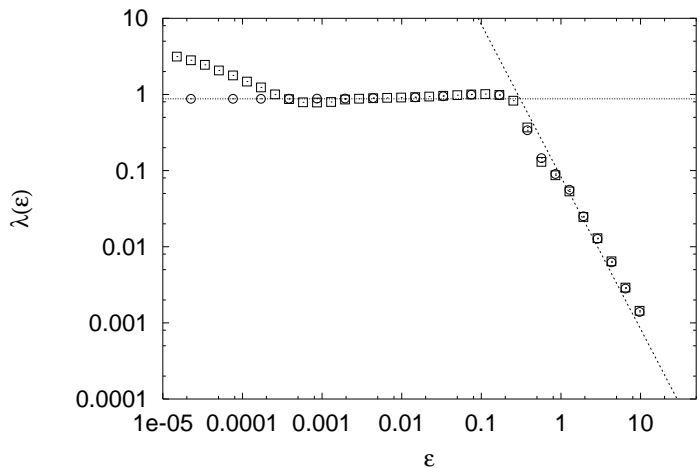,width=9.5cm,angle=0}}
\vspace{.5cm}
\caption{$\lambda(\epsilon)$ versus $\epsilon$ obtained with the map
$F(y)$ (\ref{eq:3-1.1}) with $\alpha=0.4$ ($\circ$) and with the noisy
(regular) map ($\Box$) (\ref{eq:3-5}) with $10000$ intervals of slope
$0.9$ and $\sigma=10^{-4}$. The straight lines indicates the Lyapunov
exponent $\lambda=\ln 2.4$ and the diffusive behavior
$\lambda(\epsilon) \sim \epsilon^{- 2}$.}
\label{fslediff}
\end{figure}
Now we compare the finite size Lyapunov exponent for the chaotic map
(\ref{eq:3-1}) and for the noisy one (\ref{eq:3-5}). In the latter the
FSLE has been computed using two different realizations of the noise.
In Fig.~\ref{fslediff} we show $\lambda(\epsilon)$ versus $\epsilon$
for the two cases.  The two curves are practically indistinguishable
in the region $\epsilon >\sigma$. The differences appear only at very
small scales $\epsilon < \sigma$ where one has a $\lambda(\epsilon)$
which grows with $\epsilon$ for the noisy case, remaining at the same
value for the chaotic deterministic case.

Both the FSLE and the $(\epsilon,\tau)$-entropy analysis show that we
can distinguish three different regimes observing the dynamics of
(\ref{eq:3-5}) on different length scales. On the large length scales
$\epsilon > 1$ we observe diffusive behavior in both models. On
length scales $\sigma < \epsilon < 1$ both models show chaotic
deterministic behavior, because the entropy and the FSLE are
independent of $\epsilon$ and larger than zero. Finally on the
smallest length scales $\epsilon < \sigma $ we see stochastic
behavior for the system (\ref{eq:3-5}) while the system
(\ref{eq:3-1}) still shows chaotic behavior.

\subsection{On the distinction between chaos and noise}
\label{sec:7.2.4}
The above examples show that the distinction between chaos and noise
can be a high non trivial task, which makes sense only in very
peculiar cases, e.g., very low dimensional systems.  Nevertheless,
even in this case, the entropic analysis can be unable to recognize
the ``true'' character of the system due to the lack of
resolution. Again, the comparison between the diffusive map
(\ref{eq:3-1}) and the noisy map (\ref{eq:3-5}) is an example of these
difficulties.  For $\sigma \leq \epsilon \leq 1$ both the system
(\ref{eq:3-1}) and (\ref{eq:3-5}), in spite of their ``true''
character, will be classified as chaotic, while for $\epsilon \geq 1$
both can be considered as stochastic.

In high-dimensional chaotic systems, with $N$ degrees of freedom, one
has typically $h(\epsilon)=h_{KS}\sim O(N)$ for $\epsilon \leq
\epsilon_c$ (where $\epsilon_c \to 0$ as $N\to \infty$) while for
$\epsilon\geq \epsilon_c$, $h(\epsilon)$ decreases, often with a power
law \cite{GW93}.  Since also in some stochastic processes the
$\epsilon$-entropy obeys a power law, this can be a source of
confusion.

These kind of problems are not abstract ones, as a recent debate on
``microscopic chaos'' demonstrates \cite{GBFSGDC98,DCB99,GS99}.  The
detection of microscopic chaos by data analysis has been recently
addressed in a work of Gaspard et al.~\cite{GBFSGDC98}.  These
authors, from an entropic analysis of an ingenious experiment on the
position of a Brownian particle in a liquid, claim to give an
empirical evidence for microscopic chaos.  In other words, they state
that the diffusive behavior observed for a Brownian particle is the
consequence of chaos at a molecular level.  Their work can be briefly
summarized as follows: from a long ($\approx 1.5 \times 10^5$ data)
record of the position of a Brownian particle they compute the
$\epsilon$-entropy with the Cohen-Procaccia method \cite{CP85}
(Sect.~\ref{sec:2}) from which they obtain:
\begin{equation} 
h(\epsilon) \sim {D \over \epsilon^2}\,,
\label{eq:gasp}
\end{equation} 
where $D$ is the diffusion coefficient. Then, {\em assuming} that
the system is deterministic, and making use of the inequality
$h(\epsilon >0) \leq h_{KS}$, they conclude that the system is
chaotic. However, their result does not give a direct evidence that the system is
deterministic and chaotic.  Indeed, the power law (\ref{eq:gasp}) can
be produced with different mechanisms:
\begin{enumerate}
\item a genuine chaotic system with diffusive behavior, 
      as the map (\ref{eq:3-1.1});
\item a non chaotic system with some noise, as the map (\ref{eq:3-5}), 
     or a genuine Brownian system;
\item a deterministic linear non chaotic system with many degrees of
freedom (see for instance \cite{MM60});
\item a ``complicated'' non chaotic system as the Ehrenfest wind-tree
      model where a particle diffuses in a plane due to collisions with
      randomly placed, fixed oriented square scatters, as discussed by Cohen
      et al. \cite{DCB99} in their comment to Ref.~\cite{GBFSGDC98}.
\end{enumerate}
It seems to us that the weak points of the analysis in
Ref.~\cite{GBFSGDC98} are:
\begin{description}
\item{a)} the explicit assumption that the system is deterministic;
\item{b)} the limited number of data points and therefore 
limitations in both resolution and block length.
\end{description}
The point (a) is crucial, without this assumption (even with an
enormous data set) it is not possible to distinguish between 1) and
2).  One has to say that in the cases 3) and 4) at least in principle
it is possible to understand that the systems are ``trivial''
(i.e. not chaotic) but for this one has to use a huge number of
data. For example Cohen et al.~\cite{DCB99} estimated that in order to
distinguish between 1) and 4) using realistic parameters of a typical
liquid, the number of data points required has to be at least $\sim
10^{34}$.

Concluding, we have the apparently paradoxical result that
``complexity'' helps in the construction of models. Basically, in the
case in which one has a variety of behaviors at varying the scale
resolution, there is a certain freedom on the choice of the model to
adopt. In Sect.~\ref{sec:7.2.3} one can see that, for some systems,
the behavior at large scales can be realized both with chaotic
deterministic models or suitable stochastic processes.  From a
pragmatic point of view, the fact that in certain stochastic processes
$h(\epsilon) \sim \epsilon^{-\alpha}$ can be indeed extremely useful
for modeling such high-dimensional systems. Perhaps, the most
relevant case in which one can use this freedom in modeling is the
fully developed turbulence whose non infinitesimal (the so-called
inertial range) properties can be successfully mimicked in terms of
multiaffine stochastic process (see Ref.~\cite{BBCCV98,ABCFVV00b}
Sect.~\ref{sec:4.5} and Appendix~\ref{app:synth}).  
\section{Concluding Remarks}
\label{sec:8}
\setcounter{equation}{0}
The guideline of this review has been how to interpret the different
aspects of the predictability of a system as a way to characterize its
complexity.

We have discussed the relation between the Kolmogorov-Sinai entropy
and the algorithmic complexity (Sect.~\ref{sec:1}). As clearly exposed
in the seminal works of Alekseev and Yakobson \cite{AY81} and Ford
\cite{F83,F86}, the time sequences generated by a system with
sensitive dependence on initial conditions have non-zero algorithmic
complexity. A relation exists  between the maximal compression of
a sequence and its KS-entropy.  Therefore, one can give a definition
of complexity, without referring to a specific description, as an
intrinsic property of the system.

In presence of intrinsic randomness (Sect.~\ref{sec:5.3}), one can
introduce two different Lyapunov exponents, $\lambda_{\sigma}$ in the
case of trajectories with the same realization of noise and
$K_{\sigma}$ for different realizations.  In general
$\lambda_{\sigma}$ and $K_{\sigma}$ do not coincide and characterize
different aspects of the system.  Both quantities have their
own relevance, the comparison between $\lambda_{\sigma}$ and
$K_{\sigma}$ has shown to be useful in the understanding of apparently
intricate phenomena, such as noise-induced order and noise-induced
instability.

As an example of system with many degrees of freedom and
characteristic times scales, we investigated fully developed
turbulence (Sect.~\ref{sec:4}).  In this case the Lyapunov exponent
and the KS-entropy are somehow of limited relevance because they only
characterize small scales properties.  On the other hand there exist
suitable generalizations -- the finite size Lyapunov exponent,
$\lambda (\epsilon)$, and $\epsilon$-entropy, $h(\epsilon)$ -- which
characterize the predictability properties at different scales.  The
scaling of the predictability time with the resolution $\epsilon$,
$\lambda (\epsilon) \sim \epsilon ^{-2}$, has an algorithmic
correspondence in the behavior of the $\epsilon$-entropy of the
signal measured in one point, $h (\epsilon) \sim \epsilon^{-3}$.  In
the words of Lorenz, one can say that the butterfly effect is not so
terrible for $\epsilon$-resolution in the inertial range.

Complexity in a system can also manifest in the spatial properties as,
for example, in open flows with convective chaos but with negative
Lyapunov exponents (Sect.~\ref{sec:3}). The presence of convective
chaos implies a sensitivity on the boundary conditions. An
uncertainty, $\delta x_0$, on the boundary condition is exponentially
amplified with the distance, $n$, from the boundary as $\delta x_n
\sim \delta x_0 e^{\Gamma n}$.  The ``spatial'' Lyapunov exponent
$\Gamma$ is related with the comoving Lyapunov exponent and gives a
characterization of the spatial ``complexity''.

The study of these different aspects of predictability constitutes a
useful method for a quantitative characterization of ``complexity'',
suggesting the following equivalences:
\begin{figure}[!h]
\centerline{\epsfig{figure=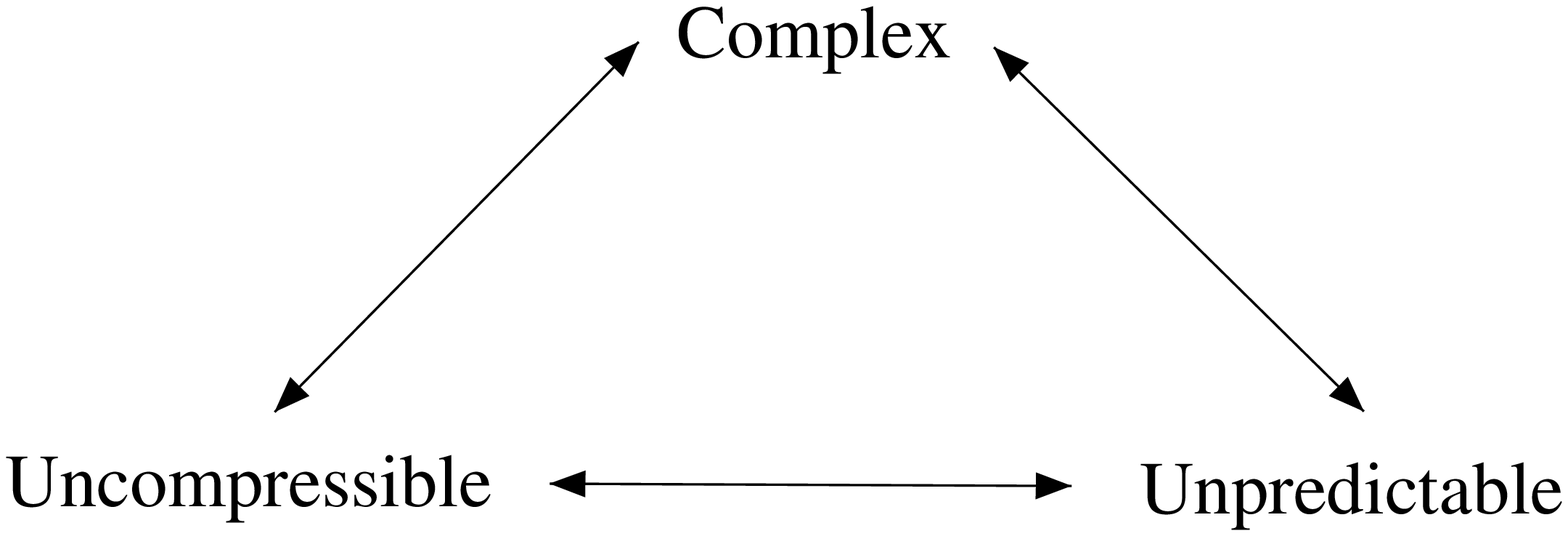,width=8cm, angle=0}}
\end{figure}

The above point of view, based on dynamical systems and information
theory, quantifies the complexity of a sequence considering each
symbol relevant but it does not capture the structural level.  Let us
clarify this point with the following example.  A binary sequence
obtained with a coin tossing is, from the point of view adopted in
this review, complex since it cannot be compressed (i.e. it is
unpredictable). On the other hand such a sequence is somehow trivial,
i.e.  with low ``organizational'' complexity. It would be important to
introduce a quantitative measure of this intuitive idea. The
progresses of the research on this intriguing and difficult issue are
still rather slow.  We just mention some of the most promising
proposals as the logical depth \cite{B90} and the sophistication
\cite{KA91}, see Ref.~\cite{BP97}.

As a final question one can wonder what can one learn by the presented
material for practical prediction problems (e.g.  weather forecast).
The main lesson concerns the framework and limitations about the
possible well posed questions in prediction and modeling. The first
relevant fact, now well established, is that in presence of
deterministic chaos one cannot hope to reach an arbitrary accuracy in
prediction by merely refining the model.  A less recognized aspect is
that the Lyapunov exponent is usually not sufficient to characterize
the limits of predictability in real situations. An appropriate
generalization of the Lyapunov exponent is necessary to account for
the large scale properties. Moreover in weather forecast the
predictability time, which is typically of $5$ days, may be as little
as $2$ days or as much as $10$ days \cite{T91}. Thus, simply quoting
an average value does not give a satisfactory answer.  At a more
conceptual level, one has severe limitations in distinguish between
deterministic or stochastic nature of systems displaying complex
behavior. This implies a certain freedom in the choice of the details
of the model, in particular whether to adopt a deterministic or a
stochastic model.
 
\begin{ack}

We are particularly grateful to  A. Celani, for the long collaboration on
the subject here discussed and for his contributions at a first stage
of this review.  

We thank M.~Abel, K.H.~Andersen, A.~Baldassarri,
E.~Aurell, F.~Bagnoli, R.~Baviera, A.~Bigazzi, R.~Benzi, L.~Biferale,
M.~Casartelli,
P.~Castiglione, A.~Crisanti, D.~P.~Eckmann, 
U.~Frisch, P.~Grassberger, K.~Kaneko,
H.~Kantz, M.H. Jensen, G.~Lacorata, R.~Livi, V. Loreto, R.~Mantegna,
G.~Mantica, U.~Marini Bettolo Marconi, A.~Mazzino, P.~Muratore-Ginanneschi,
E.~Olbrich, G.~Parisi, R.~Pasmanter, M.~Pasquini, A.~Pikovsky,
A.~Politi, A.~Provenzale, A.~Puglisi, S.~Ruffo, M.~Serva, A.~Torcini,
M. Vergassola, D.~Vergni and E. Zambianchi for collaborations,
correspondence and discussions in the last years.

We are grateful to F.~Benatti, P.~Castiglione, A.~Politi and A.~Torcini
for useful remarks and detailed comments on the manuscript.

We acknowledge  F. di Carmine, R.~T. Lampioni, B.~Marani and I. Poeta
for their friendly and warm encouragement.
\end{ack}

\begin{appendix}
\section{On the computation of the Finite Size Lyapunov Exponent}
\setcounter{equation}{0}
\label{app:fsle}
This Appendix is devoted to the methods for
computing the Finite Size Lyapunov Exponent. As we will  see there
are mainly three possible ways to compute the FSLE.
Let us start with a modification of the standard technique for
computing the largest Lyapunov exponent\cite{BGGS80,WSSV85}.

Suppose to integrate on a computer the equations of motion of a
system. After a long integration time, in order for the motion to
settle onto the attractor of the system, we introduce a very small
perturbation, i.e.  we consider the ``reference'' trajectory ${\bf
x}(0)$, which is supposed to be on the attractor, and generate a
``perturbed'' trajectory starting from ${\bf x}^{'}(0)={\bf
x}(0)+\delta{\bf x}(0)$.  We need the perturbation to be initially
very small in some chosen norm $\delta(t=0)=||\delta {\bf
x}(t=0)||=\delta_{min} \ll 1$. Then, in order to study the
perturbation growth through different scales, one defines a set of
thresholds $\delta_n$, e.g.: $\delta_n=\delta_0 r^{(n)}$ with $1 \gg
\delta_0 \gg \delta_{min}$ and $n=0,\dots,N$. To avoid saturation on
the maximum allowed separation (i.e. the attractor size) one has to
choose $\delta_N < \langle ||{\bf x}-{\bf y}||\rangle_{\mu}$ with
${\bf x},{\bf y}$ generic points on the attractor.  Note that $r$
should be larger than $1$ but not too large in order to avoid
interferences of different length scales.  Typically, one chooses
$r=2$ or $r=\sqrt{2}$.

In order to measure the perturbation growth rate at scale $\delta_n$,
one lets the system to evolve from $\delta_{min}$ up to the desired
scale $\delta_n$ ensuring the perturbation to be on the attractor and
aligned along the maximally expanding direction.  After $\delta_n$ is
reached, one computes the first time, $\tau_1 (\delta_n,r)$, to reach
the following threshold, $\delta_{n+1}$, and after that the
perturbation is rescaled to $\delta_{n}$, keeping the direction ${\bf
x}'-{\bf x}$ constant. This procedure is repeated ${\cal N}$ times for
each thresholds obtaining the set of the doubling times
$\{\tau_i(\delta_n,r)\}$ for $i=1,\dots,{\cal N}$
error-doubling experiments.  
Now if we introduce the effective doubling rates:
\begin{equation}
\gamma_i(\delta_n,r)={1 \over \tau_i(\delta_n,r)} \ln r\,,
\end{equation}
we can define their time averages as the the effective LEs on the scale 
$\delta_n$.
Therefore, we have
\begin{equation}
\lambda(\delta_n)=\langle \gamma(\delta_n,r) \rangle_{t} 
= {1 \over T} \int_0^T \, \gamma dt =
{\sum_i \gamma_i \tau_i \over \sum_i \tau_i} =
 {1 \over \langle \tau(\delta_n,r) \rangle_{e}} \ln r \, .
\label{eq:app_fsle.2}
\end{equation}
where $\langle \tau(\delta_n,r) \rangle_{e}=\sum \tau_i/{\cal N}$
and $T=\sum_i \tau_i$.

To obtain Eq.~(\ref{eq:app_fsle.2}) we assumed the distance between the
two trajectories to be continuous in time. This is not true for maps
or for discrete sampling in time and the method has to be slightly
modified.  In this case the doubling time, $\tau(\delta_n,r)$, is
defined as the minimum time such that $\delta(\tau) \ge r
\delta_n$. Because now $\delta(\tau)$ is a fluctuating quantity, from
(\ref{eq:app_fsle.2}) we have
\begin{equation}
\lambda(\delta_n) = {1 \over \langle \tau(\delta_n,r) \rangle_{e}}
\left\langle \ln \left( {\delta(\tau(\delta_n,r)) \over \delta_n} \right) 
\right\rangle_{e} \, .
\label{eq:app_fsle.4}
\end{equation}

Let us stress some points.

The computation of the FSLE is not more expensive than the one
of the Lyapunov exponent by standard algorithm.  One has simply to
integrate two copies of the system (or two different systems for
second kind predictability) and this can be done without particular
problems.

At difference with $\lambda$, $\lambda(\delta)$ may also depend on the
norm one chooses. This fact, apparently disturbing, is however
physically reasonable: when one looks at the non linear regime, for
instance, for the predictability problem the answer may depend on the
involved observable.  A similar problem appears in infinite
dimensional system where the norms are not equivalent
\cite{KF76}.

A possible problem with the above described method is  that 
we have implicitly assumed that the statistically stationary state 
of the system is homogeneous with respect to finite perturbations. 
Actually one may plausibly expect the attractor to be fractal, 
i.e., not at all equally dense at all distances, this may cause an 
incorrect sampling of the doubling times at large $\delta_n$.

A possible way to overcome such a problem is to compute the FSLE
avoiding to rescale the perturbation at finite $\delta_n$.
This can be accomplished by the following modification of the
previous method.
One defines the thresholds $\{\delta_n\}$ and initializes the 
perturbation at $\delta_{min}\ll \delta_0$ as before.
Then one lets the system to reach the first threshold, $\delta_0$.
Hence, one starts to measure the doubling time $\tau(\delta_n,r)$
following the perturbation growth from $\delta_0$ up to $\delta_N$.
In practice, one register the time $\tau(\delta_n,r)$ for going from
$\delta_n$ to $\delta_{n+1}$ for each value of $n$.
The evolution of the error from the initial value
$\delta_{min}$ to the largest threshold $\delta_{N}$ carries out a
single error-doubling experiment. When the largest threshold,
$\delta_{N}$ has been reached the ``perturbed'' trajectory is rescaled
at the initial distance, $\delta_{min}$, with respect to the
``reference'' trajectory and one starts another experiment measuring a
second set of doubling times, $\{\tau_2(\delta_n,r)\}$. The
procedure is then repeated ${\cal N}$ times to have statistics.
In this way one obtains the set of the doubling 
times $\{\tau_i(\delta_n,r)\}$ 
for $i=1,\dots,{\cal N}$. The FSLE is finally obtained by using
Eq.~(\ref{eq:app_fsle.2}) or Eq.~(\ref{eq:app_fsle.4}), which  are accurate
also in this case, according to
the continuous time and discrete time nature of the system respectively.
One understands that with this method, since finite perturbations are
realized by the dynamics (i.e.  the perturbed trajectory is on the
attractor) and not imposed by hand, the problems related to the attractor
inhomogeneity are not present.

In any case, in most numerical experiments, one does not find
significant differences between the two numerical methods.

A further possibility to compute the FSLE is to remove the threshold
condition used for defining $\tau(\delta_n,r)$ and simply compute the
average error growth rate at every time step. In other words, at every
time step $\Delta t$ in the integration, the perturbed trajectory
${\bf x}'(t)$ is rescaled to the original distance $\delta$, keeping
the direction ${\bf x}-{\bf x}'$ constant.  The FSLE is then obtained
by the average of the one-step exponential divergence:
\begin{equation}
\label{eq:app_fsle.5}
  \lambda(\delta) = \frac{1}{\Delta t}\, \left\langle
                    \ln\left(\frac{||\delta {\bf x}(t+\Delta t)||}
{||\delta {\bf x}(t)||}
                       \right)
                    \right\rangle_{t},
\end{equation}
which, if non negative, 
is equivalent to the definition (\ref{eq:app_fsle.2}). 
Let us note that the above procedure is nothing but the finite scale version
of the usual algorithm of Benettin et al.~\cite{BGGS80} for the 
LE. The one-step method (\ref{eq:app_fsle.5}) can be, in principle,
generalized to compute the sub-leading finite-size Lyapunov exponent
following the standard ortho-normalization method \cite{BGGS80}.  One
introduces $k$ perturbed trajectories ${\bf x}^{(1)},\ldots,{\bf
x}^{(k)}$ each at distance $\delta$ from ${\bf x}$ and such that ${\bf
x}^{(k)}-{\bf x}$ are orthogonal each to the others.  At every time
step, any difference ${\bf x}^{(k)}-{\bf x}$ is rescaled at the
original value and orthogonalized, while the corresponding finite size
Lyapunov exponent is accumulated according to (\ref{eq:app_fsle.5}).

Here we have again the problem of the implicitly assumed homogeneity of
the attractor, but also a problem of isotropy when we re-orthogonalize
the perturbations. We note that this could be a more serious problem.
\section{The multifractal model of turbulence}
\setcounter{equation}{0}
\label{app:mf}
The multifractal model of turbulence \cite{PF84,PV87,F95} assumes that
the velocity has a local scale-invariance, i.e. it does not have a
unique scaling exponent $h$ such that $\delta v_{\ell} \sim \ell^h$,
but a continuous spectrum of exponents, each of which belonging to a
given fractal set. In other words, in the inertial range one has
\begin{equation}
\delta v_{\ell}(x) \sim \ell^h\,,
\label{eq:app.mf-1}
\end{equation}
if $x \in S_h$, and $S_h$ is a fractal set with dimension $D(h)$ and
$h \in$ ($h_{min}$, $h_{max}$).  The probability to observe a given
scaling exponent $h$ at the scale $\ell$ is thus $P_{\ell}(h) \sim
\ell^{3-D(h)}$, so the scaling of the structure function assumes the
form:
\begin{equation}
S_p(\ell) = \langle\delta v_{\ell}^p\rangle \sim
\int_{h_{min}}^{h_{max}} \ell^{h p} \ell^{3-D(h)} {\rm d} h 
\sim \ell^{\zeta_p}\,,
\label{eq:app.mf-2}
\end{equation}
where in the last equality, being $\ell \ll 1$, a steepest 
descent estimation gives
\begin{equation}
\zeta_p = \min_{h}\left\{hp+3-D(h)\right\} = h^*p + 3 - D(h^*)
\label{eq:app.mf-3}
\end{equation}
where $h^*=h^*(p)$ is the solution of the equation $D'(h^*(p))=p$.
The Kolmogorov ``4/5'' law \cite{F95}
\begin{equation}
S_3(\ell) = - {4 \over 5} \epsilon \ell
\label{eq:app.mf-4}
\end{equation}
imposes $\zeta_3=1$ which implies that
\begin{equation}
3 h + 2 - D(h) \le 0\,,
\label{eq:app.mf-5}
\end{equation}
the equality is realized by $h^*(3)$.  The Kolmogorov similarity
theory corresponds to the case of only one singularity exponent
$h=1/3$ with $D(h=1/3)=3$.

A nontrivial consequence of the intermittency in the turbulent cascade
is the fluctuations of the dissipative scale which implies the
existence of an intermediate region between the inertial and
dissipative range \cite{FV91}.  The local dissipative scale $\ell_{D}$
is determined by imposing the effective Reynolds number to be of order
unity:
\begin{equation}
Re(\ell_D)={\delta v_{D} \ell_{D} \over \nu} \sim 1\,,
\label{eq:app.mf-6}
\end{equation}
therefore the dependence of $\ell_{D}$ on $h$ is thus
\begin{equation}
\ell_{D}(h) \sim L Re^{-{1 \over 1+h}}
\label{eq:app.mf-7}
\end{equation}
where $Re=Re(L)$ is the large scale Reynolds number.  The fluctuations
of $\ell_{D}$ modifies the evaluation of the structure functions
(\ref{eq:app.mf-2}): for a given $\ell$, the saddle point evaluation
of (\ref{eq:app.mf-2}) remains unchanged if, for the selected exponent
$h^*(p)$, one has $\ell_{D}(h^*(p)) < \ell$.  If, on the contrary, the
selected exponent is such that $\ell_{D}(h^*(p)) > \ell$ the saddle
point evaluation is not consistent, because at scale $\ell$ the
power--law scaling (\ref{eq:app.mf-1}) is no longer valid.  In this
intermediate dissipation range \cite{FV91} the integral in
(\ref{eq:app.mf-2}) is dominated by the smallest acceptable scaling
exponent $h(\ell)$ given by inverting (\ref{eq:app.mf-7}), and the
structure function of order $p$ a pseudo--algebraic behavior, i.e. a
power law with exponent $p h(\ell) + 3 - D(h(\ell))$ which depends on
the scale $\ell$.  Taking into account the fluctuations of the
dissipative range, one has for the structure functions
\begin{equation}
S_p(\ell) \sim \left\{
\begin{array}{lll}
\ell^{\zeta_p} & \mbox{if} & \ell_{D}(h^*(p)) < \ell \\
\ell^{h(\ell) p + 3 -D(h(\ell))} & \mbox{if} & \ell_{D}(h_{min} < \ell < 
\ell_{D}(h^*(p))\,.
\end{array}
\right.
\label{eq:app.mf-8}
\end{equation}

A simple calculation \cite{FV91,F95} shows that it is possible to find
a universal description valid both in the inertial and in the
intermediate dissipative ranges.  Let us discuss this point for the
energy spectrum $E(k)$.  Introducing the rescaled variables
\begin{equation}
F(\theta)={\ln E(k) \over \ln Re} \qquad {\mbox {and}}\qquad 
\theta={\ln k \over \ln Re} 
\label{eq:app.mf-9}
\end{equation}
one obtains the following behavior
\begin{equation}
F(\theta) = \left\{
\begin{array}{lll}
-(1+\zeta_2) \theta & \mbox{for} & \theta < {1 \over 1+h^*(2)} \\
-2 -2 \theta + \theta D(\theta^{-1}-1) & \mbox{for} &
{1 \over 1+h^*(2)} < \theta < {1 \over 1+h_{min}}
\end{array}
\right.
\label{eq:app.mf-10}
\end{equation}
The prediction of the multifractal model is that $\ln E(k)/\ln Re$ is
an universal function of $\ln k/\ln Re$. This is in contrast with the
usual scaling hypothesis according which $\ln E(k)$ should be a
universal function of $\ln (k/k_D)$). The multifractal universality has
been tested by collapsing energy spectra obtained from turbulent flow
in a wide range of $Re$ \cite{GC91}, see also \cite{BCVV99}.

\section{How to compute the $\epsilon$-entropy with exit times}
\setcounter{equation}{0}
\label{app:eps}

The approach based on exit times differs from the usual one (see
Sect.~\ref{sec:2.4}) in the procedure to construct the coding sequence
of the signal at a given level of accuracy \cite{ABCFVV00a}.  Indeed
an efficient coding procedure reduces the redundancy and improves the
quality of the results.  The method here discussed is particularly
suited for computing the $\epsilon$-entropy in processes in which many
scales are excited as, e.g., in turbulence \cite{ABCFVV00a,ABCFVV00b}.

The coding of a signal, $x(t)$, by the exit-time approach is the
following.  Given a reference starting time $t=t_0$, one measures the
first exit-time, $t_1$, from a cell of size $\epsilon$, i.e. the first
time necessary to have $|x(t_0 + t_1)-x(t_0)| \ge \epsilon/2$.  Then
one restarts from the time $t=t_0+t_1$ to look for the next exit-time
$t_2$, i.e., the first time such that $|x(t_0 + t_1 + t_2)-x(t_0 +
t_1)| \ge \epsilon/2 $ and so on. Finally one obtains a sequence of
exit-times, $\{t_i(\epsilon)\}$, and one also records the labels $k_i=
\pm 1$, which distinguish the direction of the exit (up or down out of
a cell).

At the end of this construction, the trajectory is coded without
ambiguity, with the required accuracy $\epsilon$, by the sequence $\{
(t_i,k_i), \; i=1, \dots, M \}$, where $M$ is the total number of
exit-time events observed during the total time $T$.  Now, performing
a coarse-graining of the possible values assumed by $t(\epsilon)$ with
a resolution time $\tau_r$, we accomplish the goal of obtaining a
symbolic sequence.  One now studies the ``exit-time words'',
$\Omega^N_i$, of various lengths $n$: $\Omega^N_i (\epsilon,
\tau_r)=\left((\eta_i,k_i),(\eta_{i+1},k_{i+1}), \dots ,
(\eta_{i+N-1},k_{i+N-1}) \right)$, where $\eta_j$ labels the cell (of
width $\tau_r$) containing the exit-time $t_j$.  From the
probabilities of these words one evaluates the block entropies
(\ref{eq:block}) at the given time resolution, $H^{\Omega}_N(\epsilon,
\tau_r)$, and then the exit-time $(\epsilon, \tau_r)$-entropies:
\begin{equation}
h^\Omega(\epsilon, \tau_r) = \lim_{N \to \infty}
H^\Omega_{N+1}(\epsilon, \tau_r) - H^\Omega_N(\epsilon, \tau_r)\; .
\end{equation}
The limit of infinite time-resolution gives us the $\epsilon$-entropy
{\it per exit}, i.e.:
\begin{equation}
h^\Omega(\epsilon) = \lim_{\tau_r \to 0}
h^\Omega(\epsilon, \tau_r)\,.
\label{homegalim}
\end{equation}

The link between $h^{\Omega}(\epsilon)$ and the $\epsilon$-entropy
(\ref{def:eps}) can be obtained as follows.  We note that there is a
one-to-one correspondence between the (exit-time)-histories and the
$(\epsilon,\tau)$-histories (in the limit $\tau \to 0$) originating
from a given $\epsilon$-cell. The Shannon-McMillan theorem \cite{K57}
assures that the number of the typical $(\epsilon,\tau)$-histories of
length $N$, ${\cal N} (\epsilon,N)$, is such that: $\ln {\cal N}
(\epsilon,N) \simeq h(\epsilon) N \tau = h(\epsilon) T$. For the
number of typical (exit-time)-histories of length $M$, ${\cal M}
(\epsilon,M)$, we have: $\ln {\cal M} (\epsilon,M) \simeq
h^{\Omega}(\epsilon) M$. If we consider $T=M \langle t(\epsilon)
\rangle$, where $\langle t(\epsilon)\rangle =1/M\,\sum_{i=1}^M t_i$,
 we must obtain the same number of (very long)
histories. Therefore, from the relation $M = T/{\langle t(\epsilon)
\rangle}$
we obtain finally for the $\epsilon$-entropy per unit time:
\begin{equation}
h(\epsilon) = {M  h^\Omega(\epsilon) \over  T} =
\frac{h^\Omega(\epsilon)}{\langle t(\epsilon) \rangle}  
\simeq {h^\Omega(\epsilon,\tau_r) 
\over \langle t(\epsilon)\rangle }\,.
\label{epsent}
\end{equation}
Where the last equality is valid at least for small enough $\tau_r$
\cite{ABCFVV00a}.  In most of the cases, the leading
$\epsilon$-contribution to $h(\epsilon)$ in (\ref{epsent}) is given by
the mean exit-time $\langle t(\epsilon) \rangle$ and not by
$h^\Omega(\epsilon,\tau_r)$. Anyhow, the computation of
$h^\Omega(\epsilon,\tau_r)$ is compulsory in order to recover, e.g., a
zero entropy for regular (e.g. periodic) signals.

One can easily estimate an upper and a lower bound for $h(\epsilon)$
which can be computed in the exit time scheme \cite{ABCFVV00a}.  We
use the following notation: for given $\epsilon$ and $\tau_r$,
$h^\Omega(\epsilon,\tau_r) \equiv h^\Omega(\{ \eta_i,k_i \})$, and we
indicate with $h^\Omega(\{k_i\})$ and $h^\Omega(\{ \eta_i\})$
respectively the Shannon entropy of the sequence $\{k_i\}$ and
$\{\eta_i\}$.  By applying standard results of information theory
\cite{S48} one obtains the inequalities (see
\cite{ABCFVV00a,ABCFVV00b} for more details):
\begin{equation}
h^\Omega(\{ k_i \}) \leq h^\Omega(\{ \eta_i,k_i \})
\leq h^\Omega(\{ \eta_i\}) + h^\Omega(\{k_i\}).
\end{equation} 
Moreover, $h^\Omega(\{ \eta_i\}) \leq H^\Omega_1(\{ \eta_i\})$,
where $H^\Omega_1(\{\eta_i\})$ is the one-symbol
entropy of $\{ \eta_i\}$, (i.e. the entropy of the probability
distribution of the exit-times measured on the scale $\tau_r$) which
can be written as
$$
H^\Omega_1(\{ \eta_i\}) =  c(\epsilon) +
\ln \left( {\langle t(\epsilon)\rangle \over \tau_r} \right)\,\,,
$$
where $c(\epsilon) = -\int p(z)\ln p(z) {\mathrm d}z$, and $p(z)$ is
the probability distribution function of the rescaled exit-time
$z(\epsilon) = t(\epsilon)/\langle t(\epsilon)\rangle$.  Finally,
using the previous relations, one obtains the following bounds for the
$\epsilon$-entropy:
\begin{equation}
\label{bound-entro}
{h^\Omega(\{ k_i \}) \over \langle t(\epsilon) \rangle} \leq
h(\epsilon) \leq
{h^\Omega(\{ k_i \}) + c(\epsilon) + \ln
(\langle t(\epsilon)\rangle / \tau_r)
\over \langle t(\epsilon) \rangle} \,.
\end{equation}
Note that such bounds are relatively easy to compute and give a good
estimate of $h(\epsilon)$. In particular, as far as the scaling
behavior of $h(\epsilon)$ is concerned, the exit-time method allows
for very efficient and good estimates of the scaling exponent. The
reason is that at fixed $\epsilon$, $\langle t(\epsilon) \rangle$
automatically selects the typical time at that scale. Consequently, it
is not necessary to reach very large block sizes -- at least if
$\epsilon$ is not too small. So that the leading contribution is given
by $\langle t(\epsilon) \rangle$, and $h^{\Omega}(\epsilon,\tau_r)$
introduces, at worst, a sub-leading logarithmic contribution
$h^\Omega(\epsilon,\tau_r) \sim \ln (\langle t(\epsilon) \rangle
/\tau_r)$ (see Eq.~(\ref{bound-entro})).

In Ref.~\cite{ABCFVV00a,ABCFVV00b} one can found the details of the
derivation and some applications.

\section{Synthetic signals for turbulence}
\setcounter{equation}{0}
\label{app:synth}
In this Appendix we recall some recently introduced methods for the
generation of multi-affine stochastic signals~\cite{BBCPVV93,BBCCV98},
whose scaling properties are fully under control.  The first step
consists in generating a $1$-dimensional signal, and the second in
decorating it such as to build the most general $(d+1)$-dimensional
process, $v({\bf x},t)$, with given scaling properties in time and in
space.  \\ For the $1$-dimensional case there are at least two
different kind of algorithms. One is based on a dyadic decomposition
of the signal in a wavelet basis with a suitable assigned series of
stochastic coefficients \cite{BBCPVV93}. The second is based on a
multiplication of sequential Langevin-processes with a hierarchy of
different characteristic times \cite{BBCCV98}.  \\ The first procedure
suits particularly appealing for modeling of spatial turbulent
fluctuations, because of the natural identification between wavelets
and eddies in the physical space. The second one looks more
appropriate for mimicking the turbulent time evolution in a fixed
point of the space.
\subsection{Reproducing the spatial properties or the temporal ones }
\label{sec:synth.1}
A non-sequential algorithm for $1$-dimensional multi-affine signal in
$[0,1]$, $v(x)$, can be introduced as~\cite{BBCPVV93}:
\begin{equation}
v(x) = \sum_{n=1}^N\sum_{k=1}^{2^{(n-1)}} a_{n,k}\,
                   \varphi\!\left(\frac{x-x_{n,k}}{\ell_n}\right)
\label{diadic1}
\end{equation}
where we have a set of reference scales $\ell_n=2^{-n}$ and
$\varphi(x)$ is a wavelet-like function \cite{F92}, i.e. of zero mean
and rapidly decaying in both real space and Fourier-space.  The signal
$v(x)$ is built in terms of a superposition of fluctuations,
$\varphi((x-x_{n,k})/\ell_n)$ of characteristic width $\ell_n$ and
centered in different points of $[0,1]$, $x_{n,k} = (2k+1)/2^{n+1}$.
In \cite{BBCCV98} it has been proved that provided the coefficients
$a_{n,k}$ are chosen by a random multiplicative process, i.e.  the
daughter is given in terms of the mother by a random process,
$a_{n+1,k'} = X a_{n,k}$ with $X$ a random number identical,
independent distributed for any $\{n,k\}$, then the result of the
superposition is a multi-affine function with given scaling exponents,
namely:
$$
\langle \langle |v(x+R)-v(x)|^p \rangle\rangle \sim
R^{\,\zeta(p)}\,\,,
$$ 
with $\zeta(p) = -p/2 - \log_2 \langle X^p \rangle$ and $ \ell_N \leq
R \leq 1$.  In this Appendix $\langle \cdot \rangle$ indicates the
average over the probability distribution of the multiplicative
process.

Besides the rigorous proof, the rationale for the previous result is
simply that due to the hierarchical organization of the fluctuations,
one may easily see that the term dominating the expression of a
velocity fluctuation at scale $R$, in (\ref{diadic1}) is given by the
couple of indices $\{n,k\}$ such that $n \sim log_2(R)$ and $x \sim
x_{n,k}$, i.e.  $v(x+R)-v(x) \sim a_{n,k}$.  The generalization
(\ref{diadic1}) to d-dimension is given by:
$$
v({\bf x}) = \sum_{n=1}^N\sum_{k=1}^{2^{d(n-1)}} a_{n,k}\,
 \varphi\!\left(\frac{{\bf x}-{\bf x}_{n,k}}{\ell_n}\right)\,\,,
$$
where now the coefficients $\{a_{n,k}\}$ are given in terms of a
d-dimensional dyadic multiplicative process.  \\ On the other hand, as
previously said, sequential algorithms look more suitable for
mimicking temporal fluctuations. Let us now discuss how to construct
these stochastic multi-affine fields. With the application to
time-fluctuations in mind, we will denote now the stochastic
1-dimensional functions with $u(t)$. The signal $u(t)$ is obtained by
a superposition of functions with different characteristic times,
representing eddies of various sizes~\cite{BBCCV98}:
\begin{equation}
u(t)=\sum_{n=1}^N u_n(t) \; .
\label{eq:decomp}
\end{equation}
The functions $u_n(t)$ are defined by the multiplicative process
\begin{equation}
u_n(t)=g_n(t)x_1(t)x_2(t)\ldots x_n(t) \; ,
\label{def:mult}
\end{equation}
where the $g_n(t)$ are independent stationary random processes, whose
correlation times are supposed to be $\tau_n=(\ell_n)^\alpha$, where
$\alpha = 1-h$ (i.e. $\tau_n$ are the eddy-turn-over time at scale
$\ell_n$) in the quasi-Lagrangian frame of reference \cite{LPP97} and
$\alpha = 1$ if one considers $u(t)$ as the time signal in a given
point, and $\langle g_n^2 \rangle = (\ell_n)^{2h}$, where $h$ is the
H\"older exponent. For a signal mimicking a turbulent flow, ignoring
intermittency, we would have $h=1/3$.  Scaling will appear for all
time delays larger than the UV cutoff $\tau_N$ and smaller than the IR
cutoff $\tau_1$.  The $x_j(t)$ are independent, positive defined,
identical distributed random processes whose time correlation decays
with the characteristic time $\tau_j$. The probability distribution of
$x_j$ determines the intermittency of the process.

The origin of (\ref{def:mult}) is fairly clear in the context of fully
developed turbulence. Indeed we can identify $u_n$ with the velocity
difference at scale $\ell_n$ and $x_j$ with
$(\varepsilon_j/\varepsilon_{j-1})^{1/3}$, where $\varepsilon_j$ is
the energy dissipation at scale $\ell_j$ \cite{BBCCV98}.

The following arguments show, that the process defined according to
(\ref{eq:decomp},\ref{def:mult}) is multi-affine.  Because of the fast
decrease of the correlation times $\tau_j=(\ell_j)^\alpha$, the
characteristic time of $u_n(t)$ is of the order of the shortest one,
i.e., $\tau_n=(\ell_n)^\alpha$.  Therefore, the leading contribution
to the structure function $\tilde{S}_q(\tau) =
\langle\langle|u(t+\tau)-u(t)|^q\rangle\rangle$ with $\tau \sim
\tau_n$ stems from the $n$-th term in (\ref{eq:decomp}).  This can be
understood nothing that in $u(t+\tau)-u(t) = \sum_{k=1}^N
[u_k(t+\tau)-u_k(t)]$ the terms with $k \le n$ are negligible because
$u_k(t+\tau) \simeq u_k(t)$ and the terms with $k \ge n$ are
sub-leading.  Thus one has:
\begin{equation}
\tilde{S}_q(\tau_n) 
\sim \langle |u_n|^q \rangle \sim 
\langle |g_n|^q \rangle \langle x^q \rangle^n
\sim \tau_n^{\frac{h q}{\alpha} - \frac{\log_2\langle x^{q} \rangle}{\alpha}} 
\end{equation}
and therefore for the scaling exponents:
\begin{equation}
\zeta_q={h q \over \alpha} - {\log_2\langle x^{q} \rangle \over \alpha} \; .
\label{eq:zq}
\end{equation}
The limit of an affine function can be obtained when all the $x_j$ are
equal to $1$. A proper proof of these result can be found in
\cite{BBCCV98}.  \\ Let us notice at this stage that the previous
``temporal'' signal for $\alpha = 1 - h$ is a good candidate for a
velocity measurements in a Lagrangian, co-moving frame of reference
\cite{LPP97}.  Indeed, in such a reference frame the temporal
decorrelation properties at scale $\ell_n$ are given by the
eddy-turn-over times $\tau_n=(\ell_n)^{1-h}$.  On the other hand, in
the laboratory reference frame the sweeping dominates the time
evolution in a fixed point of the space and we must use as
characteristic times of the processes $x_n(t)$ the sweeping times
$\tau_n^{(s)} = \ell_n$, i.e., $\alpha=1$.
\subsection{Reproducing  both the spatial and the temporal  properties }
\label{sec:synth.2}

We have now all the ingredients to perform a merging of temporal and
spatial properties of a turbulent signal in order to define stochastic
processes able to reproduce in a realistic way both spatial and
temporal fluctuations (\ref{eq:aggiunta1})-(\ref{eq:aggiunta2}).  We
just have to merge in a proper way the two previous algorithms. \\ For
example, for a d-dimensional multi-affine field such as, say, one of
the three components of a turbulent field in a Lagrangian reference
frame we can use the following model:
\begin{equation}
v_L({\bf x},t) = \sum_{n=1}^N\sum_{k=1}^{2^{d(n-1)}} a_{n,k}(t)\,
 \varphi \! \left(\frac{{\bf x}-{\bf x}_{n,k}}{\ell_n}\right).
\label{lagrangian_field1}
\end{equation}
where the temporal dependency of $ a_{n,k}(t)$ is chosen following the
sequential algorithm while its spatial part are given by the dyadic
structure of the non-sequential algorithm.  In
(\ref{lagrangian_field1}) we have used the notation $v_L({\bf x},t)$
in order to stress the typical Lagrangian character of such a field.
\\ We are now able to guess a good candidate for the same field
measured in the laboratory-reference frame, i.e. where the time
properties are dominated by the sweeping of small scales by large
scales.  Indeed, in order to reproduce the sweeping effects one needs
that the centers $\{{\bf x}_{n,k}\}$ of the wavelets-like functions
move according a swept-dynamics.

To do so, let us define the Eulerian model:
\begin{equation}
\label{eulerian_field}
v_E({\bf x},t) = \sum_{n=1}^N\sum_{k=1}^{2^{d(n-1)}} a_{n,k}(t)
 \varphi\left(\frac{{\bf x}-{\bf x}_{n,k}(t)}{\ell_n}\right).
\end{equation}
where the difference with the previous definition is in the temporal
dependency of the centers of the wavelets, ${\bf x}_{n,k}(t)$.
According to the Richardson-Kolmogorov cascade picture, one assumes
that sweeping is present, i.e., ${\bf x}_{n,k}={\bf
x}_{n-1,k^\prime}+{\bf r}_{n,k}$ where $(n,k^\prime)$ labels the
``mother'' of the $(n,k)$-eddy and ${\bf r}_{n,k}$ is a stochastic
vector which depends on ${\bf r}_{n-1,k^\prime}$ and evolves with
characteristic time $\tau_n \propto (\ell_n)^{1-h} $.  Furthermore,
its norm is $O(\ell_n)$: $c_1 < |{\bf r}_{n,k}|/\ell_n < c_2$ where
$c_1$ and $c_2$ are constants of order one. \\ We now see that if we
measure in one fixed spatial point a fluctuations over a time delay
$\Delta t$, is like to measure a simultaneous fluctuations at scale
separation $R=U\Delta t$, i.e. due to the sweeping the main
contribution to the sum will be given by the terms with scale-index $n
= \log_2(R=U\Delta t)$ while the temporal dependency of the
coefficients $\{a_{n,k}(t)\}$ will be practically frozen on that time
scale.  Therefore, one has the proper spatial and temporal statistics,
see Ref.~\cite{ABCFVV00b} for details.  This happens because in
presence of the sweeping the main contribution is given by the
displacement of the center at large scale, i.e.  $\delta r_0 = |{\bf
r_0}(t+\Delta t) - {\bf r_0}(t)|\sim U \Delta t$, and the eddy
turnover time at scale $\ell_n$ is $O((\ell_n)^{1-h})$ always large
that the sweeping time $O(\ell_n)$ at the same scale.  \\ In the
previous discussion for the sake of simplicity we did not consider the
incompressibility condition.  However one can take into account this
constraint by the projection on the solenoidal space.  \\ In
conclusion we have a way to build up a synthetic signal with the
proper Eulerian (laboratory) properties, i.e. with sweeping, and also
with the proper Lagrangian properties.

\end{appendix}

\end{document}